\documentclass[aps,floatfix,prd,amsmath,amssymb,superscriptaddress,nofootinbib,showpacs,preprintnumbers,notitlepage]{revtex4-1}

\usepackage[english]{babel}
\usepackage{graphicx,subfigure}
\usepackage[]{hyperref} 
\usepackage[german]{datetime}
\usepackage[usenames,dvipsnames]{color}
\usepackage{upgreek}
\usepackage{todonotes}
\usepackage{ctable}
\usepackage{footnote}
\usepackage{units}
\usepackage{listings}
\usepackage{simplewick}
\usepackage{setspace}
\usepackage{amsmath} 
\usepackage{amsfonts} 
\usepackage{amssymb} 
\usepackage{multirow} 
\usepackage{booktabs} 
\usepackage{lmodern} 
\usepackage{slashed} 
\usepackage{makeidx} 
\usepackage{braket} 
\usepackage{fancyhdr}
\usepackage{url} 
\usepackage{cancel}

\newcommand{\nn}{\nonumber}
\newcommand{\be}{\begin{eqnarray}}
\newcommand{\ee}{\end{eqnarray}}
\newcommand{\beq}{\begin{equation}}
\newcommand{\eeq}{\end{equation}}

\newcommand{\mainz}{Institut f\"ur Kernphysik \& PRISMA$^+$ Cluster of Excellence,
Johannes Gutenberg Universit\"at, D-55099 Mainz, Germany}
\newcommand{\jlab}{Thomas Jefferson National Accelerator Facility, Newport News, VA 23606, USA}
\newcommand{\iu}{Center for Exploration of Energy and Matter, Indiana University, Bloomington, IN 47403, USA}
\newcommand{\msu}{Skobeltsyn Nuclear Physics Institute and Physics Department at Moscow State University, 
119899 Moscow, Russia}
\newcommand{\unam}{Instituto de Ciencias Nucleares, Universidad Nacional Aut\'onoma de M\'exico,  
Ciudad de M\'exico 04510, Mexico}
\newcommand{\ect}{European Centre for Theoretical Studies in Nuclear Physics and related Areas (ECT$^*$) and Fondazione Bruno Kessler, Villazzano (Trento), I-38123, Italy}
\newcommand{\genova}{INFN Sezione di Genova, Genova, I-16146, Italy}

\begin{document}

\title{Nucleon resonance contributions to unpolarised inclusive electron scattering}
\author{A.~N.~\surname{Hiller Blin}}
\affiliation{\mainz}
\author{V.~\surname{Mokeev}}
\affiliation{\jlab}
\author{M.~\surname{Albaladejo}}
\affiliation{\jlab}
\author{C.~Fern\'andez-Ram\'{\i}rez}
\affiliation{\unam}
\author{V.~\surname{Mathieu}}
\affiliation{\jlab}
\author{A.~\surname{Pilloni}}
\affiliation{\ect}
\affiliation{\genova}
\author{A.~\surname{Szczepaniak}}
\affiliation{\jlab}
\affiliation{\iu}

\collaboration{Joint Physics Analysis Center}
\author{V.~D.~\surname{Burkert}}
\affiliation{\jlab}
\author{V.~V.~\surname{Chesnokov}}
\affiliation{\msu}
\author{A.~A.~\surname{Golubenko}}
\affiliation{\msu}
\author{M.~\surname{Vanderhaeghen}}
\affiliation{\mainz}

\preprint{JLAB-THY-19-2913}

\begin{abstract}
The first CLAS12 experiments will provide high-precision
data on inclusive electron scattering observables at a photon virtuality $Q^2$ ranging from 0.05~GeV$^2$ to 12~GeV$^2$ and center-of-mass energies $W$ up to 4~GeV. In view of this endeavour, we present the modeling of the resonant contributions to the inclusive electron scattering observables. As input, we use the existing CLAS electrocoupling results obtained from exclusive meson electroproduction data off protons, and evaluate for the first time the resonant contributions based on the experimental results on the nucleon resonance electroexcitation. The uncertainties are given by the data and duly propagated through a Monte Carlo approach. In this way, we obtain estimates for the resonant contributions, important for insight into the nucleon parton distributions in the resonance region and for the studies of quark-hadron duality.
\end{abstract}
\date{\today}
\maketitle

\section{Introduction}
Studies of inclusive electron scattering off nucleons represent an important avenue in the exploration of the  nucleon structure. The global analysis~\cite{Harland-Lang:2014zoa,Dulat:2015mca,Accardi:2016qay,Alekhin:2017kpj,Ball:2017nwa} (see also the reviews in Refs.~\cite{Jimenez-Delgado:2013sma,Gao:2017yyd,Deur:2018roz}) has delivered detailed information on the parton distribution functions of the 
nucleon  (PDFs) for all quark flavors and gluons in the Bjorken variable range of $10^{-4}<x< 1$. In particular, the 
 Jefferson Lab inclusive electron scattering data~\cite{Osipenko:2003bu,Malace:2009kw,Christy:2007ve,Prok:2014ltt} had a major impact on the contemporary knowledge of the PDFs contributing to the large-$x$ data in the nucleon resonance region~\cite{Osipenko:2003bu,JLab:E00-002,Liang:2004tj}. 
Due to its near 4$\pi$ angular coverage, the CLAS detector offers a unique possibility of obtaining the inclusive structure functions $F_2(x,Q^2)$ in a very broad range of $x$ (or $W$) at a given 
 photon virtuality $Q^2$. This is particularly important in the resonance region, due to the presence of several resonant structures in the observable kinematics, which makes it challenging to use the typical interpolation of $F_2(x,Q^2)$ structure functions over $x$ at fixed $Q^2$. A compilation of the data for the unpolarized structure functions and  inclusive cross sections together with a tool for the interpolation 
  between bins is available online from the CLAS database~\cite{Golubenko:2019gxz,CLAS:DB,CLAS:SFDB}. This tool covers the  range $1.07~\text{GeV}\leq W\leq 4~\text{GeV}$ and $0.5~\text{GeV}^2\leq Q^2\leq 7~\text{GeV}^2$. 
   It uses the data and fits thereof~\cite{Osipenko:2003bu,Christy:2007ve} for the interpolation and extrapolation to regions outside the data coverage. The structure function $F_1(x,Q^2)$ was obtained from the data on $F_2(x,Q^2)$, assuming the parametrization in Ref.~\cite{Ricco:1998yr,Osipenko:2003bu} for the longitudinal over transverse cross section ratio $R_{LT}$. In this work, we update $R_{LT}$ to the more recent parametrization described in Ref.~\cite{Tomalak:2015hva}, based on data from the ZEUS and H1 experiments~\cite{Whitlow:1990gk,Dasu:1993vk,Arneodo:1996qe}. The online tool is particularly useful for the analyses of the  CLAS12 experiments, which will reach the largest $Q^2$ coverage ever achieved in the resonance region, also for the inclusive data~\cite{Burkert:2018nvj}.

There is a strong interest in studying the structure functions at large $x$ within the resonance region, since this range is dominated by contributions from valence quarks and there are 
 several theoretical predictions 
  for the PDF's in the limit of $x\to 1$ that need to be validated~\cite{Feynman:1972xm,Farrar:1975yb,Close:1988br,Leader:2001kh,Nocera:2014uea}. Furthermore, recent 
   developments of novel approaches  to PDFs
using Euclidean concepts~\cite{Ji:2013dva,Radyushkin:2017cyf,Ma:2017pxb,Lin:2017snn} make it possible to calculate in lattice QCD quantities that should converge to the measured PDFs. 
The PDFs in the resonance region were also evaluated 
 with the help of quark-hadron duality~\cite{Bloom:1970xb,Melnitchouk:2005ye,Melnitchouk:2005zr,Malace:2009kw,Christy:2011cv}, which relates the inclusive structure functions averaged over 
  individual resonance widths to the continuum  evaluated using the 
   deep inelastic scattering (DIS) region. Quark-hadron duality has been shown to
approximately work even at $Q^2$ as low as 1~GeV$^2$~\cite{Malace:2009kw}.

 In the past years, the CLAS experiments on exclusive meson electroproduction~\cite{Aznauryan:2011qj,Park:2014yea,Mokeev:2015moa,Mokeev:2016hqv,Burkert:2019bhp,Mokeev:2018zxt,Burkert:2019opk}  coupled with sophisticated 
 analysis models~\cite{Aznauryan:2002gd,Mokeev:2008iw,Aznauryan:2009mx,Mokeev:2012vsa,Mokeev:2015lda} 
  enabled to determine individual nucleon resonance contributions to electro-production by measuring,
  for the first time, nucleon resonance transverse $A_{1/2}$, $A_{3/2}$ and longitudinal $S_{1/2}$ electroexcitation amplitudes (also referred to as 
   the $\gamma_{v}pN^*$ electrocouplings).

As of now, the photocouplings of most of the excited nucleon states in the mass range up to 2~GeV have been well determined~{\cite{Dugger:2009pn,Golovatch:2018hjk}}. Furthermore, the $\gamma_{v}pN^*$ electrocouplings in the mass range up to 1.8~GeV were
determined by CLAS for a $Q^2$ range up to 5.0~GeV$^2$ from the $N\,\pi$~\cite{Aznauryan:2009mx,Park:2014yea}, $N\,\eta$~\cite{Thompson:2000by,Denizli:2007tq,Dalton:2008aa} and $\pi^+\pi^- p$~\cite{Mokeev:2008iw,Mokeev:2012vsa,Mokeev:2015lda} channels, and have been made available online~\cite{CLAS:coupsDB}. The consistency of the results in the different channels supports their extraction~\cite{Mokeev:2015lda}.  In addition, substantial evidence of a new baryon state $N^\prime(1720)~3/2^+$ has been found recently~\cite{Mokeev:2015moa,Burkert:2019opk}.

It is of ever growing importance to consistently include the  effects of the resonant and non-resonant (background) contributions into a single framework. This is possible with a combined study of exclusive and inclusive electron scattering data: the exclusive reactions offer us insight into the $Q^2$ evolution of the resonance electrocouplings; these can then be used as input for the computation of the resonant contributions to the inclusive cross sections.

 The availability of  $\gamma_{v}pN^*$ electrocoupling 
data on individual nucleon resonances 
makes it possible to evaluate the 
resonant contributions to the inclusive  electron scattering observables, which is the goal of the present work. 
We use a relativistic Breit-Wigner ansatz to estimate the resonant contributions to the inclusive electron-scattering unpolarized cross sections, which relate to the unpolarized structure functions. As input, the known masses and widths of the resonances are used, as well as the experimental information on electrocouplings from CLAS. This allows us to single out the resonant contributions to the inclusive electron scattering observables. We emphasize that this information is obtained  from the exclusive meson electroproduction data off protons and is  independent of the inclusive measurements. Therefore, it can also be used to estimate the 
non-resonant part of the inclusive electron scattering. 
In a next step, as one moves towards higher $W$ (lower $x$) this will allow to determine the transition between resonant and non-resonant contributions and consequently between resonance bound valence quarks and asymptotically free partons. 
 The results of the present work are very timely, since with the CLAS12 experiments the input from exclusive reactions will be extended further, and at the same time more data will be available on inclusive reactions, with higher precision.
 
The paper is organized as follows. 
In Sec.~\ref{S:form}, we  discuss the formalism on using exclusive reactions to estimate the resonant contributions to inclusive electron scattering. We show the results for the resonant contributions to the structure function $F_2$ from the inclusive CLAS data in Sec.~\ref{S:res}. There, we also show the resonant contributions to the transverse, longitudinal and unpolarized full inclusive cross sections, which in the present framework can be duly evaluated from the transverse and longitudinal $\gamma_{v}pN^*$ electrocoupling values without the need of parametrizing $R_{LT}$. In Sec.~\ref{S:sum}, we summarize our findings and discuss their applicability, also in view of the upcoming  experimental results.

\section{Formalism}\label{S:form}
In order to describe the contributions of the $N^*$ and $\Delta^*$ resonances to the observables in inclusive $e^-p$ scattering, we include the information on the $\gamma_{v}pN^*$ electrocouplings of the resonances with masses below 1.8~GeV from CLAS~\cite{CLAS:coupsDB,CLAS:coups}. Apart from the resonances whose existence is certain (marked as four stars  in the Review of Particle Physics (RPP)~\cite{Tanabashi:2018oca}), we include the new $N^\prime(1720)~3/2^+$ state~\cite{Mokeev:2015moa,Burkert:2019opk}, since the used CLAS electrocouplings were obtained accounting for its contribution. 
The resonances and their properties, as used in this work, are listed in Table~\ref{T:N****}.
\begin{table}[htp]
\begin{center}
\begin{tabular}{cccccccc}
\multirow{2}{*}{$N^*$}&$M_r$&$\Gamma_r$&\multirow{2}{*}{$L_r$}&\multirow{2}{*}{$\beta_{\pi N}$}&\multirow{2}{*}{$\beta_{\eta N}$}&\multirow{2}{*}{$\beta_{r.}$}&$X$\\
&[MeV]&[MeV]&&&&&[GeV]\\\hline
$\Delta(1232)~3/2^+$&1232&117&1&{1.00}&0&0&---\\
$N(1440)~1/2^+$&1430&350&1&{0.65}&0&0.35&0.3\\
$N(1520)~3/2^-$&1515&115&2&{0.60}&0&0.40&0.1\\
$N(1535)~1/2^-$&1535&150&0&{0.45}&0.42&0.13&0.5\\
$\Delta(1620)~1/2^-$&1630&140&0&{0.25}&0&0.75&0.5\\
$N(1650)~1/2^-$&1655&140&0&{0.60}&0.18&0.22&0.5\\
$N(1675)~5/2^-$&1675&150&2&{0.40}&0&0.60&0.5\\
$N(1680)~5/2^+$&1685&130&3&{0.68}&0&0.32&0.2\\
$\Delta(1700)~3/2^-$&1700&293&2&{0.10}&0&0.90&0.22\\
$N(1710)~1/2^+$&1710&100&1&{0.13}&0.30&0.57&0.5\\
$N(1720)~3/2^+$&1748&114&1&{0.14}&0.04&0.82&0.5\\
$N^\prime(1720)~3/2^+$&1725&120&1&0.38&0&0.62&0.5
\end{tabular}
\caption{Values used for resonance masses, widths, branching fractions $\beta$ and quantum numbers, based on the RPP~\cite{Tanabashi:2018oca}, with modifications to the $N(1720)~3/2^+$ and $\Delta(1700)~3/2^-$ states due to the inclusion of the new  $N^\prime(1720)~3/2^+$ state. See the text for discussion of the residual branching fractions $\beta_r$.  The values of the inverse radii $X$ for each resonance are also shown, following Refs.~\cite{Aznauryan:2002gd}.}
\label{T:N****}
\end{center}
\end{table}

 The resonant contributions to the transverse ($\sigma_{T}^R$) and longitudinal ($\sigma_{L}^R$) inclusive virtual photon-proton cross sections from a resonance of mass $M_r$, total width at the resonant point $\Gamma_r=\Gamma_\text{tot}(W=M_r)$ and spin $J_r$ can be described using the  Breit-Wigner formula~\cite{Mokeev:2012vsa}
\begin{align}
\sigma_{T,L}^R&(W,Q^2)=\frac{\pi}{q_\gamma^2}\sum_{N^*}(2J_r+1)\frac{M_r^2\Gamma_\text{tot}(W){\Gamma_\gamma}^{T,L}(M_r,Q^2)}{(M_r^2-W^2)^2+M_r^2\Gamma_\text{tot}^2(W)},\label{Eq:BW}
\end{align}
with the following kinematics
\begin{align}
q_\gamma=\sqrt{Q^2+E^2_\gamma},\quad
E_\gamma=\frac{W^2-Q^2-M_N^2}{2W},\quad
K=\frac{W^2-M_N^2}{2W}. 
\end{align}
Here $E_\gamma$ and $q_\gamma$ are the 
 virtual photon energy and magnitude  of its three-momentum in the center-of-mass frame, respectively, and $K$ is the equivalent photon energy. 
As usual, $q^2=-Q^2$ is the 4-momentum squared of the virtual photon, while $W$ is the virtual photon-proton system's  center-of-mass energy. The resonance electromagnetic decay widths to the final states with transversely  ($\Gamma_\gamma^T$) and longitudinally ($\Gamma_\gamma^L$) polarized photons at the resonant point are given by
\begin{align}
\Gamma_\gamma^T(W=M_r,Q^2)&=\frac{q^2_{\gamma,r}(Q^2)}{\pi}\frac{2M_N}{(2J_r+1)M_r}\left(|A_{1/2}(Q^2)|^2+|A_{3/2}(Q^2)|^2\right),\nn\\
\Gamma_\gamma^L(W=M_r,Q^2)&=2\frac{q^2_{\gamma,r}(Q^2)}{\pi}\frac{2M_N}{(2J_r+1)M_r}|S_{1/2}(Q^2)|^2,\label{Eq:EMWidths}
\end{align}
with $q_{\gamma,r}=\left.q_{\gamma} \right|_{W=M_r}$. The electrocouplings $A_{1/2}(Q^2)$, $A_{3/2}(Q^2)$ and $S_{1/2}(Q^2)$ are taken from the CLAS results listed in~\cite{CLAS:coupsDB,CLAS:coups}. For consistency with the electrocoupling normalization, the electrocouplings of the $\Delta(1620)~1/2^-$, $\Delta(1700)~3/2^-$, $N(1720)~3/2^+$ and $N^\prime(1720)~3/2^+$ resonances, extracted from double-pion  electroproduction off protons, need to include an additional factor $q_\gamma/K$ in both the electromagnetic widths of Eq.~\ref{Eq:EMWidths}, and a factor 1/2 in the longitudinal width.

In order to compute the energy dependence of the resonance total decay width, we split it into three pieces, to take 
 into account the decays into the two main two-body channels, $\pi N$ and $\eta N$, and the remainder, which includes $\pi\pi N$ and all other final states. 
 The hadronic decay widths to these final states at the resonant points are computed as the products of the total decay widths from the RPP~\cite{Tanabashi:2018oca}  and the respective branching fractions $\beta$, summarized in Table~\ref{T:N****}. For the resonance decays into the $\pi N$ and $\eta N$ final states, we use the central values of the respective branching fractions employed in the extraction of the $\gamma_{v}pN^*$ electrocouplings from exclusive $\pi N$ electroproduction data~\cite{Aznauryan:2009mx,Park:2014yea}. 
The branching fractions for the decays into the remaining channels $\beta_\text{r.}$ are evaluated as
\begin{align}
\beta_\text{r.}&= 1-\beta_{\pi N}-\beta_{\eta N}.
\end{align}
For those resonances which decay preferentially into the $\pi\pi N$ final states, $\Delta(1620)~1/2^-$, $\Delta(1700)~3/2^-$, $N(1720)~3/2^+$, the values of $\beta_\text{r.}$ obtained in this way are in good agreement with the branching fractions for resonance decays into the $\pi\pi N$ final states used in the extraction of the $\gamma_{v}pN^*$ electrocouplings from $\pi^+\pi^-p$ CLAS data~\cite{Mokeev:2015lda,Mokeev:2015moa,Mokeev:2016hqv}. Furthermore, as mentioned above, we include the candidate baryon state $N^\prime(1720)~3/2^+$, suggested in the analysis of the CLAS $\pi^+\pi^-p$ photo-/electroproduction data off protons~\cite{Mokeev:2015moa}. It is needed in order to enable describing the data in the third resonance region such that the masses and the hadronic decay parameters are independent of $Q^2$. For this state, the branching fraction for the decay into the $\pi\pi N$ final state is available from the analysis in Ref.~\cite{Mokeev:2015moa}. We assume the remaining decay width for this resonance to be saturated by $\pi N$. Due to the different hadronic decays of the conventional $N(1720)~3/2^+$ and the $N^\prime(1720)~3/2^+$ candidate, the interference between these states is negligible.

The $W$-dependence of the total and partial resonance decay widths is   
determined by the centrifugal barrier penetration and can be parametrized as ~\cite{Aznauryan:2002gd}
\begin{align}
\Gamma_\text{tot}(W)&=\Gamma_{\pi N}(W)+\Gamma_{\eta N}(W)+\Gamma_\text{r.}(W),\\
\intertext{where}
\Gamma_{\pi(\eta) N}(W)&=\Gamma_r\, \beta_{\pi(\eta) N}\left(\frac{p_{\pi(\eta)}(W)}{p_{\pi(\eta)}(M_r)}\right)^{2L_r+1}\left(\frac{X^2+p_{\pi(\eta)}(M_r)^2}{X^2+p_{\pi(\eta)}(W)^2}\right)^{L_r},\nn\\
\Gamma_\text{r.}(W)&=\Gamma_r\,\beta_\text{r.}\left(\frac{p_{\pi\pi}(W)}{p_{\pi\pi}(M_r)}\right)^{2L_r+4}\left(\frac{X^2+p_{\pi\pi}(M_r)^2}{X^2+p_{\pi\pi}(W)^2}\right)^{L_r+2},
\end{align}
and
\begin{align}
p_{\pi(\eta)}(W)&=\sqrt{E_{\pi(\eta)}^2(W)-m_{\pi(\eta)}^2},\nn\\
 E_{\pi(\eta)}(W)&=\frac{W^2+m_{\pi(\eta)}^2-M_N^2}{2W},\nn\\
p_{\pi\pi}(W)&=\sqrt{E_{\pi\pi}^2(W)-4m_\pi^2},\nn\\
 E_{\pi\pi}(W)&=\frac{W^2+4m_\pi^2-M_N^2}{2W}.
\end{align}
Note that an effective parametrization of the multibody decays is used~\cite{Aznauryan:2002gd}. 
The values of $L_r$ are shown in Table~\ref{T:N****}, as are the values of $X$, which are taken from the best fit in Refs.~\cite{Aznauryan:2002gd}. An exception is made for the well-isolated $\Delta(1232)~3/2^+$ resonance, whose width is fully described by the $\pi N$ decay in the following way~\cite{Mokeev:2012vsa}:
\begin{align}
\Gamma_{\pi N}(W)&=\Gamma_r\frac{M_r}{W}\frac{J_L^2[R\,p_\pi(M_r)]+N_L^2[R\,p_\pi(M_r)]}{J_L^2[R\,p_\pi(W)]+N_L^2[R\,p_\pi(W)]},
\end{align}
where the interaction radius $R$ was set to 1~fm. The functions $J_L$ and $N_L$ are the conventional Bessel and Neumann functions. In terms of the resonant electroproduction cross sections the unpolarized inclusive cross section is 
 given by~\cite{Christy:2007ve}
\begin{align} 
\label{xs}
\sigma_U^R(W,Q^2)&=\sigma_T^R(W,Q^2)+\epsilon_T\sigma_L^R(W,Q^2),\\
 \epsilon_T &=
 \left(1+ 2\,\frac{\nu^2+Q^2}{Q^2}\tan^2\frac{\theta_e}{2}\right)^{-1},\label{SigmaU1}\\ 
 \intertext{where the electron scattering angle $\theta_e$ is written in terms of the electron beam energy $E_\text{b}$ as}
\sin^2\frac{\theta_e}{2}&=\frac{Q^2}{4E_\text{b}(E_\text{b}-\nu)},\\
\intertext{and $\nu$ is the energy transferred by the virtual photon}
\nu&=\frac{W^2-M_N^2+Q^2}{2M_N}.
\end{align}
The transverse polarization parameter of the virtual photon $\epsilon_{T}$ is fully determined by the electron scattering kinematics.  Comparing Eq.~(\ref{xs}) with the standard definition of the structure functions one finally obtains~\cite{Drechsel:2002ar}
\begin{align}
\label{sf} 
F_1^R(x,Q^2)&=\frac{K W}{4\pi^2\alpha}\sigma_T^R,\nn\\
F_2^R(x,Q^2)&=\frac{K W}{4\pi^2\alpha}\frac{2x}{1+\frac{Q^2}{\nu^2}}
\left(\sigma_T^R +\sigma_L^R\right)= \frac{K W}{4\pi^2\alpha}\frac{2x}{1+\frac{Q^2}{\nu^2}} \frac{1+R_{LT}}{1+\epsilon_T R_{LT}} \sigma_U^R,
\end{align}
where $x=\frac{Q^2}{2M_N\nu}$ and $R_{LT}=\sigma_L/\sigma_T$.

\begin{figure*}
\includegraphics[width=.3\textwidth]{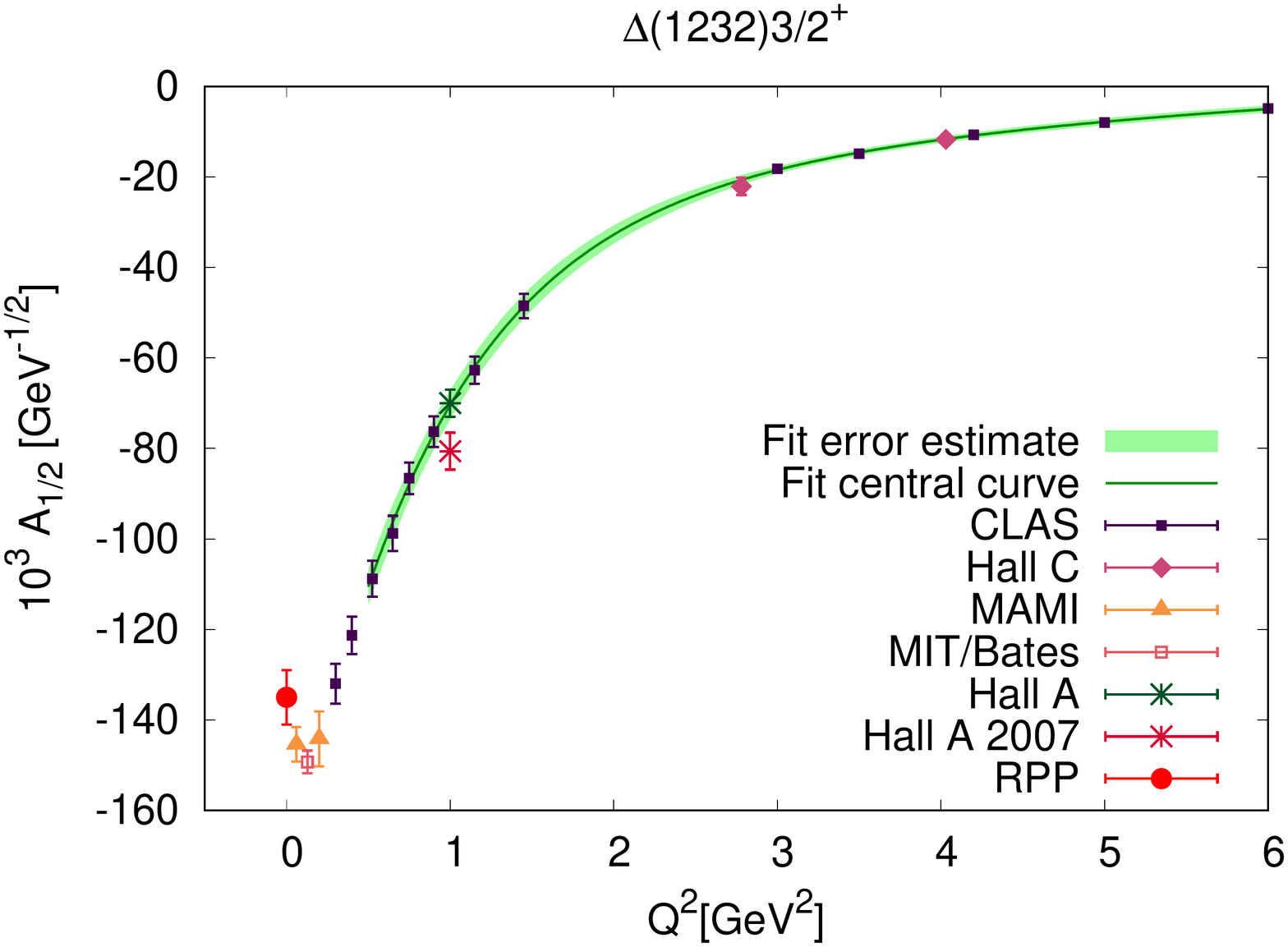}
\includegraphics[width=.3\textwidth]{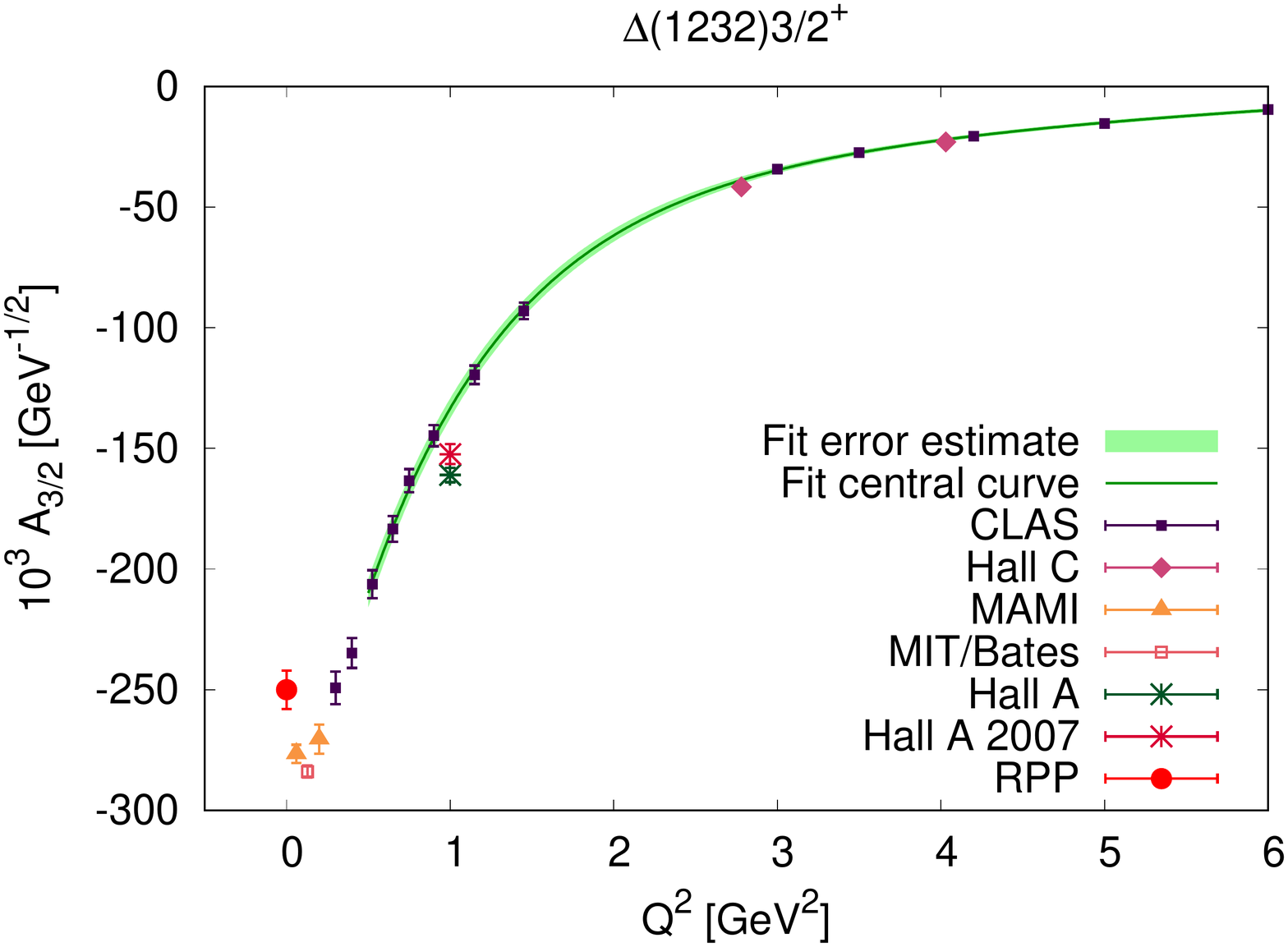}
\includegraphics[width=.3\textwidth]{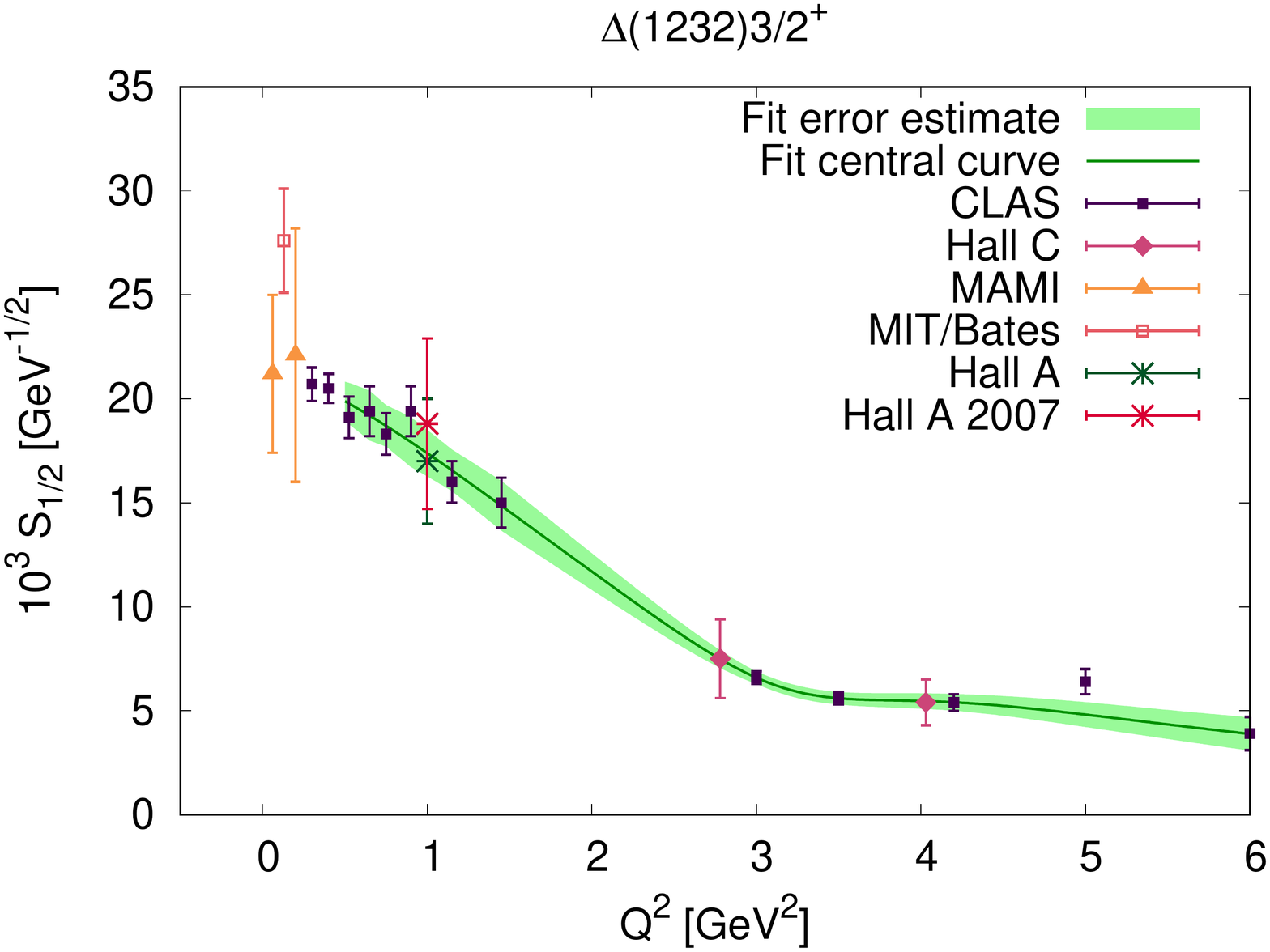}\\
\includegraphics[width=.3\textwidth]{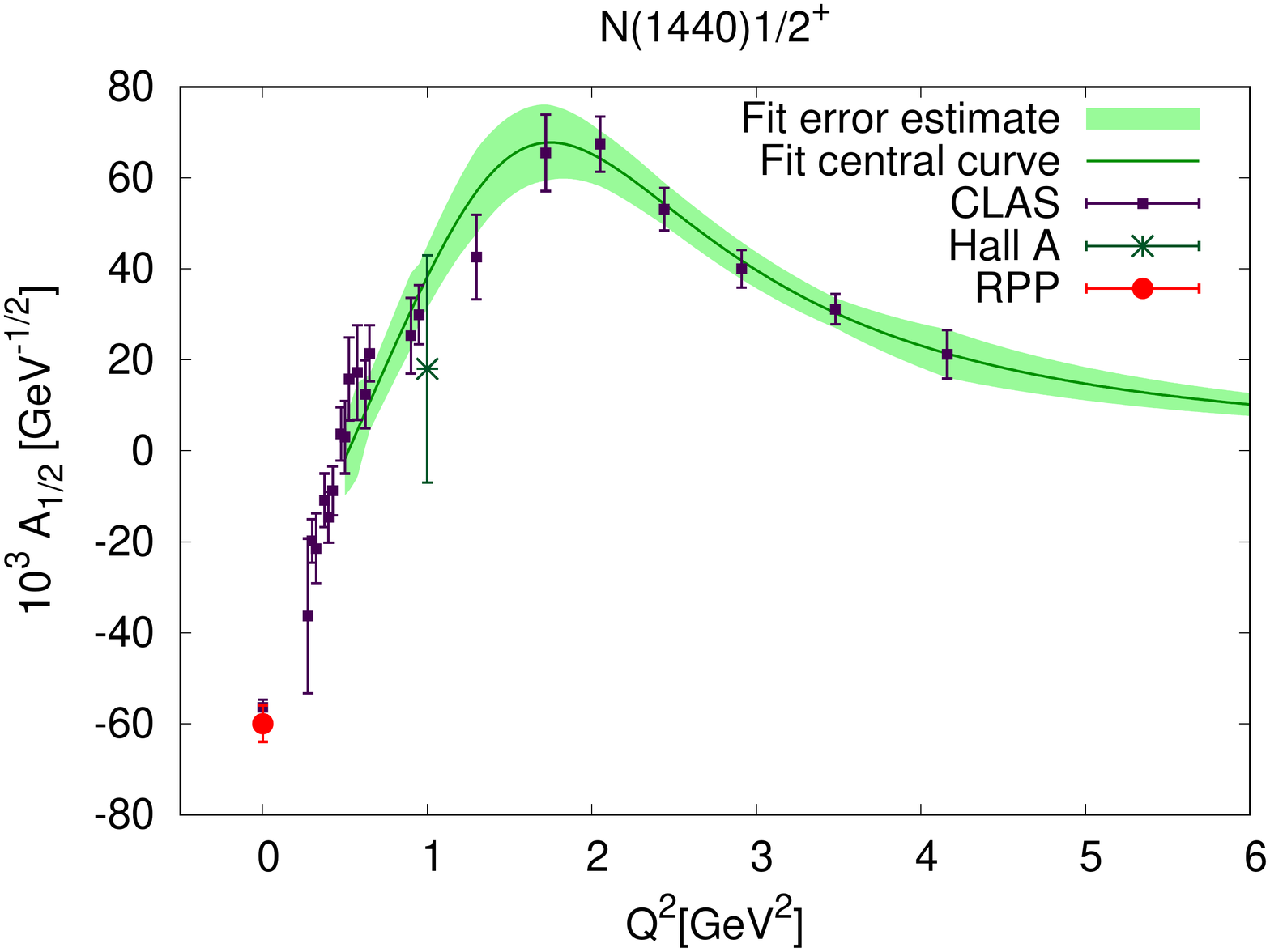}
\includegraphics[width=.3\textwidth]{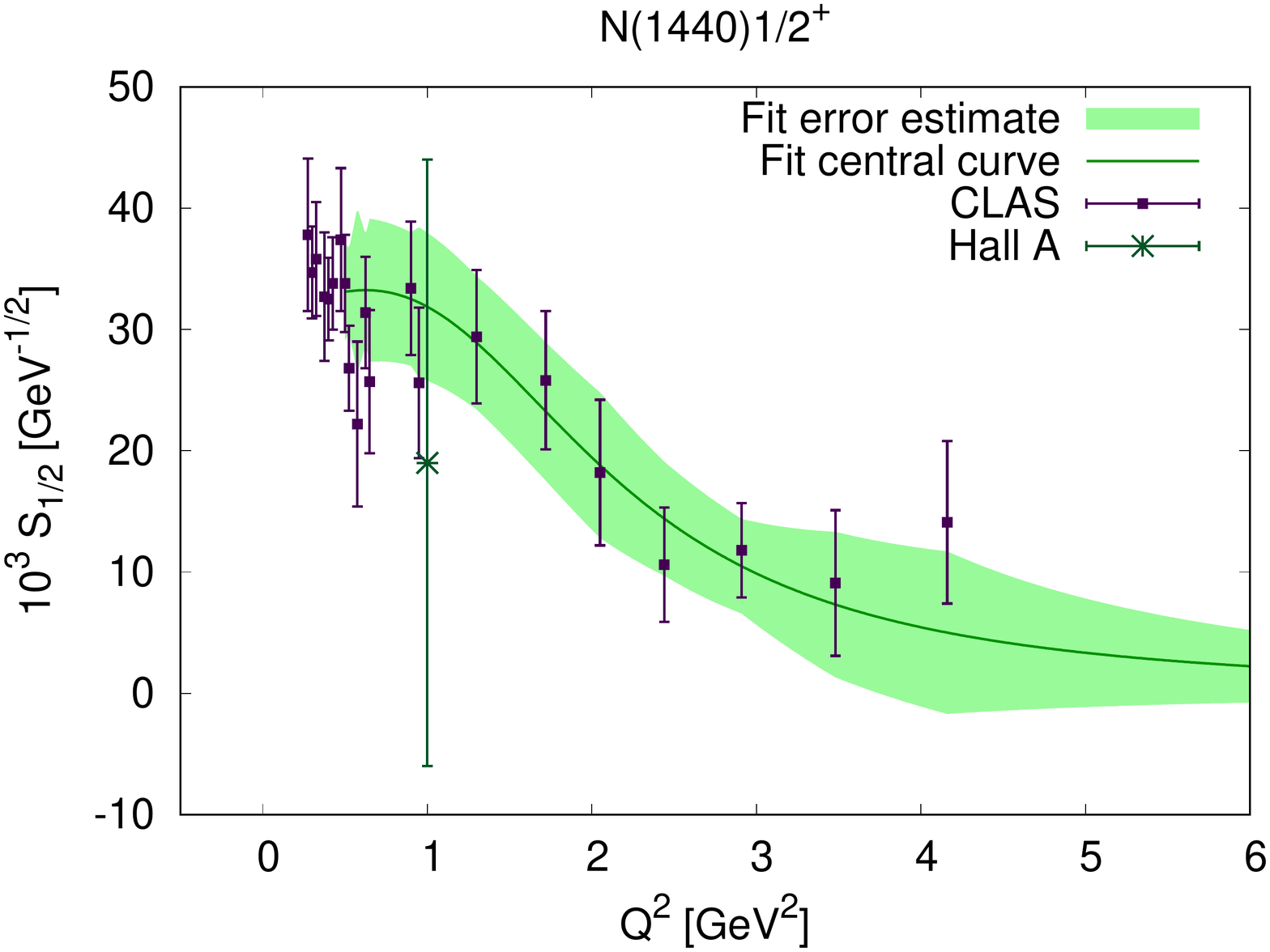}\\
\includegraphics[width=.3\textwidth]{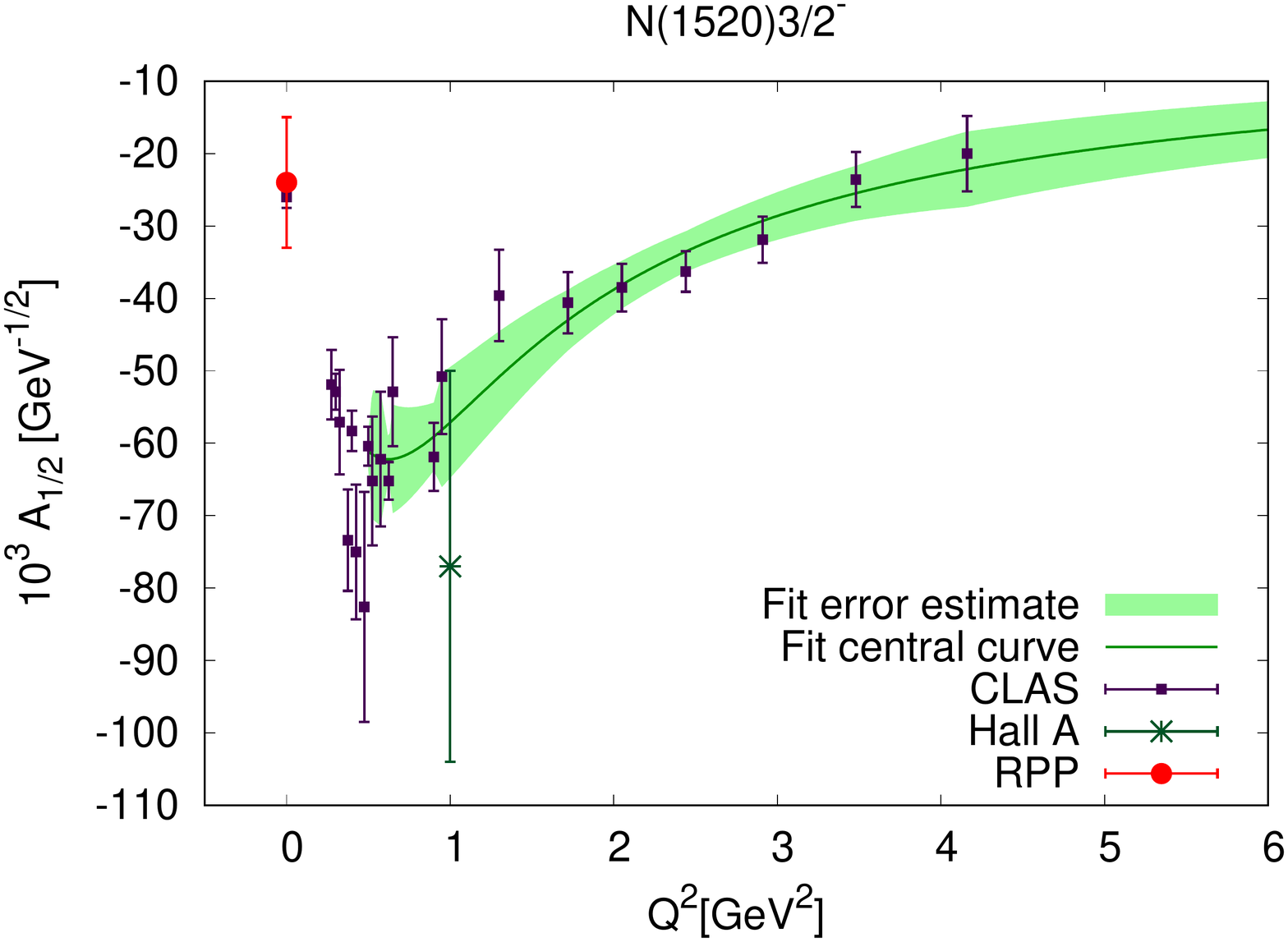}
\includegraphics[width=.3\textwidth]{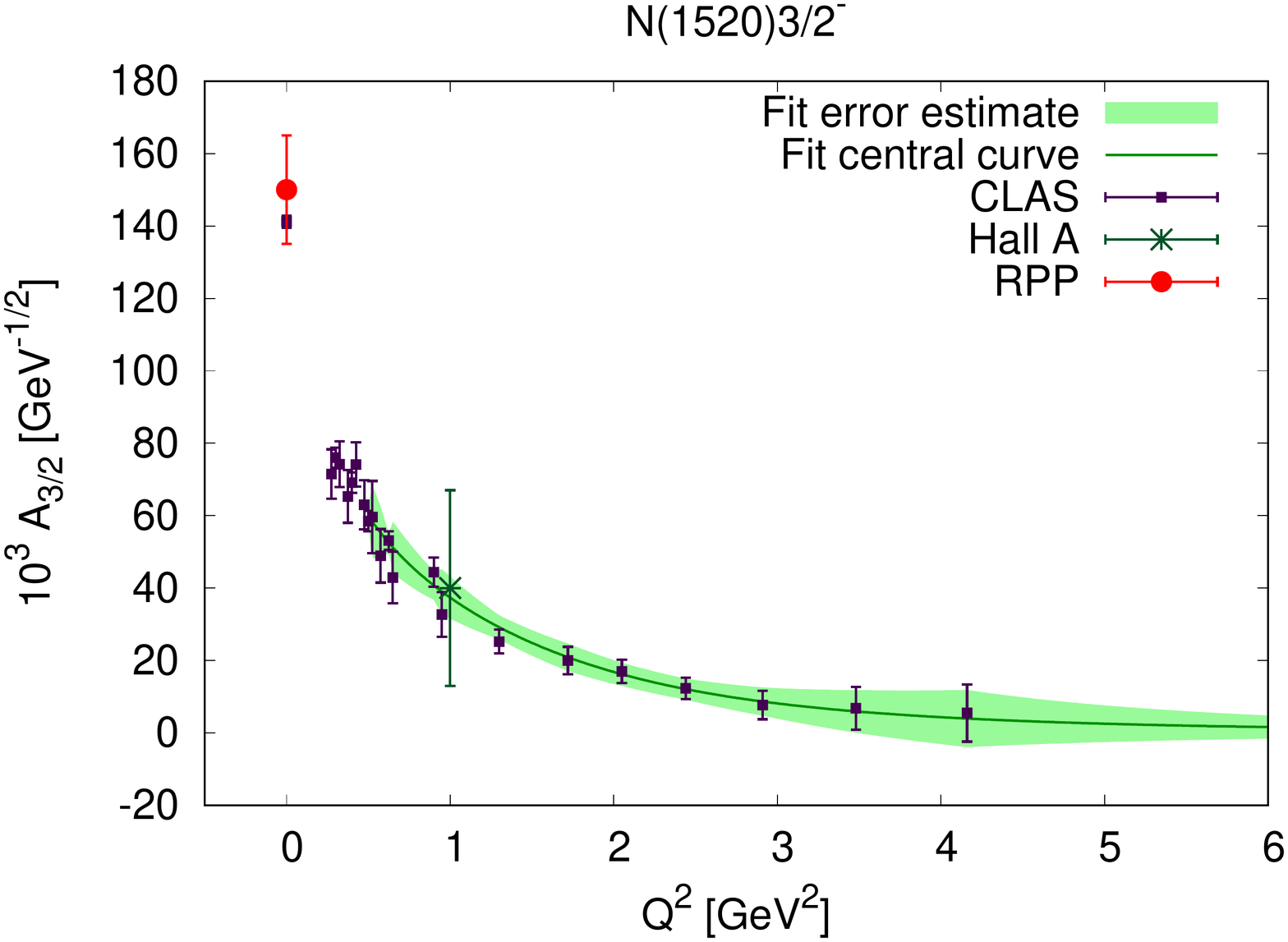}
\includegraphics[width=.3\textwidth]{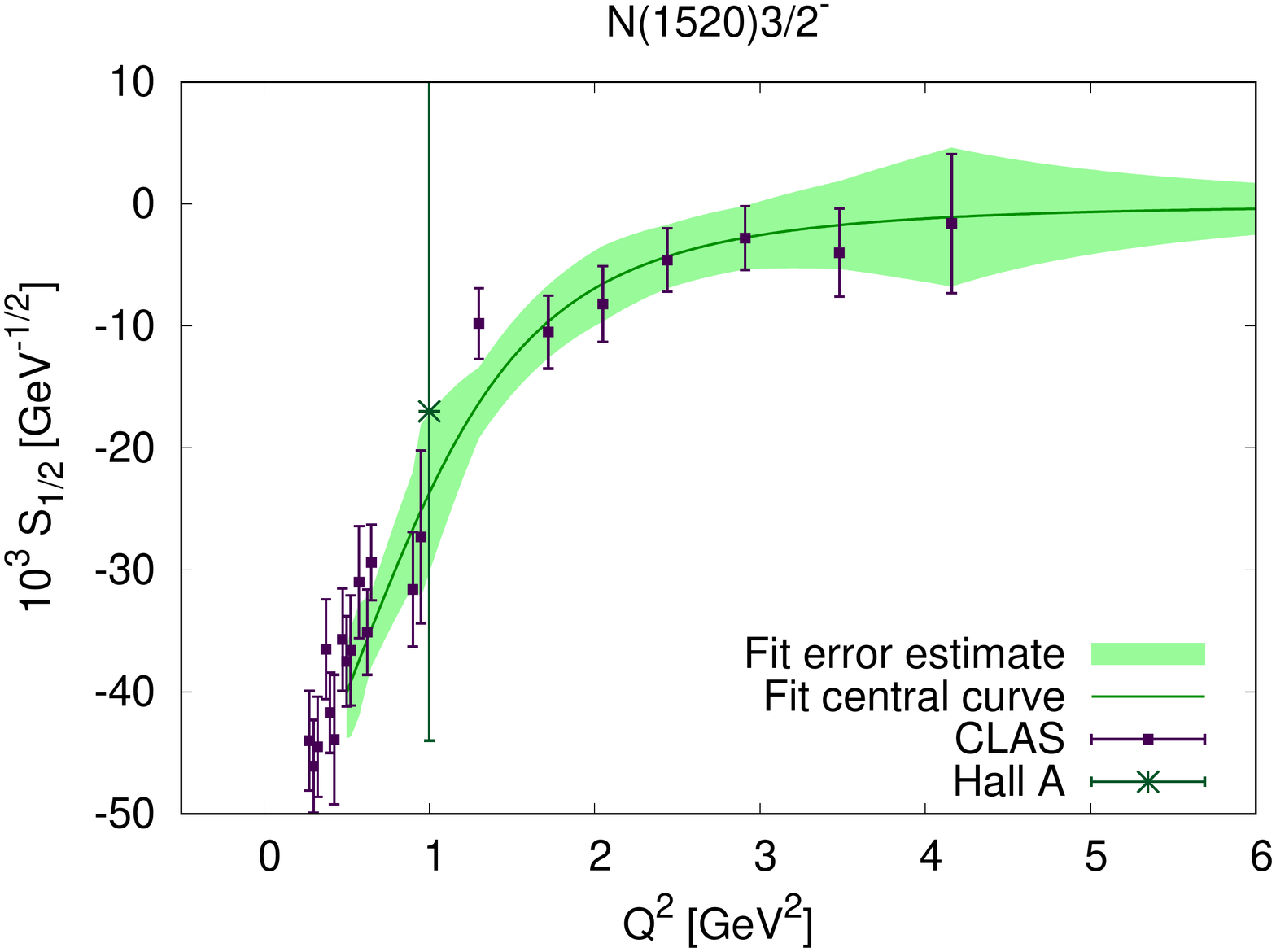}\\
\includegraphics[width=.3\textwidth]{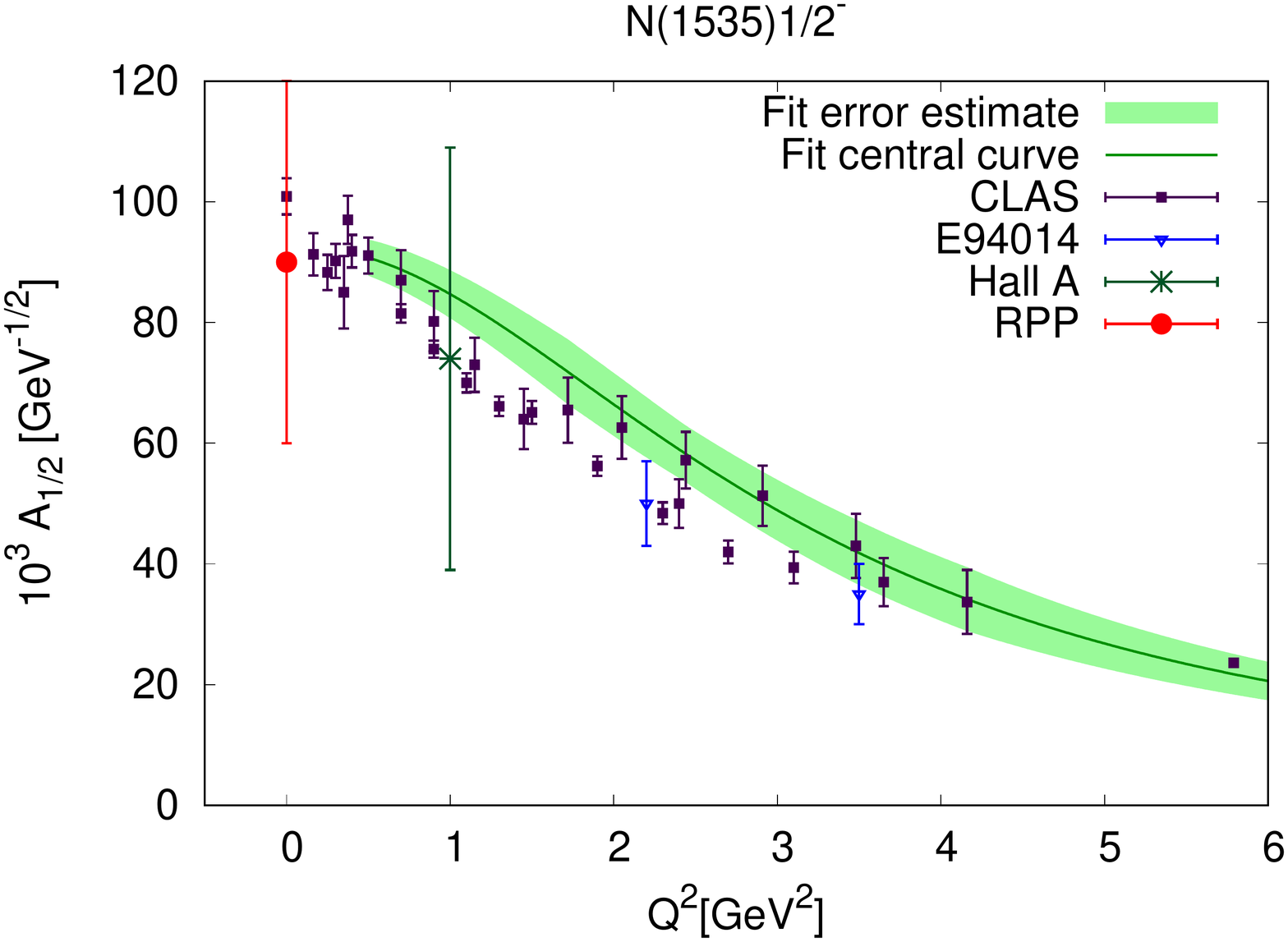}
\includegraphics[width=.3\textwidth]{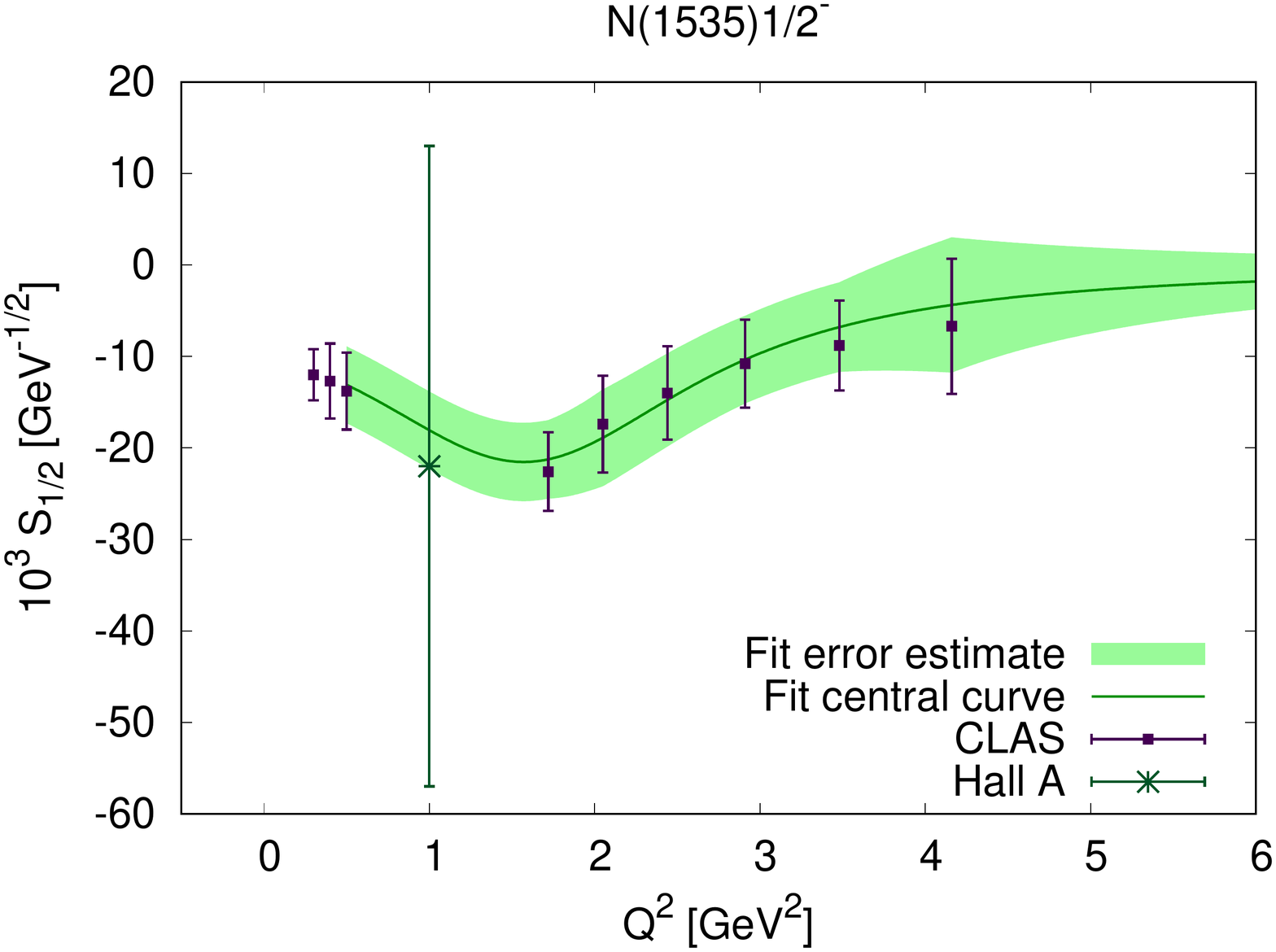}\\
\includegraphics[width=.3\textwidth]{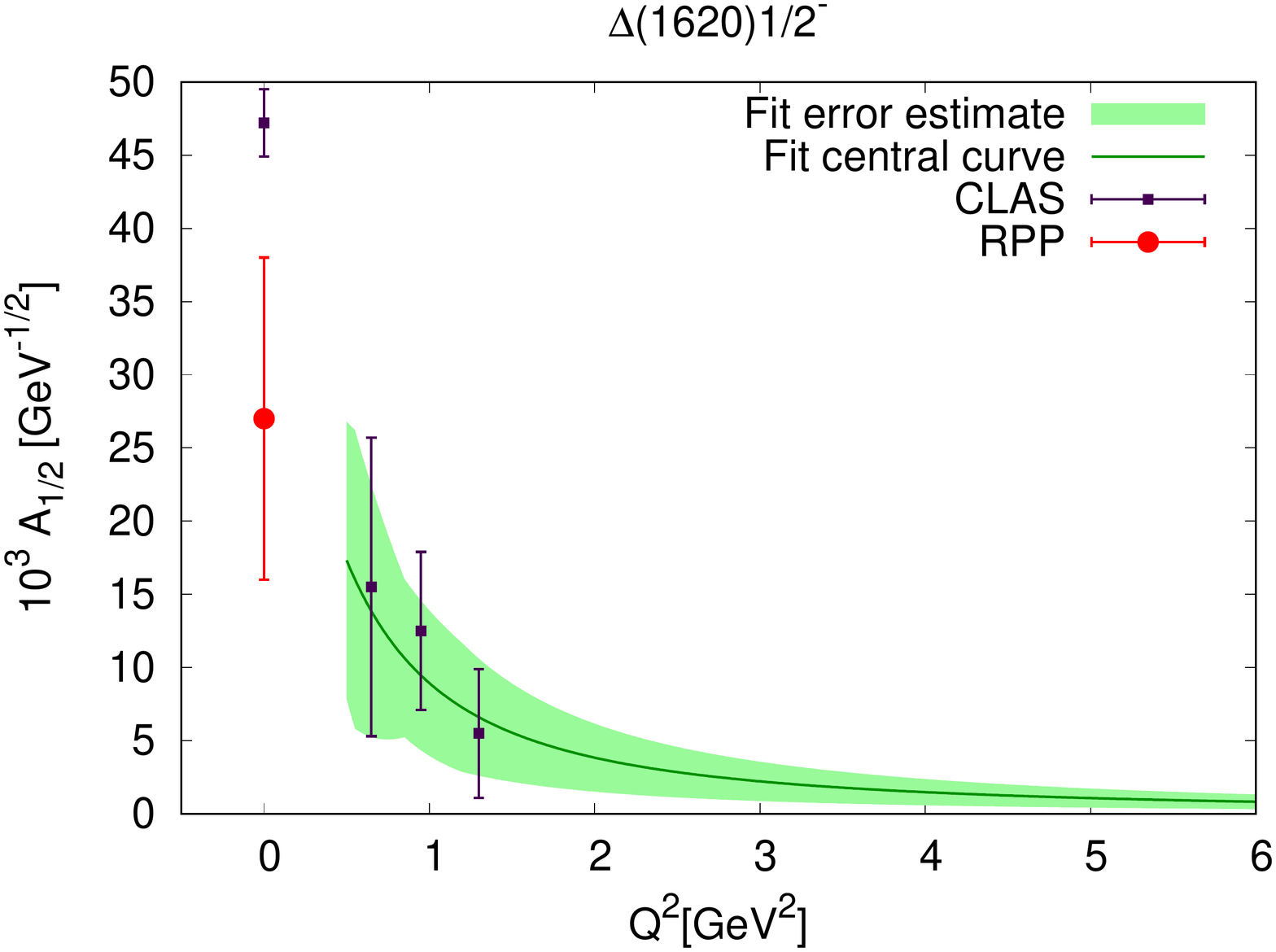}
\includegraphics[width=.3\textwidth]{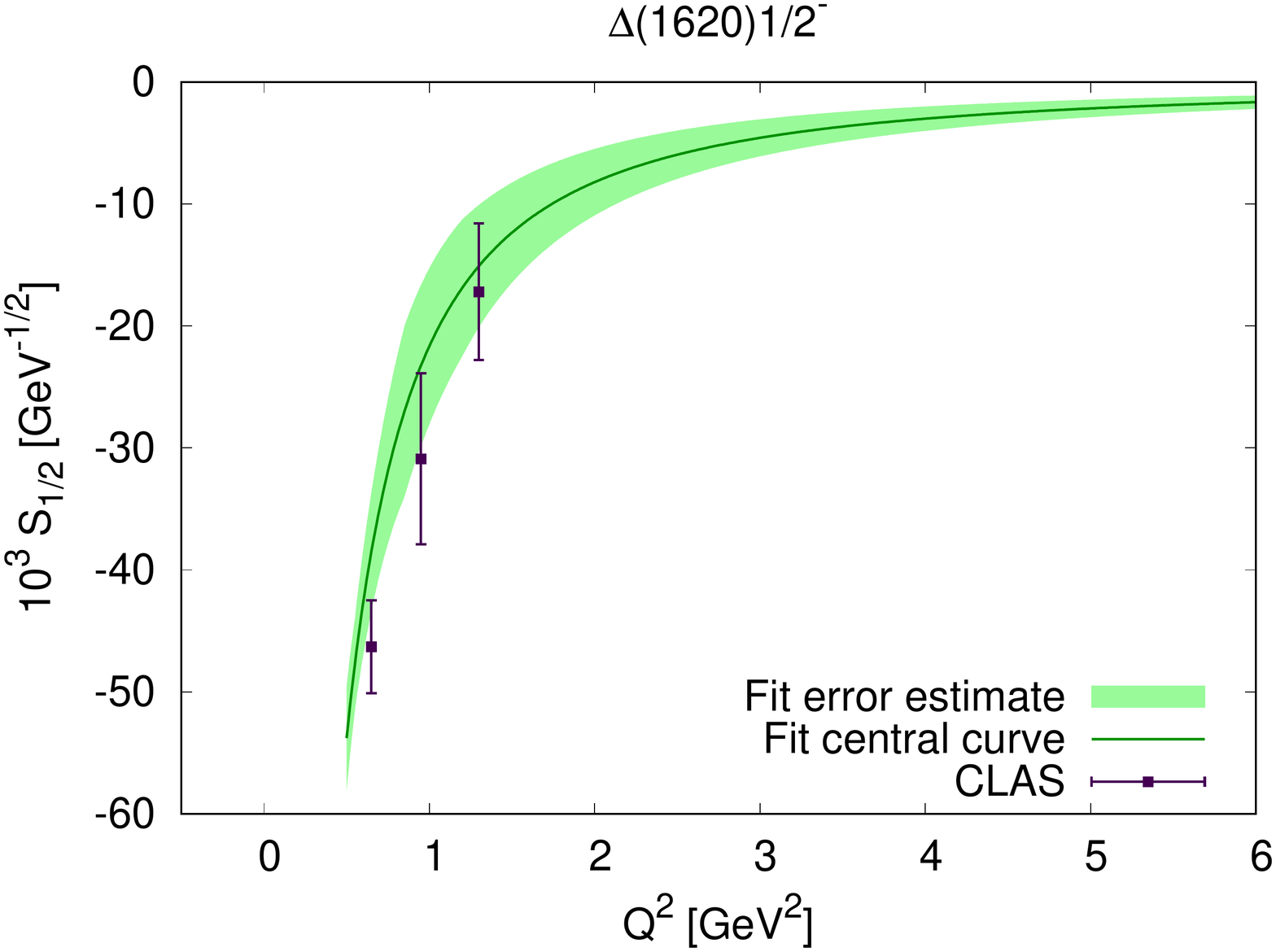}
\caption{Electrocouplings of the $\Delta(1232)~3/2^+$, $N(1440)~1/2^+$, $N(1520)~3/2^-$, $N(1535)~1/2^-$ and $\Delta(1620)~1/2^-$. The left column shows $A_{1/2}$a nd the right column shows $S_{1/2}$, while the central column shows $A_{3/2}$ when applicable. The data are from the RPP~\cite{Tanabashi:2018oca}, CLAS~\cite{Thompson:2000by,Denizli:2007tq,Dalton:2008aa,Dugger:2009pn,Aznauryan:2012ec,Mokeev:2012vsa,Mokeev:2015lda} the E94014 experiment~\cite{Armstrong:1998wg}, MAMI~\cite{Stave:2008aa}, MIT/Bates~\cite{Sparveris:2004jn}, Hall A~\cite{Laveissiere:2003jf,Kelly:2005jy} and Hall C~\cite{Villano:2009sn}.}
\label{F:Ecoup1}
\end{figure*}

\begin{figure*}
\includegraphics[width=.3\textwidth]{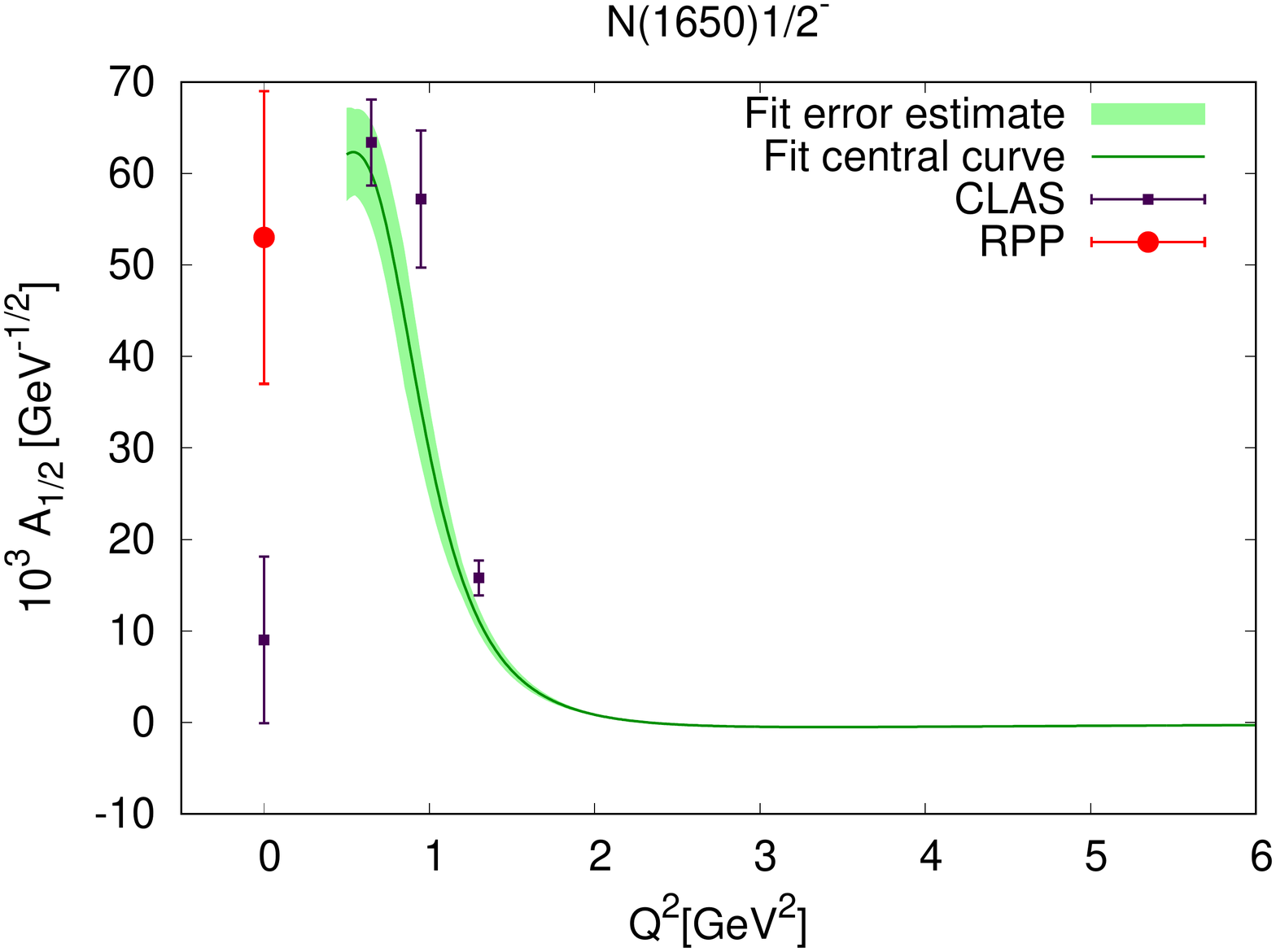}
\includegraphics[width=.3\textwidth]{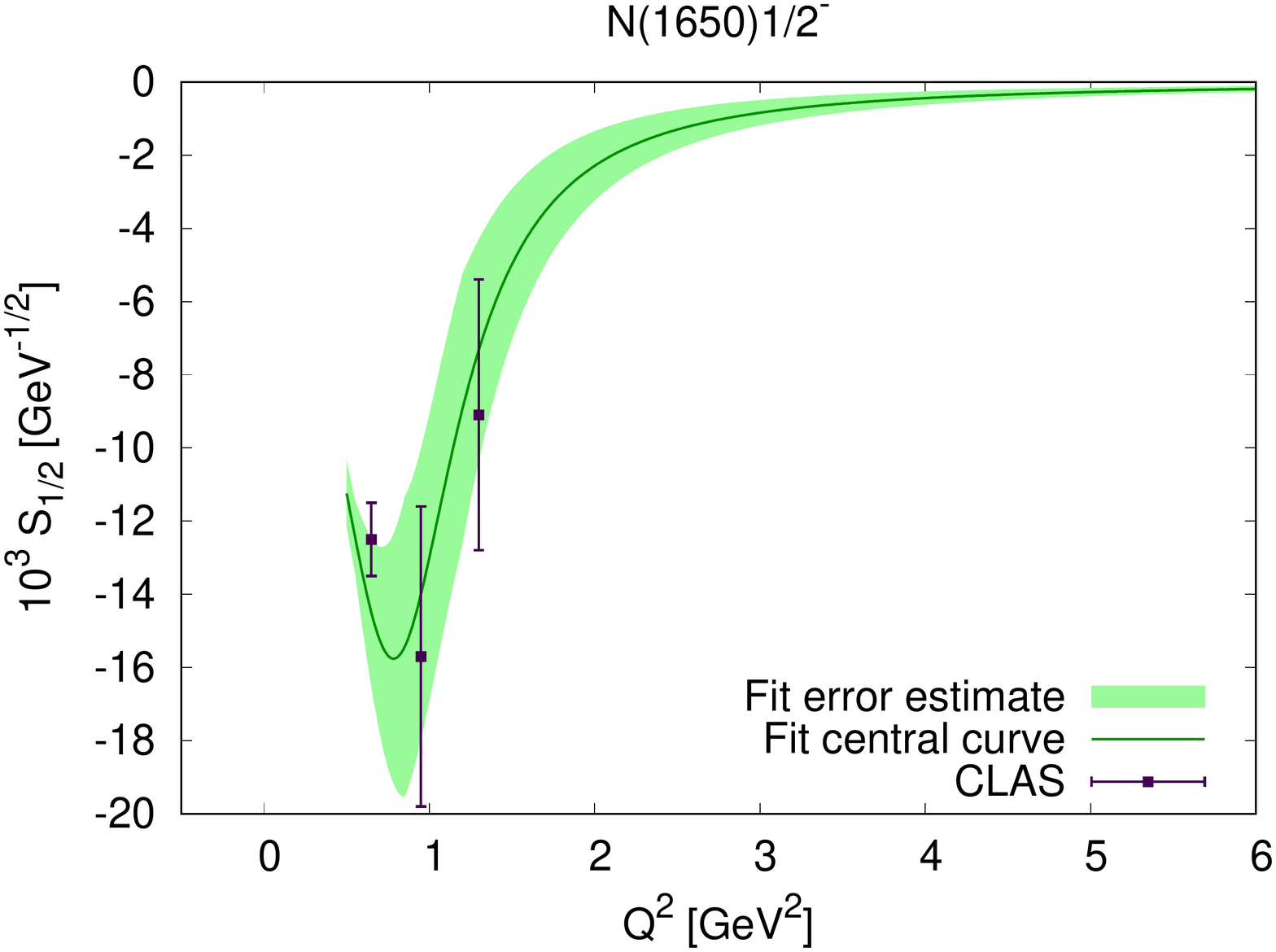}\\
\includegraphics[width=.3\textwidth]{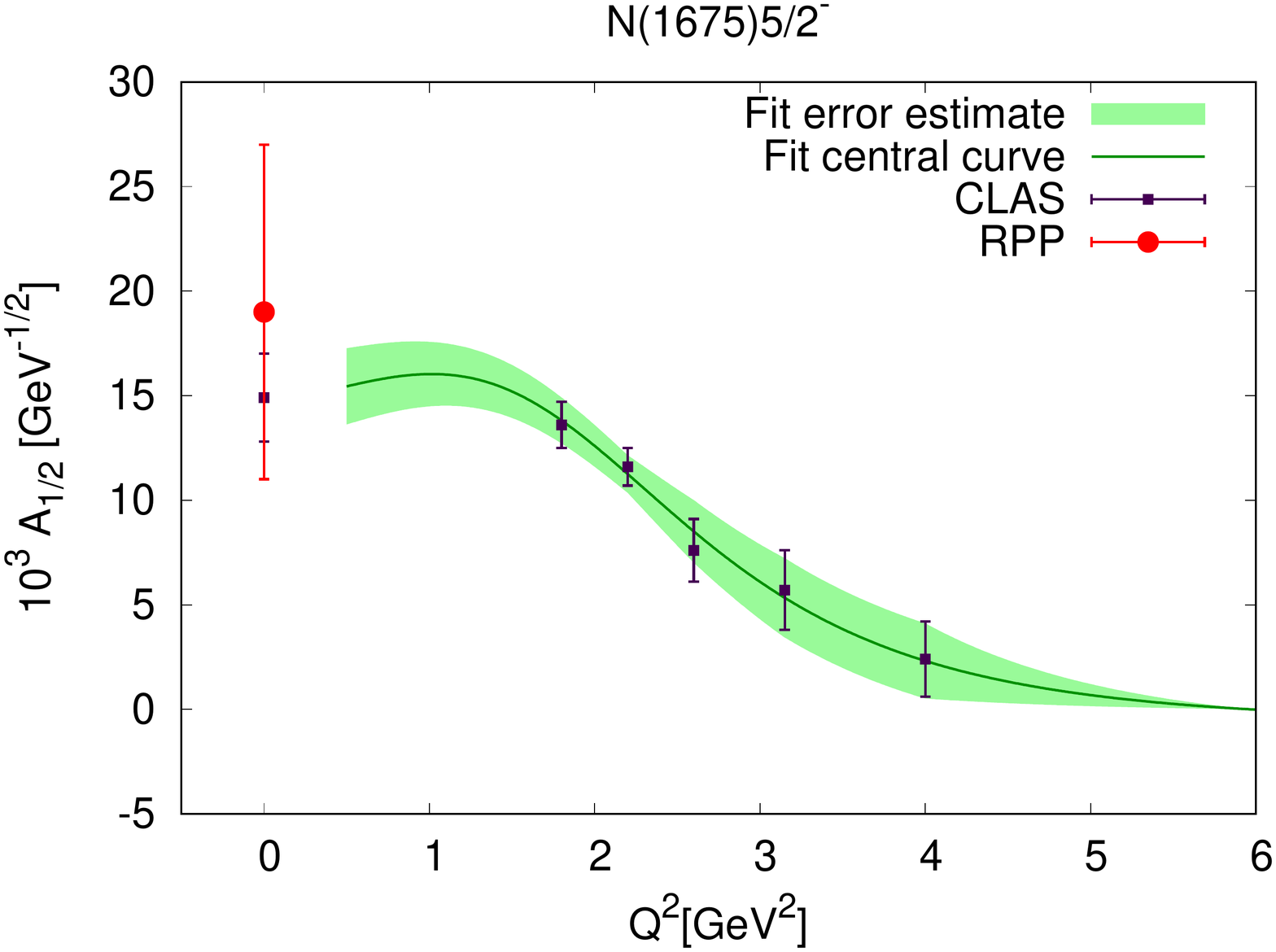}
\includegraphics[width=.3\textwidth]{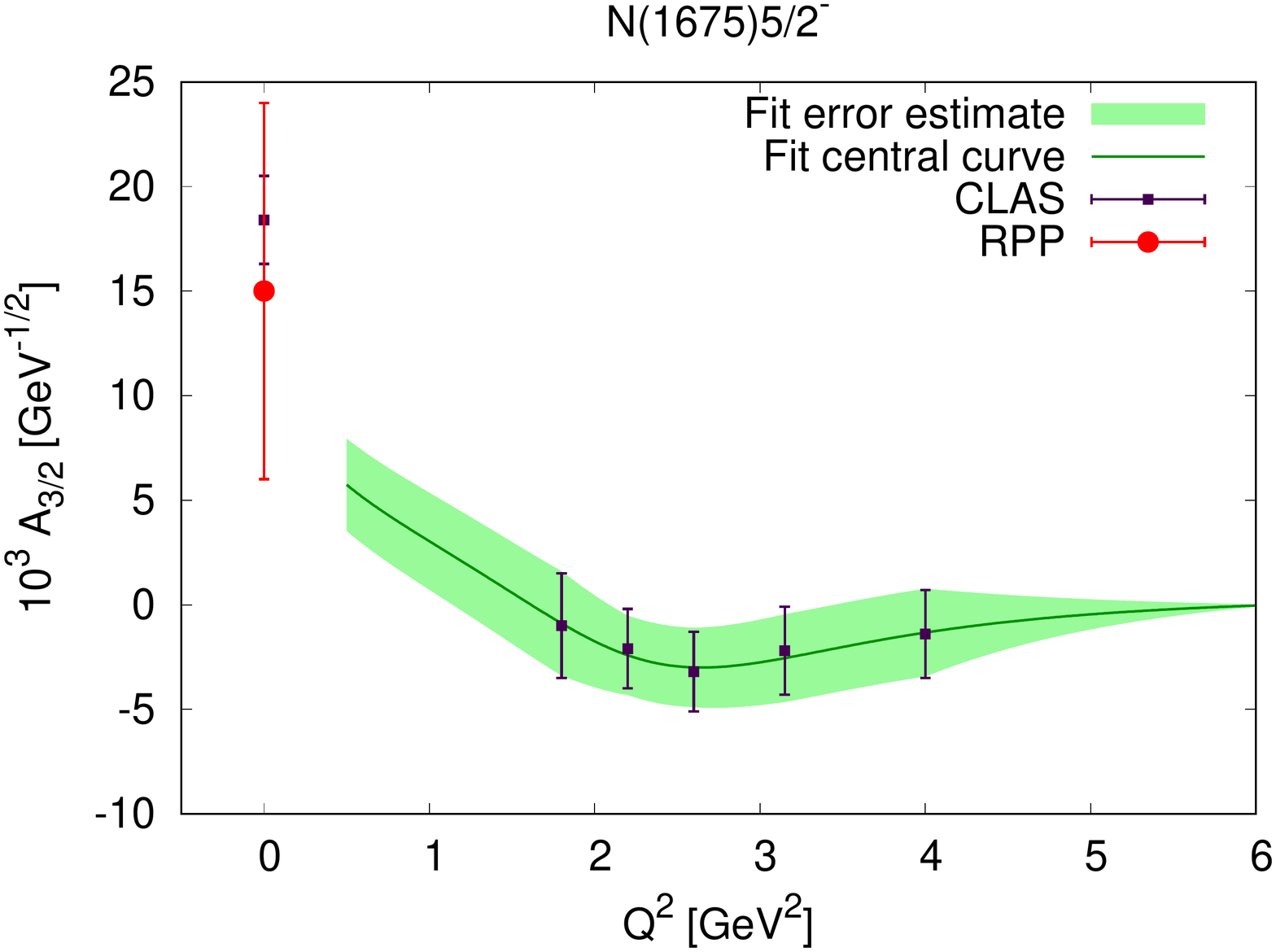}
\includegraphics[width=.3\textwidth]{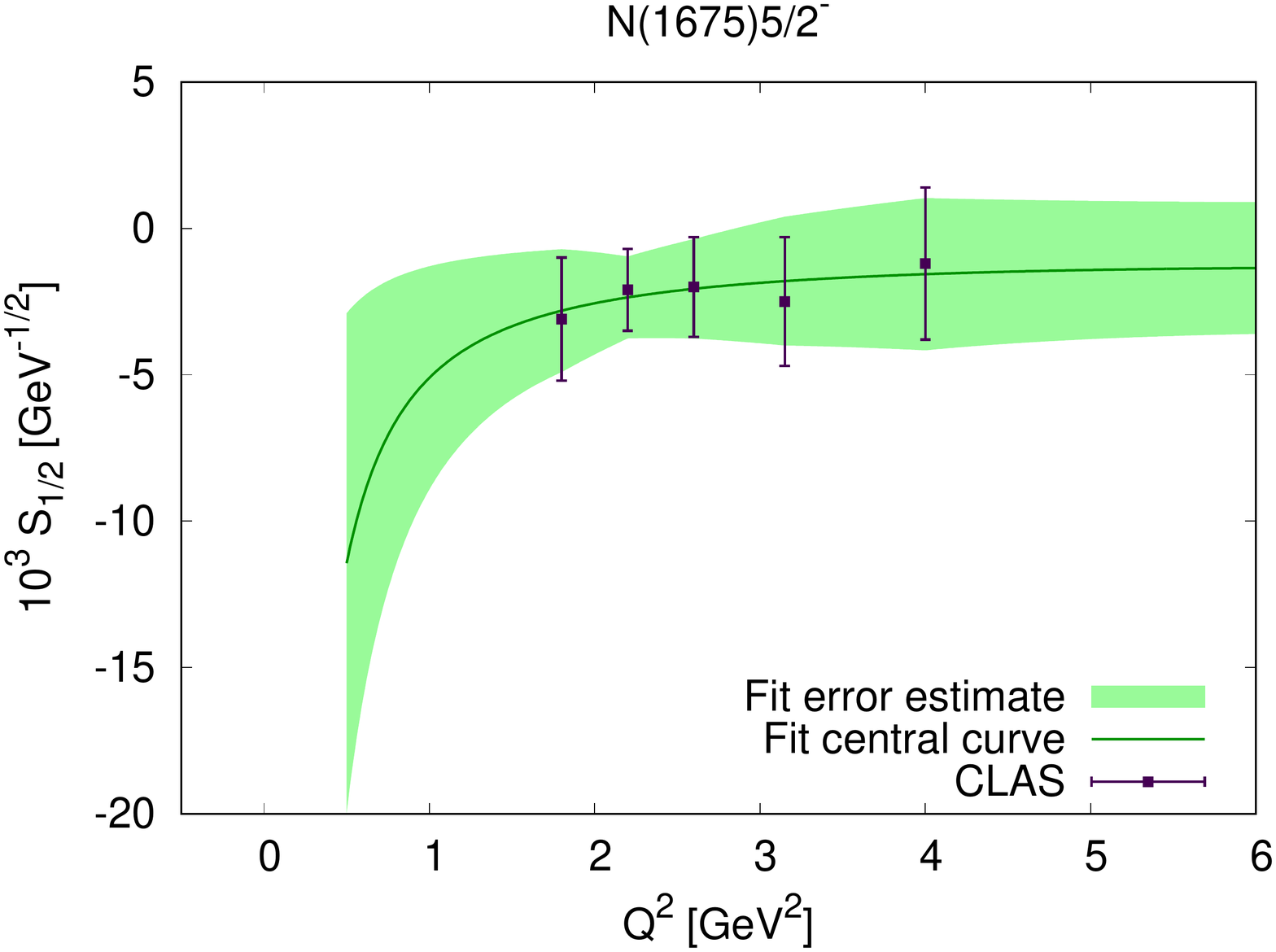}\\
\includegraphics[width=.3\textwidth]{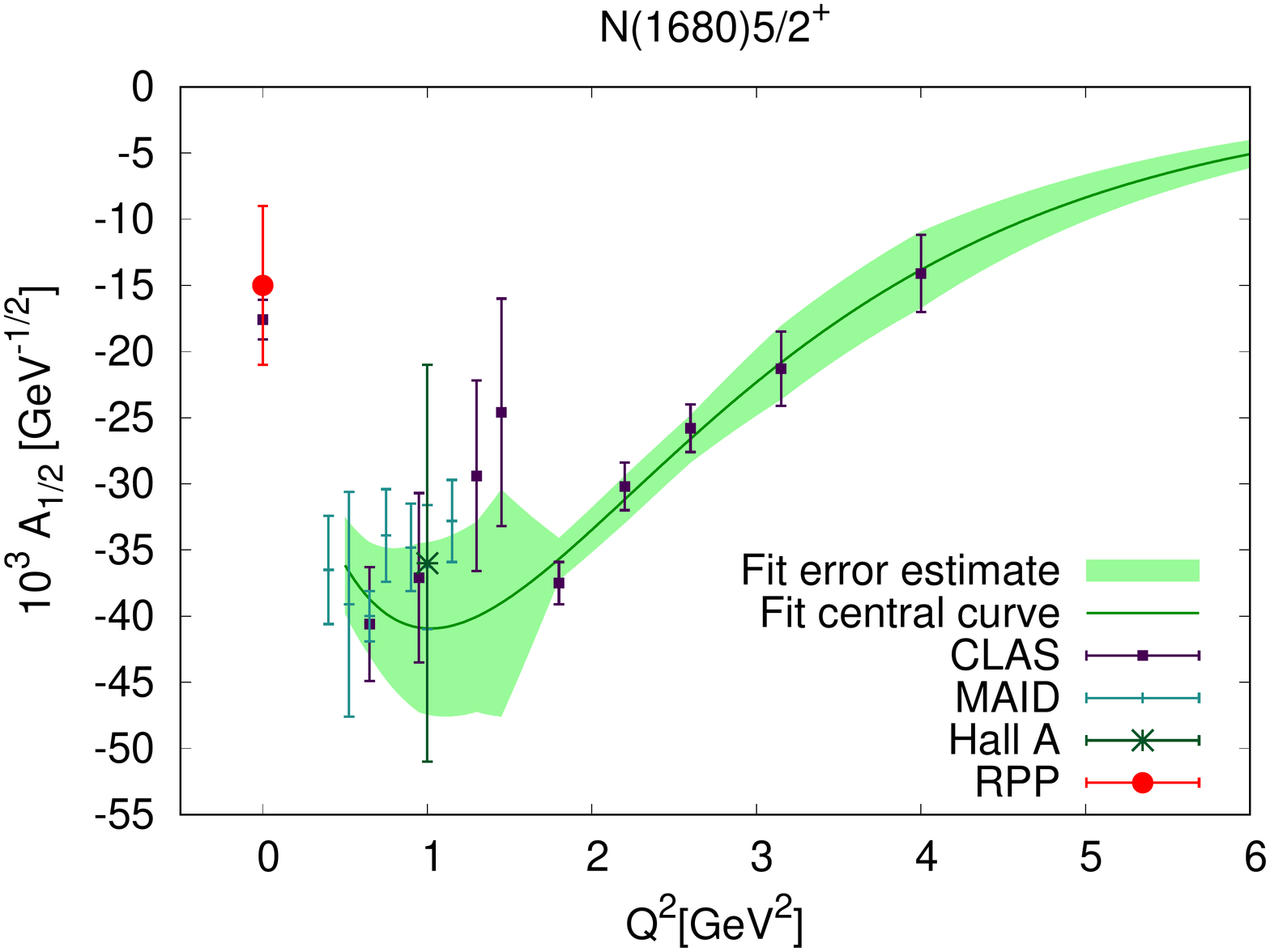}
\includegraphics[width=.3\textwidth]{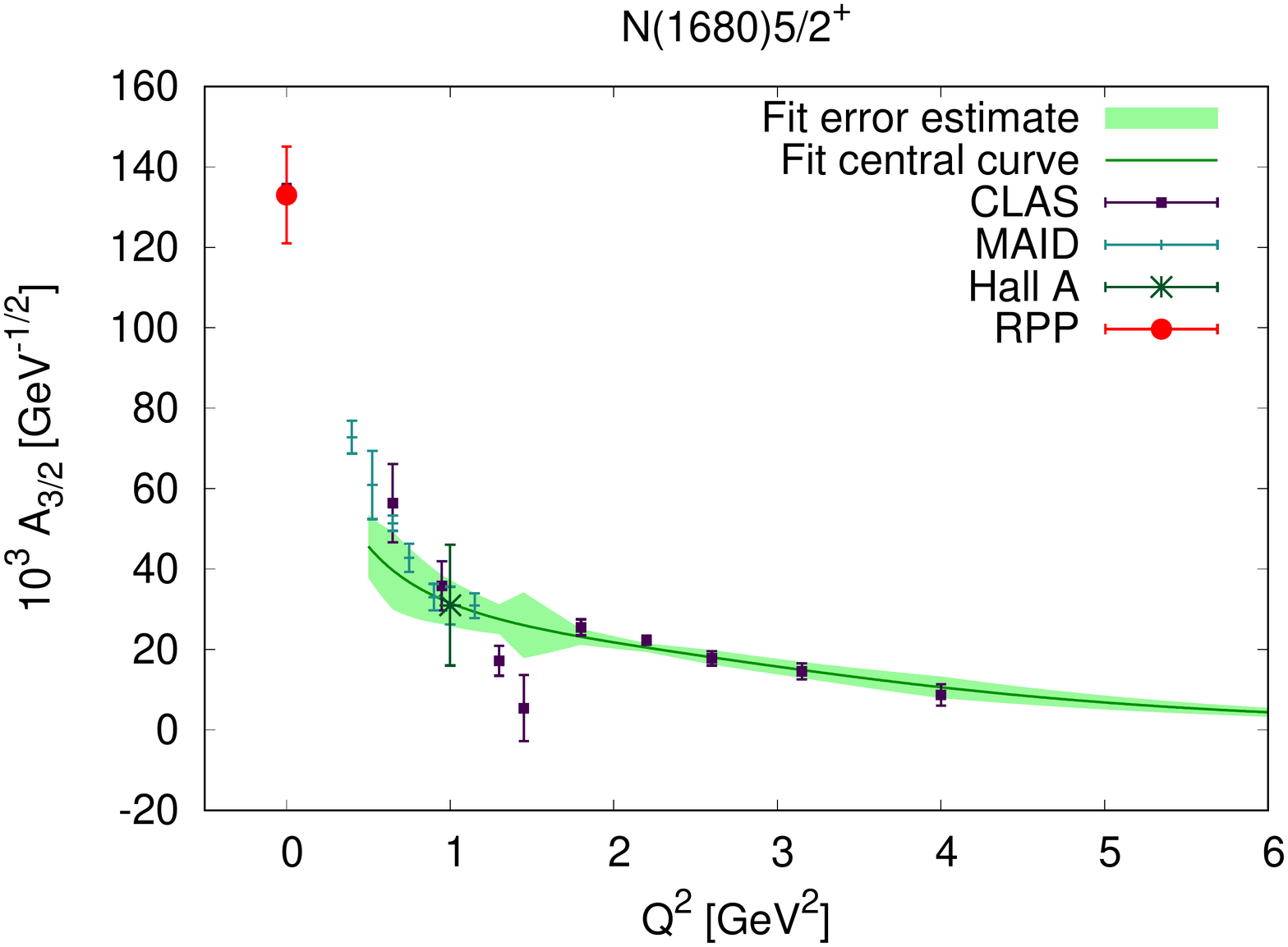}
\includegraphics[width=.3\textwidth]{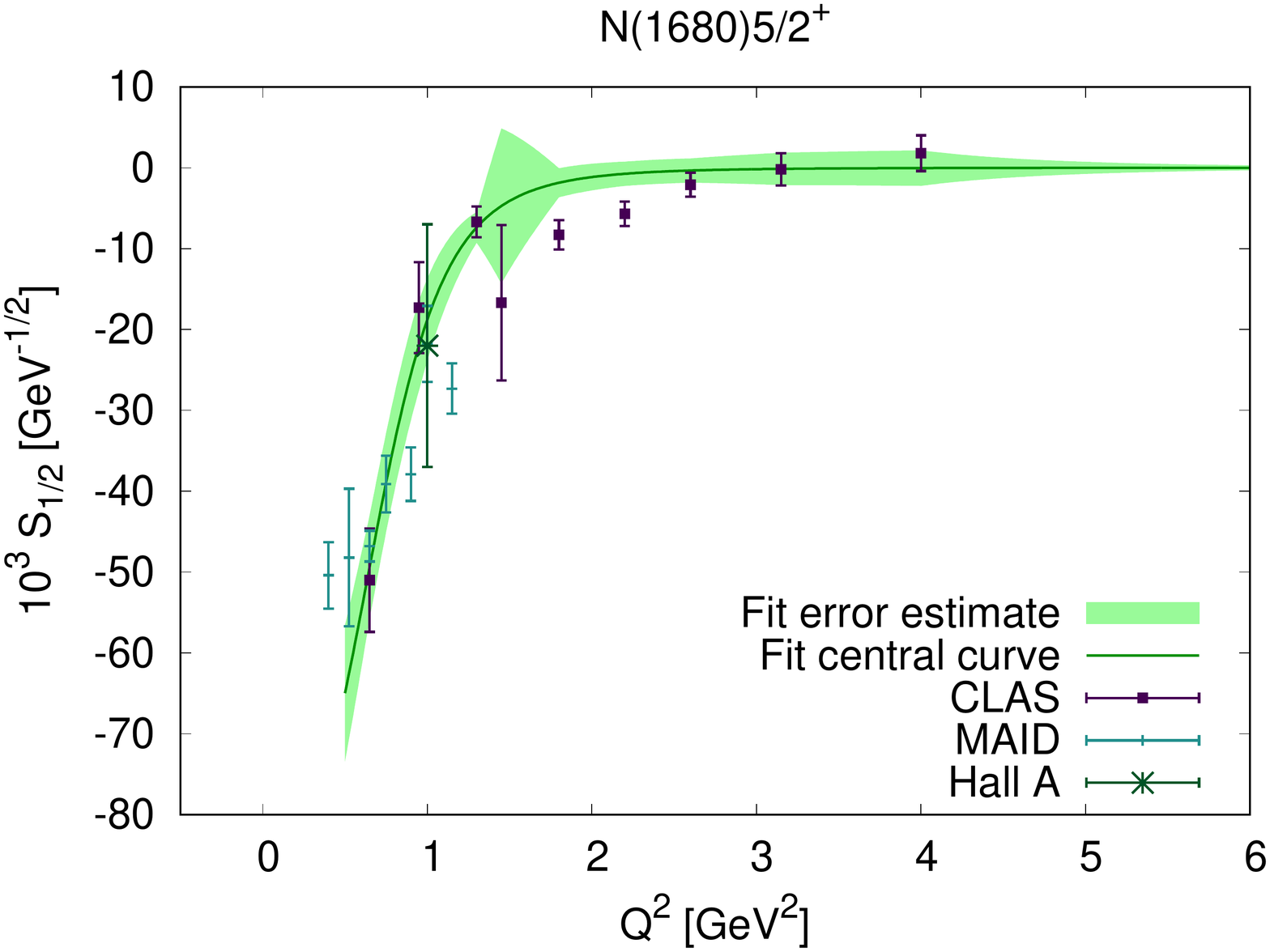}\\
\includegraphics[width=.3\textwidth]{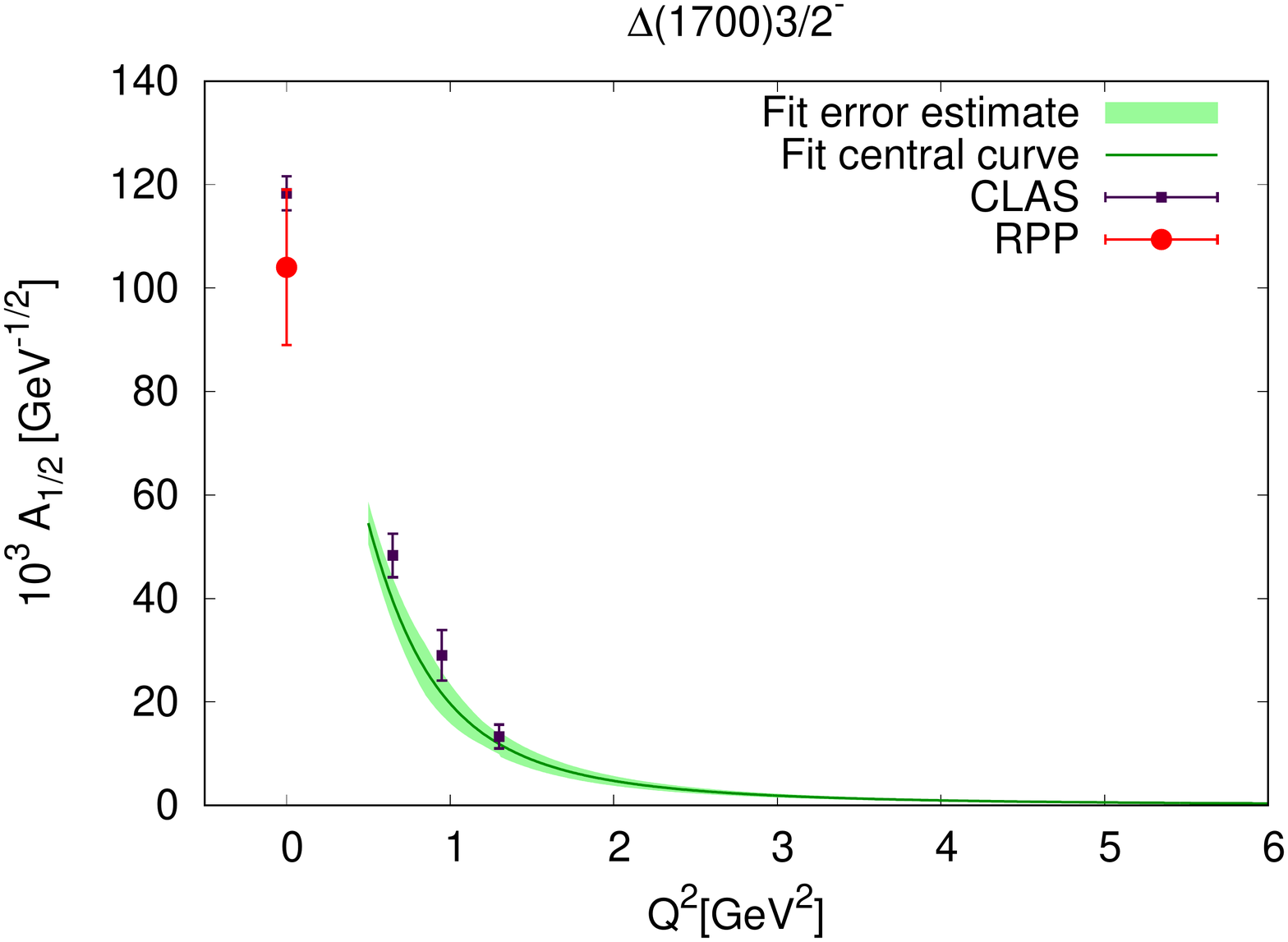}
\includegraphics[width=.3\textwidth]{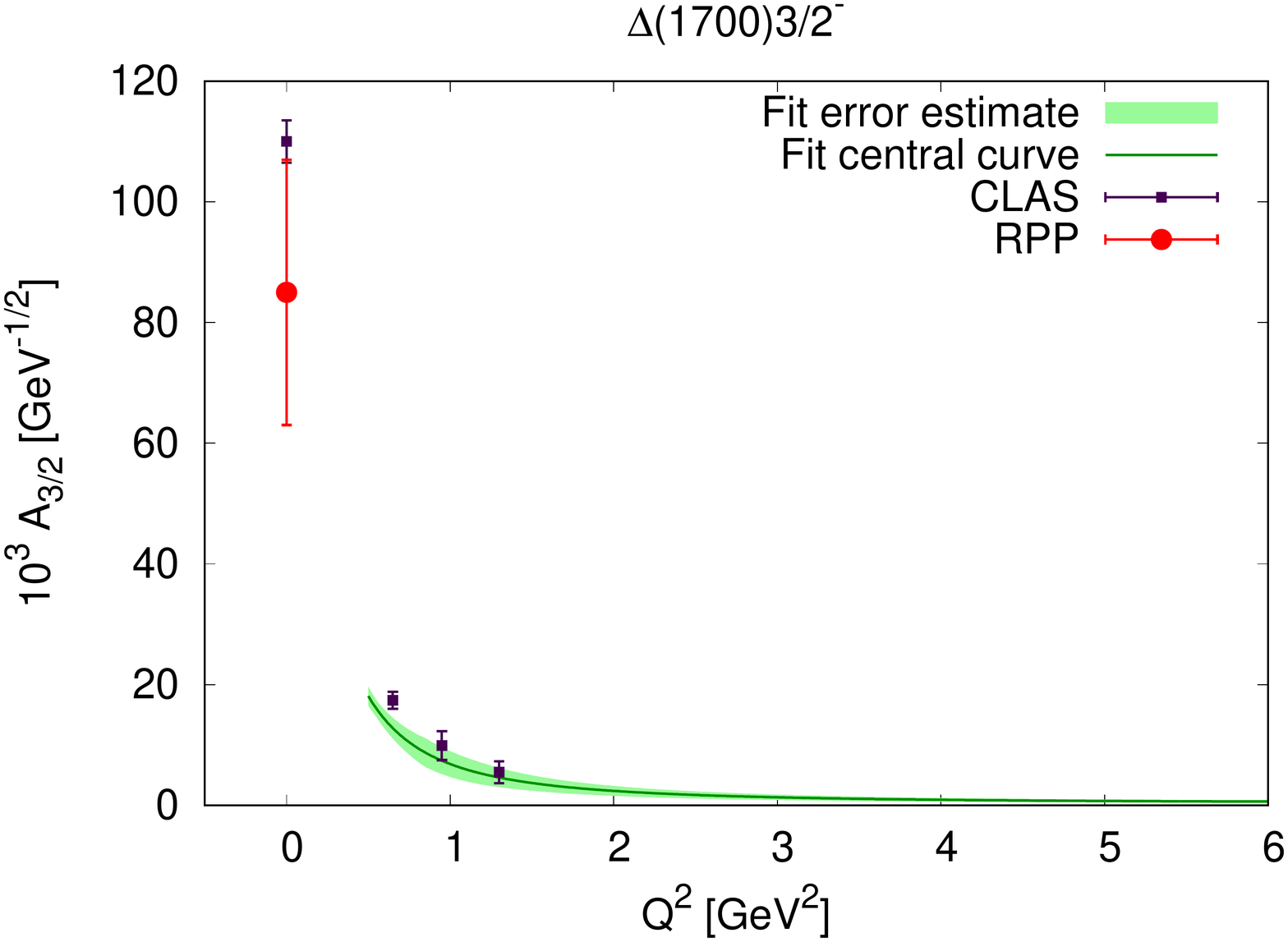}
\includegraphics[width=.3\textwidth]{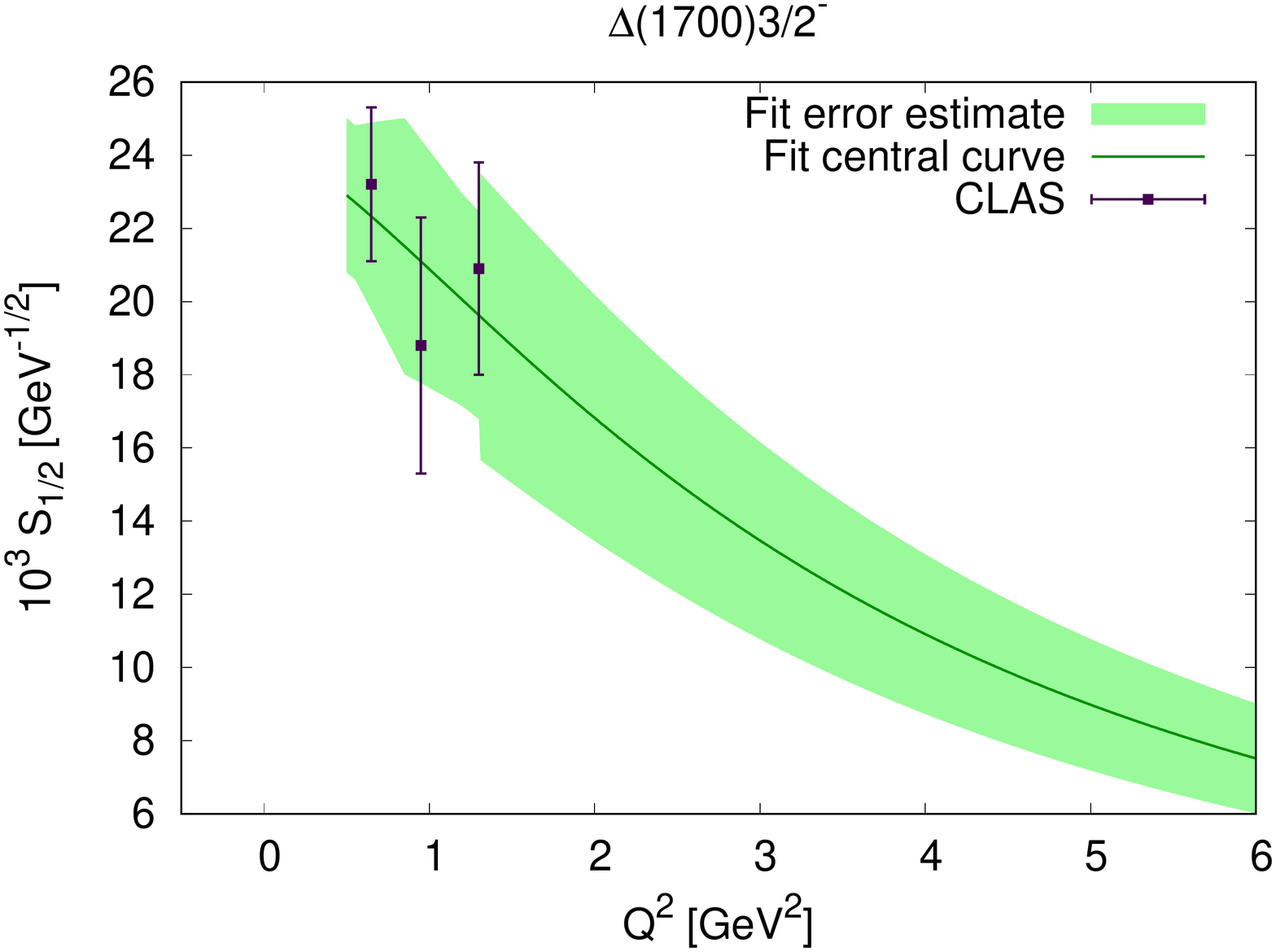}\\
\includegraphics[width=.3\textwidth]{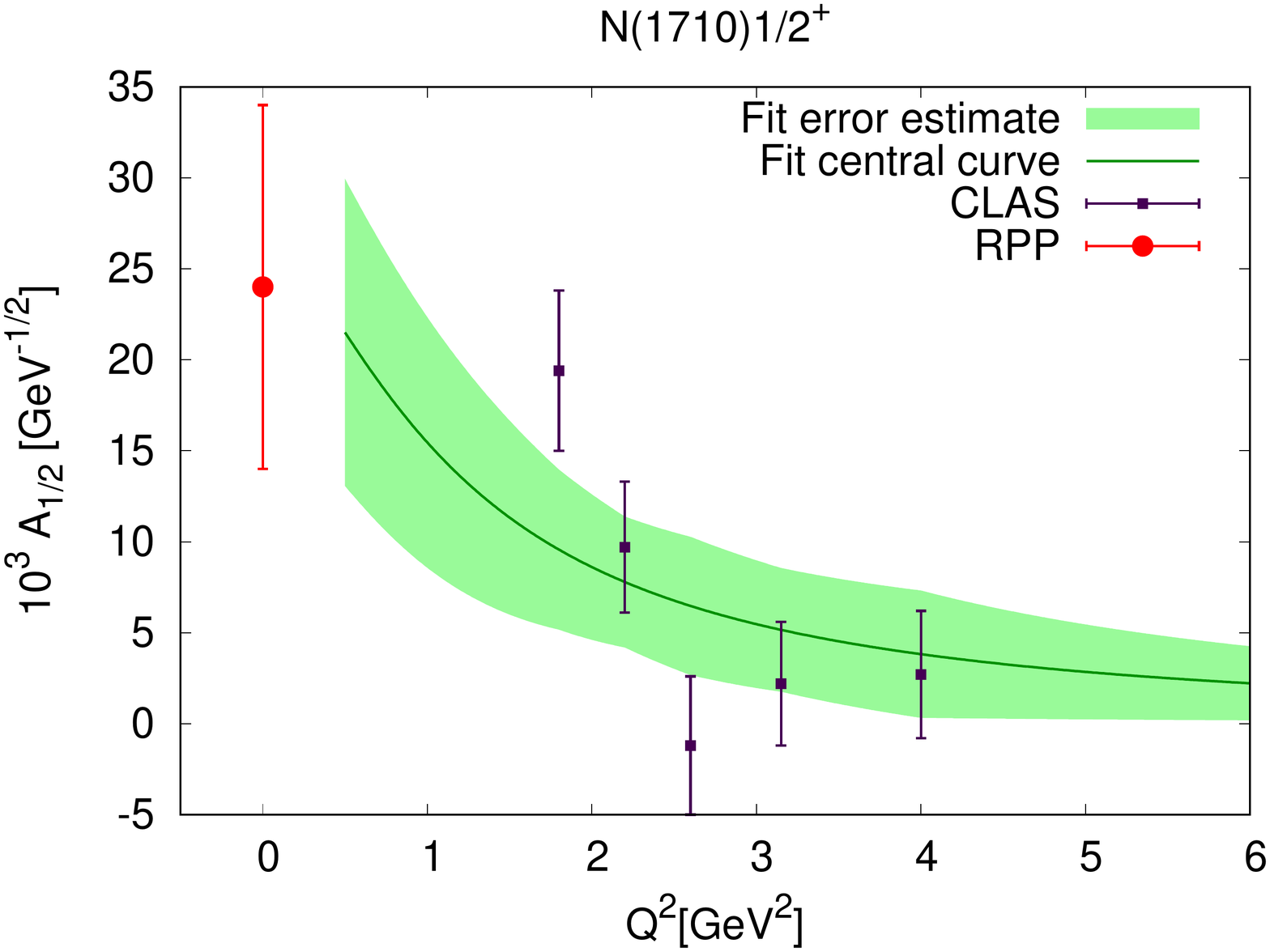}
\includegraphics[width=.3\textwidth]{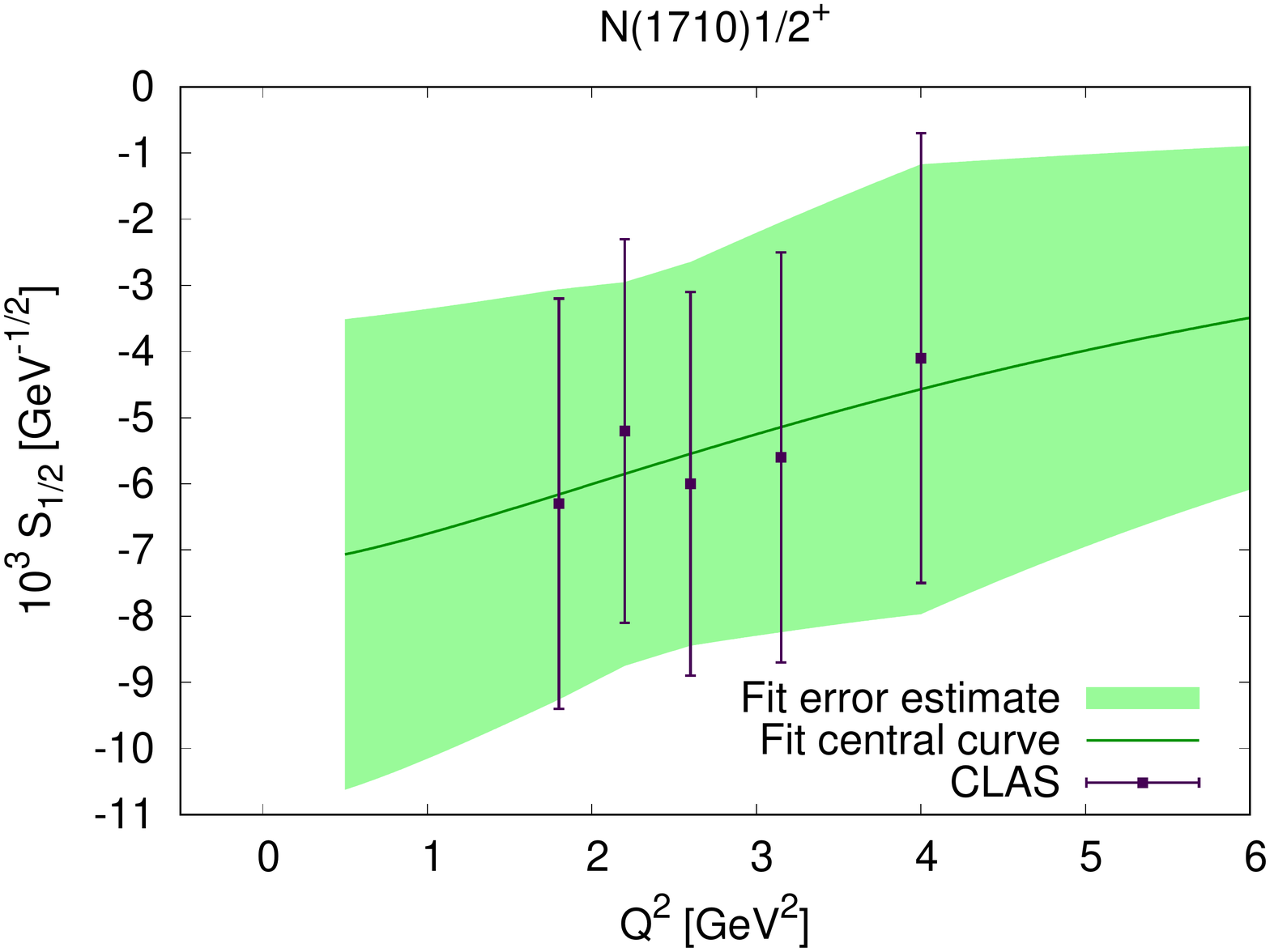}
\caption{Electrocouplings of the $N(1650)~1/2^-$, $N(1675)~5/2^-$, $N(1680)~5/2^+$, $\Delta(1700)~3/2^-$ and $N(1710)~1/2^+$. The left column shows $A_{1/2}$ and the right column shows $S_{1/2}$, while the central column shows $A_{3/2}$ when applicable. The data points are from the RPP~\cite{Tanabashi:2018oca}, CLAS~\cite{Dugger:2009pn,Mokeev:2013kka,Park:2014yea}, a MAID analysis~\cite{Tiator:2011pw} and Hall A~\cite{Laveissiere:2003jf}.}
\label{F:Ecoup2}
\end{figure*}

\begin{figure*}
\includegraphics[width=.3\textwidth]{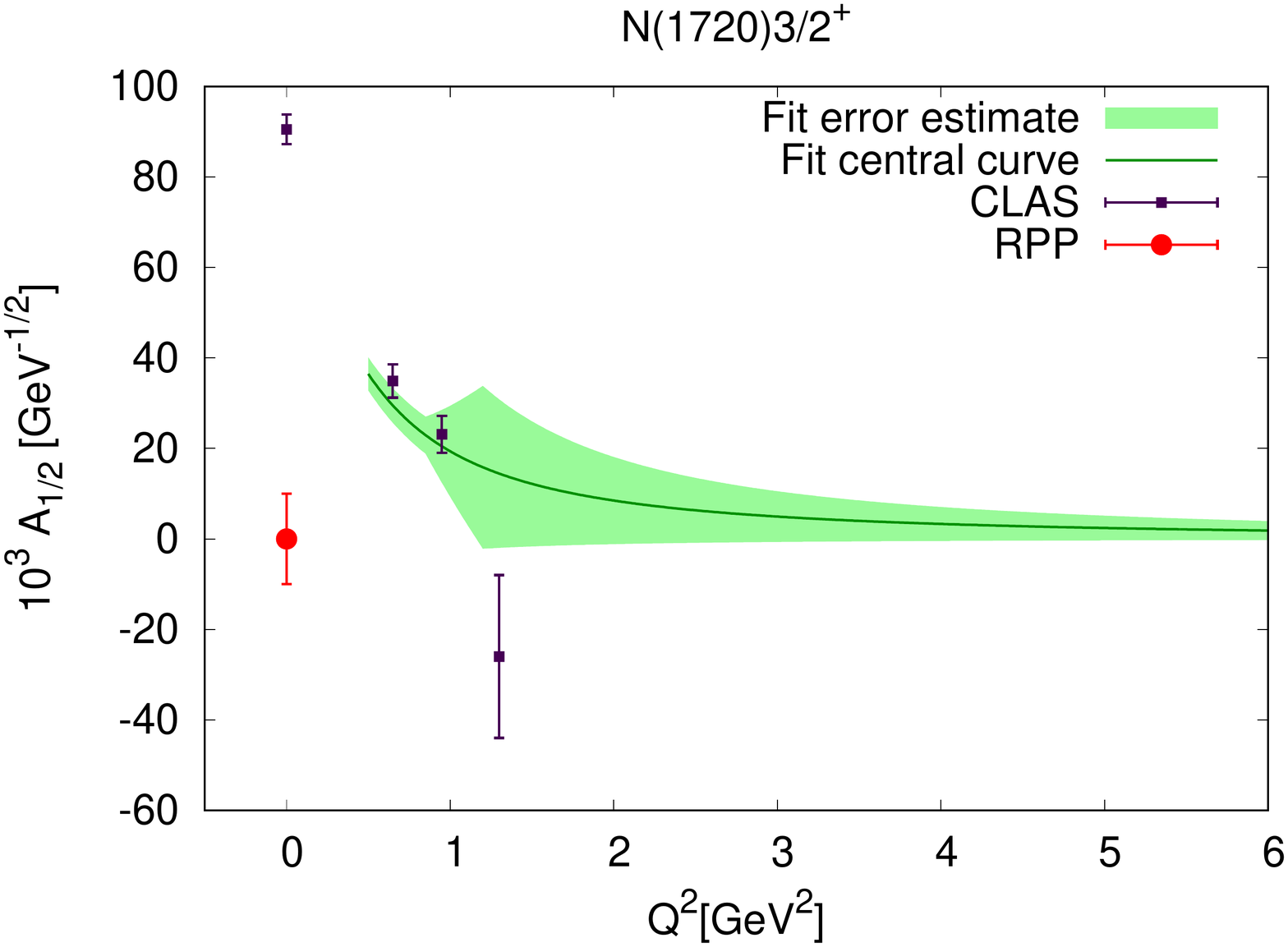}
\includegraphics[width=.3\textwidth]{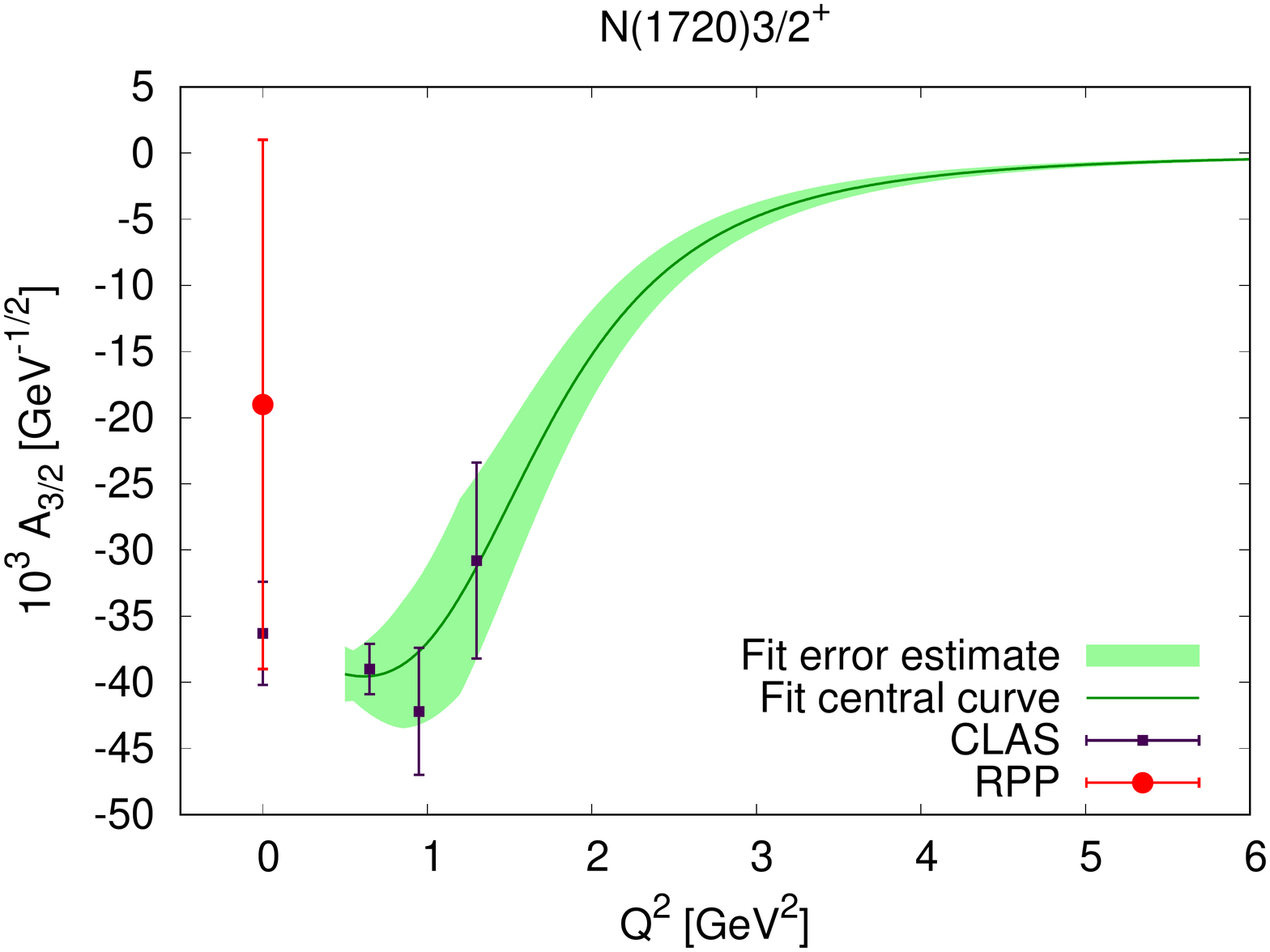}
\includegraphics[width=.3\textwidth]{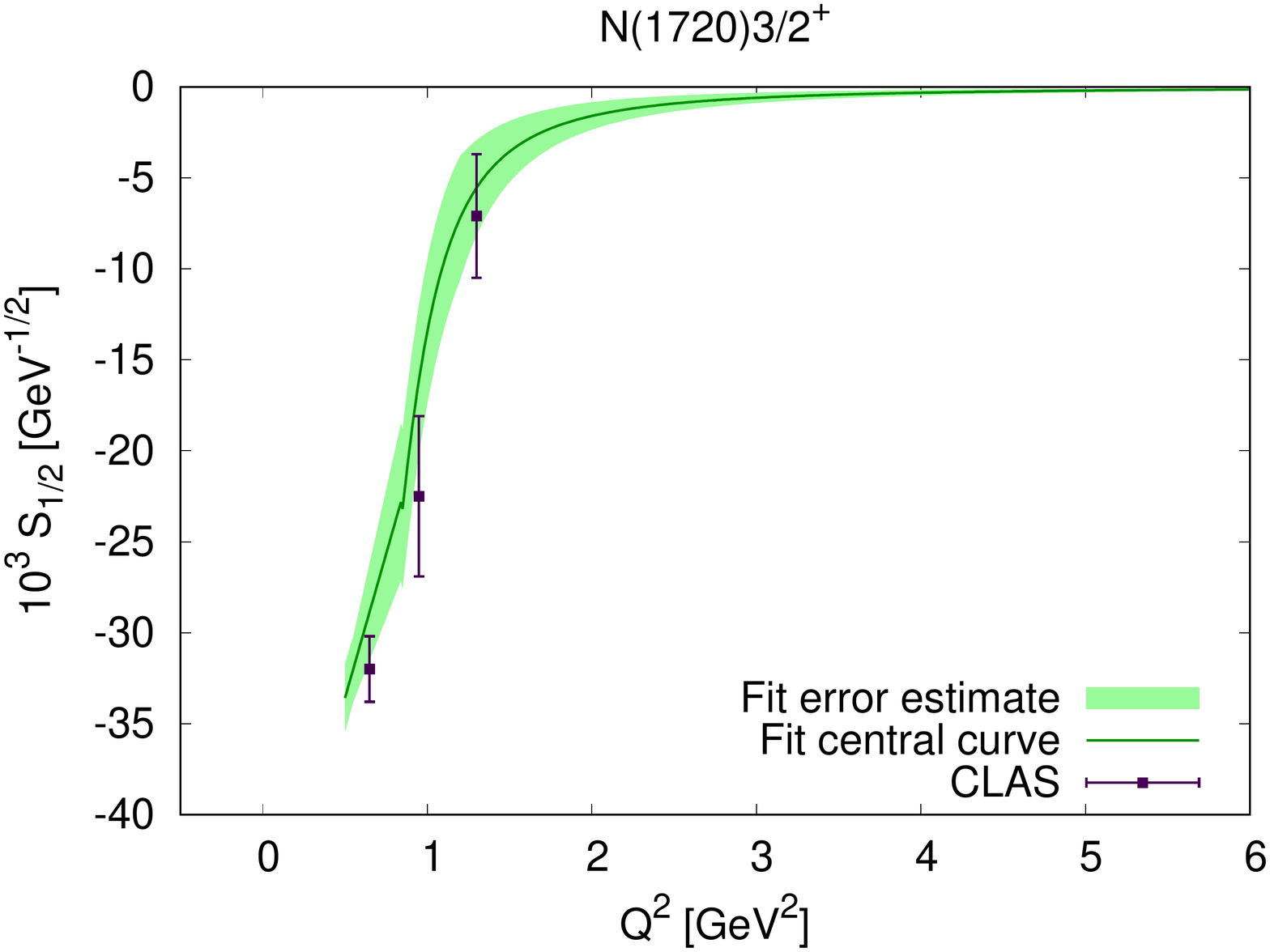}\\
\includegraphics[width=.3\textwidth]{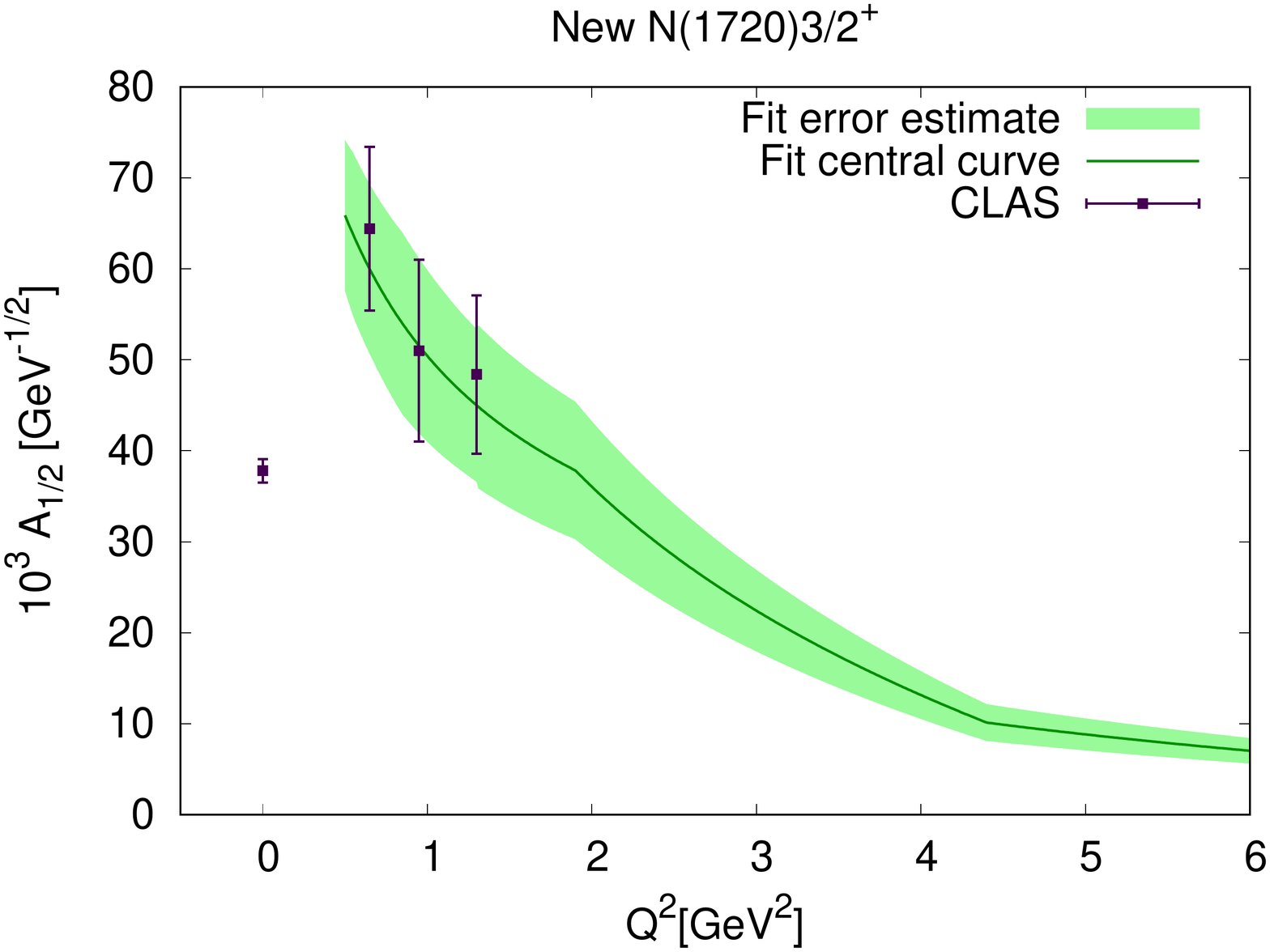}
\includegraphics[width=.3\textwidth]{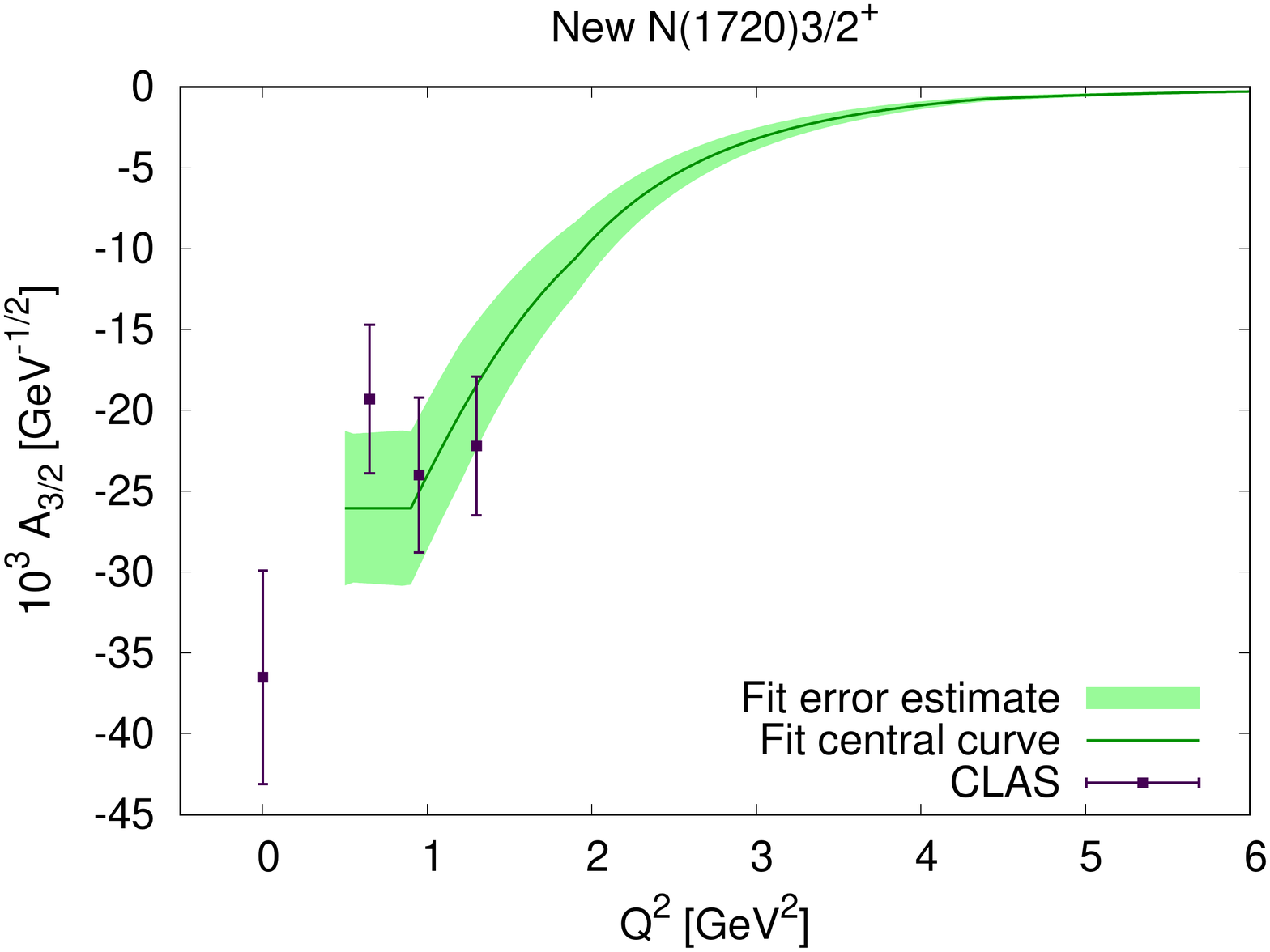}
\includegraphics[width=.3\textwidth]{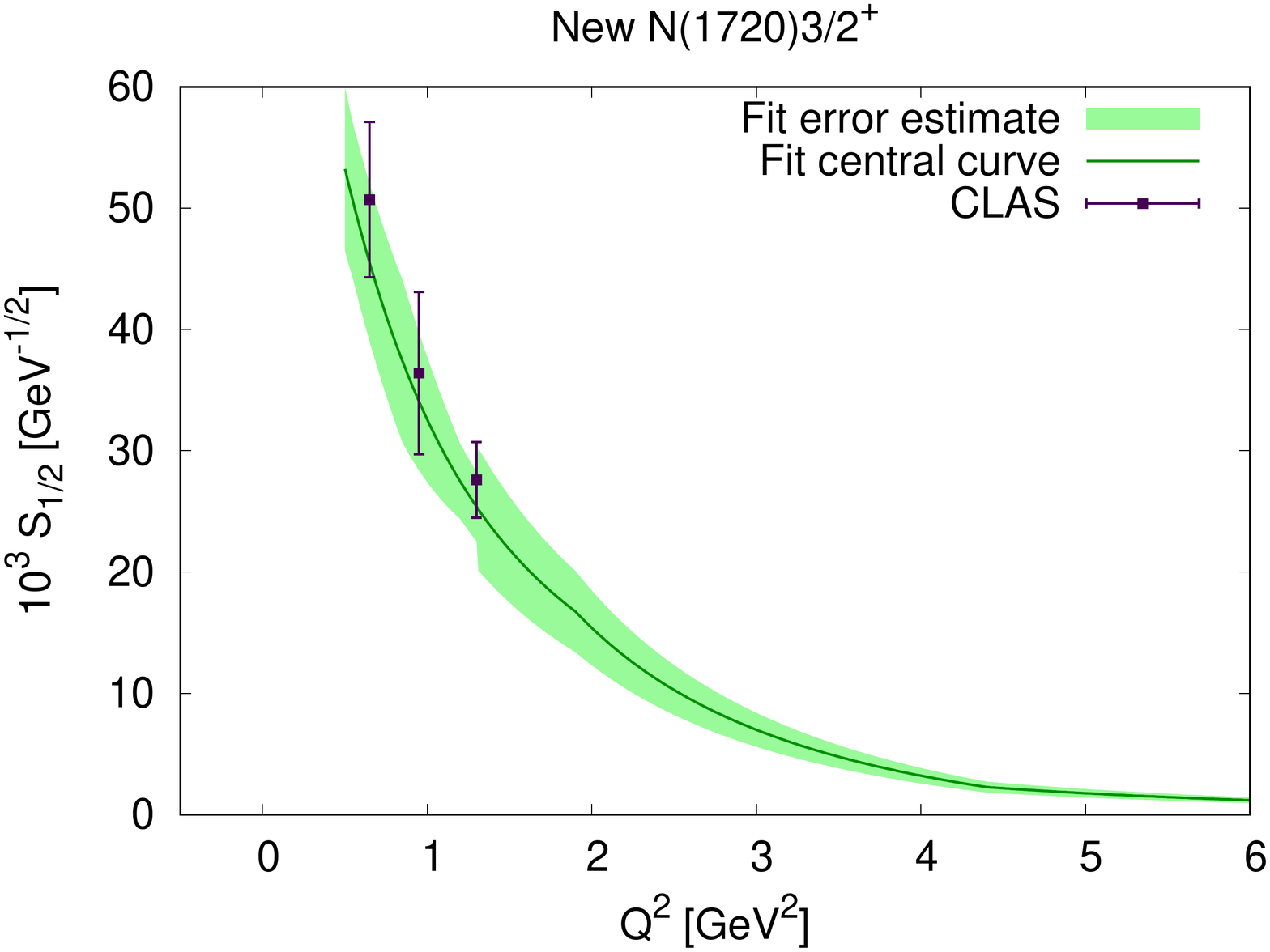}
\caption{Electrocouplings of the $N(1720)~3/2^+$ and $N^\prime(1720)~3/2^+$. The left column shows $A_{1/2}$a nd the right column shows $S_{1/2}$, while the central column shows $A_{3/2}$. The data points are from the RPP~\cite{Tanabashi:2018oca} and CLAS~\cite{Dugger:2009pn,Mokeev:2015moa}.}
\label{F:Ecoup3}
\end{figure*}

We use the interpolation/extrapolation tools developed by CLAS~\cite{CLAS:coups} for the central values of the electrocouplings. In order to estimate the uncertainties of the resonance contributions to the inclusive electron scattering observables, we interpolate between the error bars of the experimental bins in $Q^2$~\cite{CLAS:coupsDB}. As for the regions of $Q^2$ 
 that are not covered by the resonance electrocoupling data, we  assume the relative uncertainty  to be the same as the one of the data point at highest $Q^2$. 
Note that this is only an extrapolation estimate, and therefore the results outside the data range in $Q^2$ need to be taken with care. Furthermore, the data for $\Delta(1620)1/2^-$, $N(1720)~3/2^+$, $N^\prime(1720)~3/2^+$ and $\Delta(1700)~3/2^-$ cover a smaller range of $Q^2<1.5$~GeV$^2$ when compared to the other resonances, $Q^2<5$~GeV$^2$. For these resonances, we have preliminary estimates of the central electrocoupling values~\cite{Burkert:2019opk} based on the good description of the $\pi^+\pi^-p$ electroproduction data off protons from CLAS at 2.0~GeV$^2<Q^2<5.0$~GeV$^2$~ \cite{Isupov:2017lnd,Fedotov:2018oan,Trivedi:2018rgo}. We give a conservative extrapolation uncertainty  estimate for these resonance electrocouplings of at least $20\%$ (when the relative error as calculated above is smaller). The electrocoupling curves in the $Q^2$ range used in this work are shown in Figs.~\ref{F:Ecoup1} to \ref{F:Ecoup3}, and compared to electrocoupling world data~\cite{Frolov:1998pw,Armstrong:1998wg,Thompson:2000by,Laveissiere:2003jf,Sparveris:2004jn,Kelly:2005jy,Denizli:2007tq,Dalton:2008aa,Stave:2008aa,Dugger:2009pn,Villano:2009sn,Tiator:2011pw,Aznauryan:2012ec,Mokeev:2012vsa,Mokeev:2013kka,Park:2014yea,Mokeev:2015lda,Mokeev:2015moa}.

For the propagation of the electrocoupling uncertainties to the observables, we use a  bootstrap based approach. 
 First,  for each $Q^2$, resonance and electrocoupling type, we generate a set of values distributed as Gaussians according to their central value and error. The size of our set is of $10^4$ samples, in order to obtain statistical significance. Then we use these $10^4$ samples to generate $10^4$ samples of the observables at each $Q^2$ and $W$ value. The band of values within $1\sigma$ from the average value of these samples can then be determined.

\section{Results and discussion}\label{S:res}

\begin{figure*}
\includegraphics[width=0.4\textwidth]{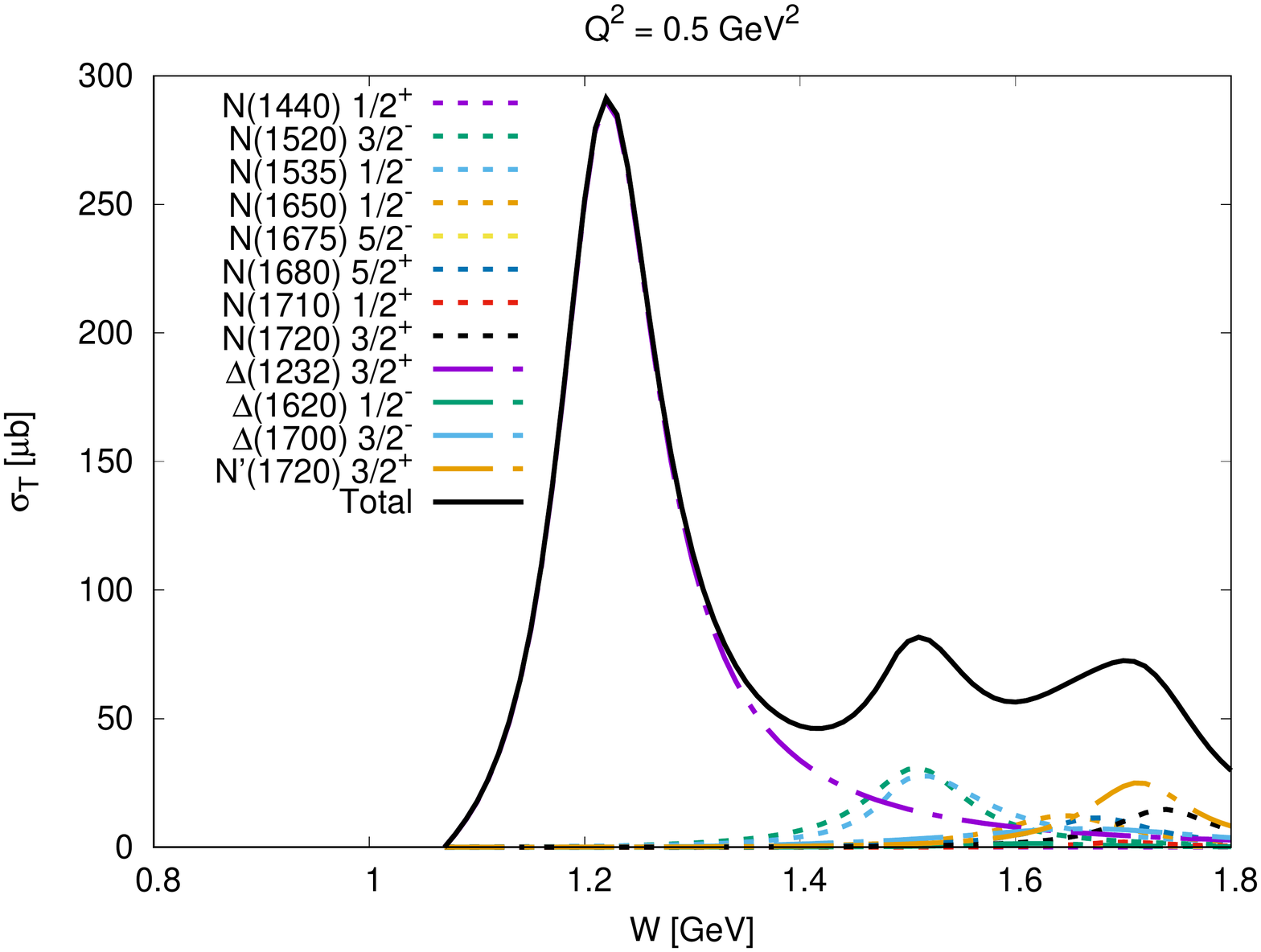}
\includegraphics[width=0.4\textwidth]{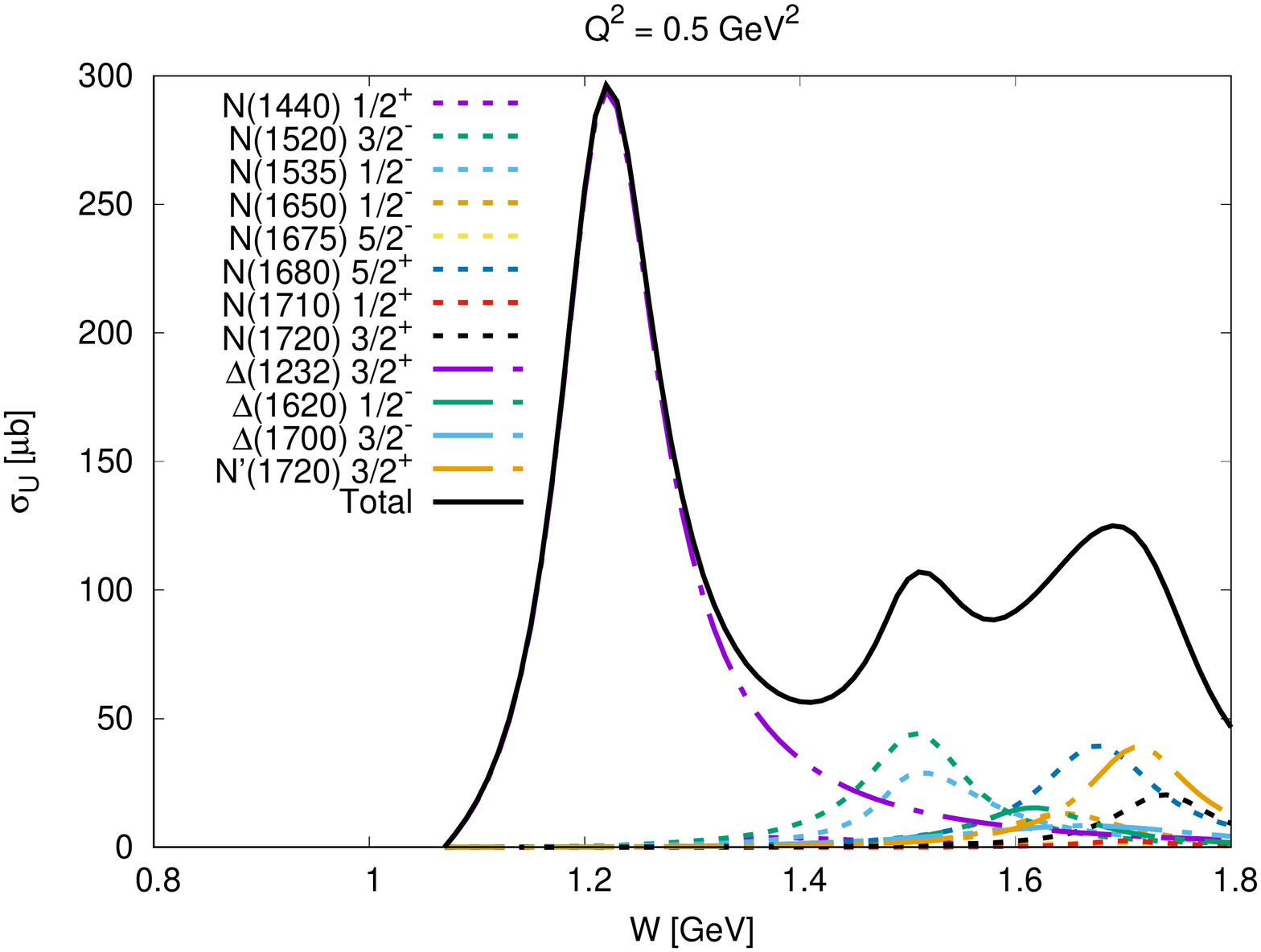}
\includegraphics[width=0.4\textwidth]{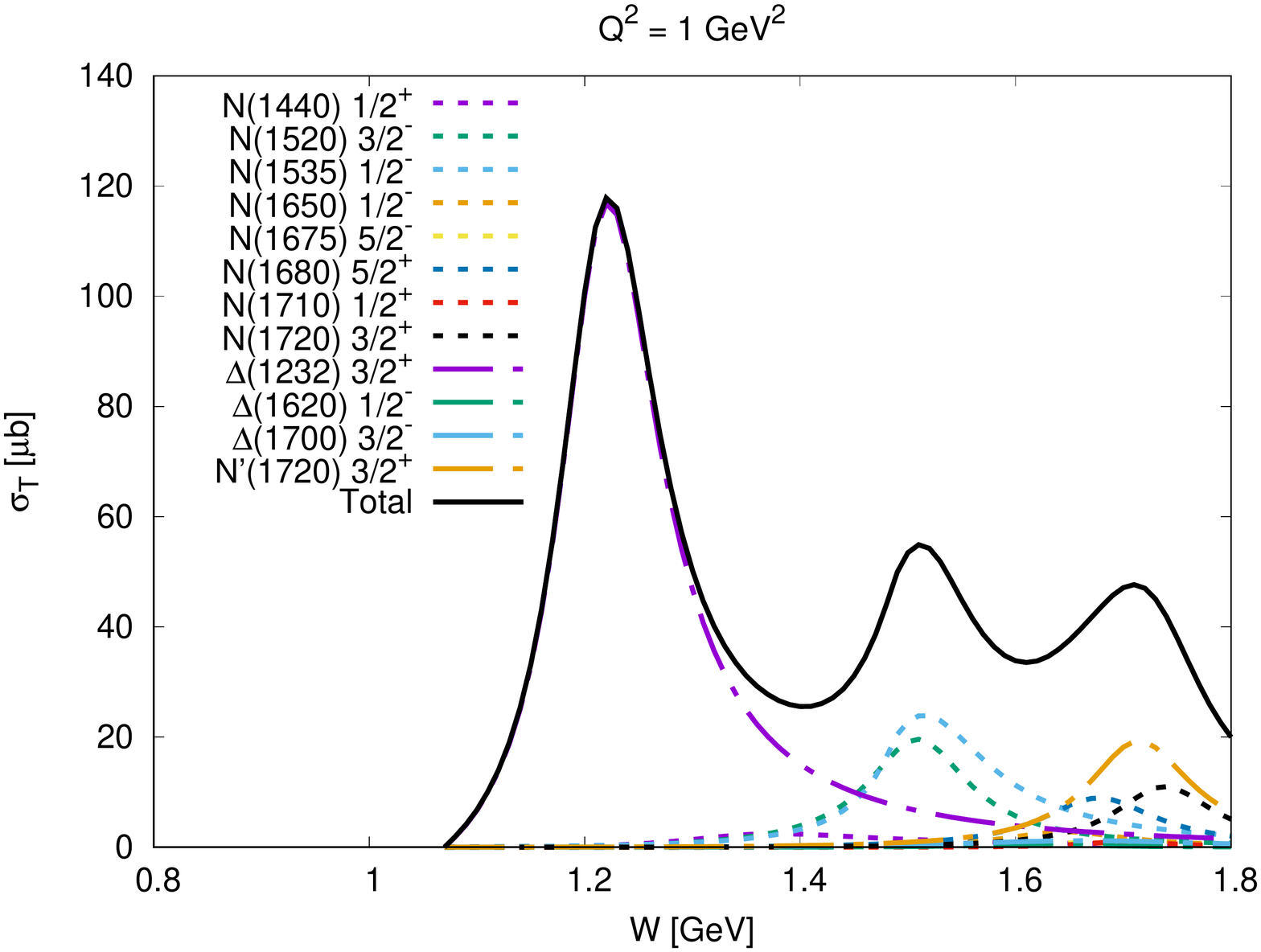}
\includegraphics[width=0.4\textwidth]{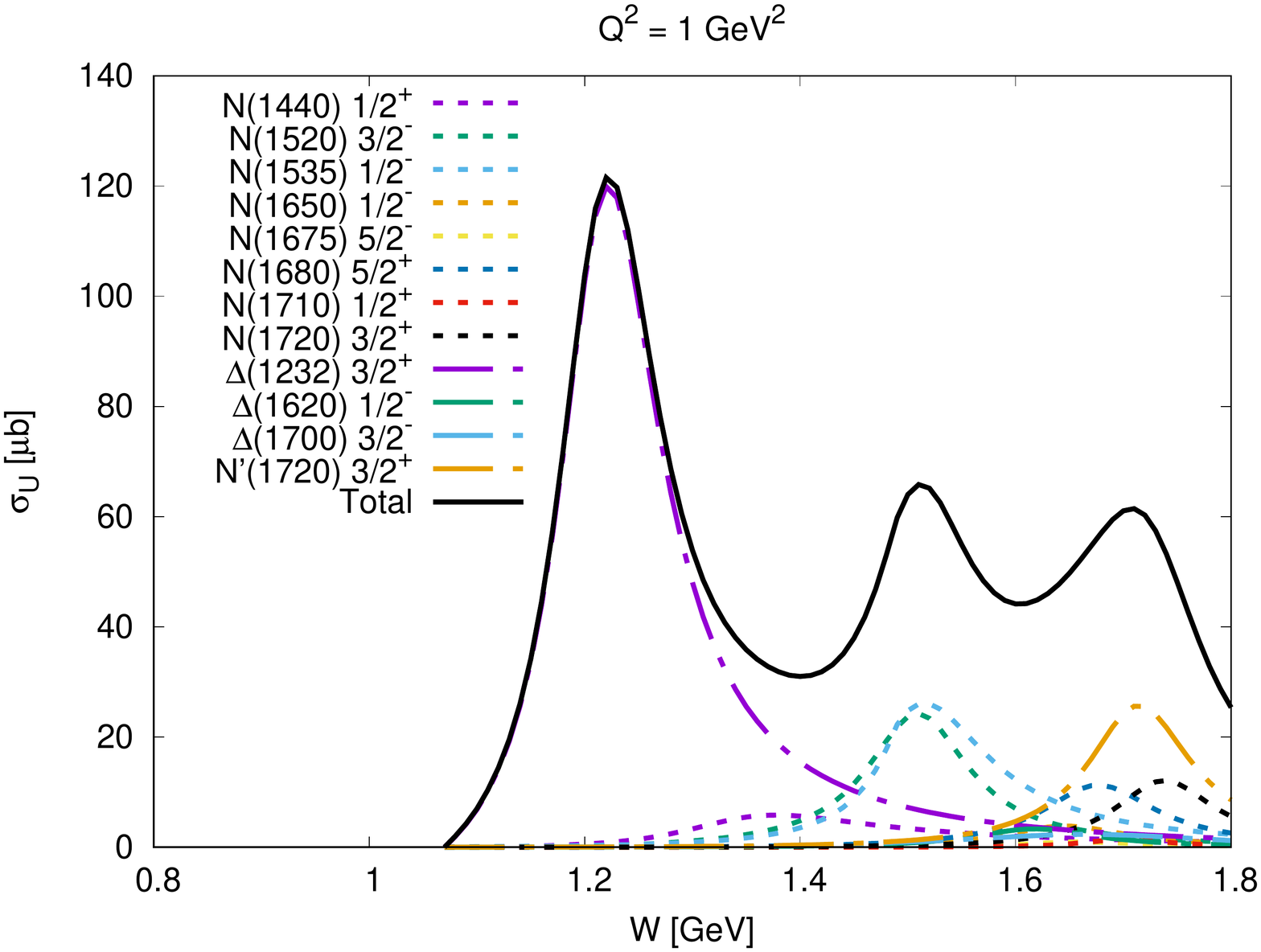}
\includegraphics[width=0.4\textwidth]{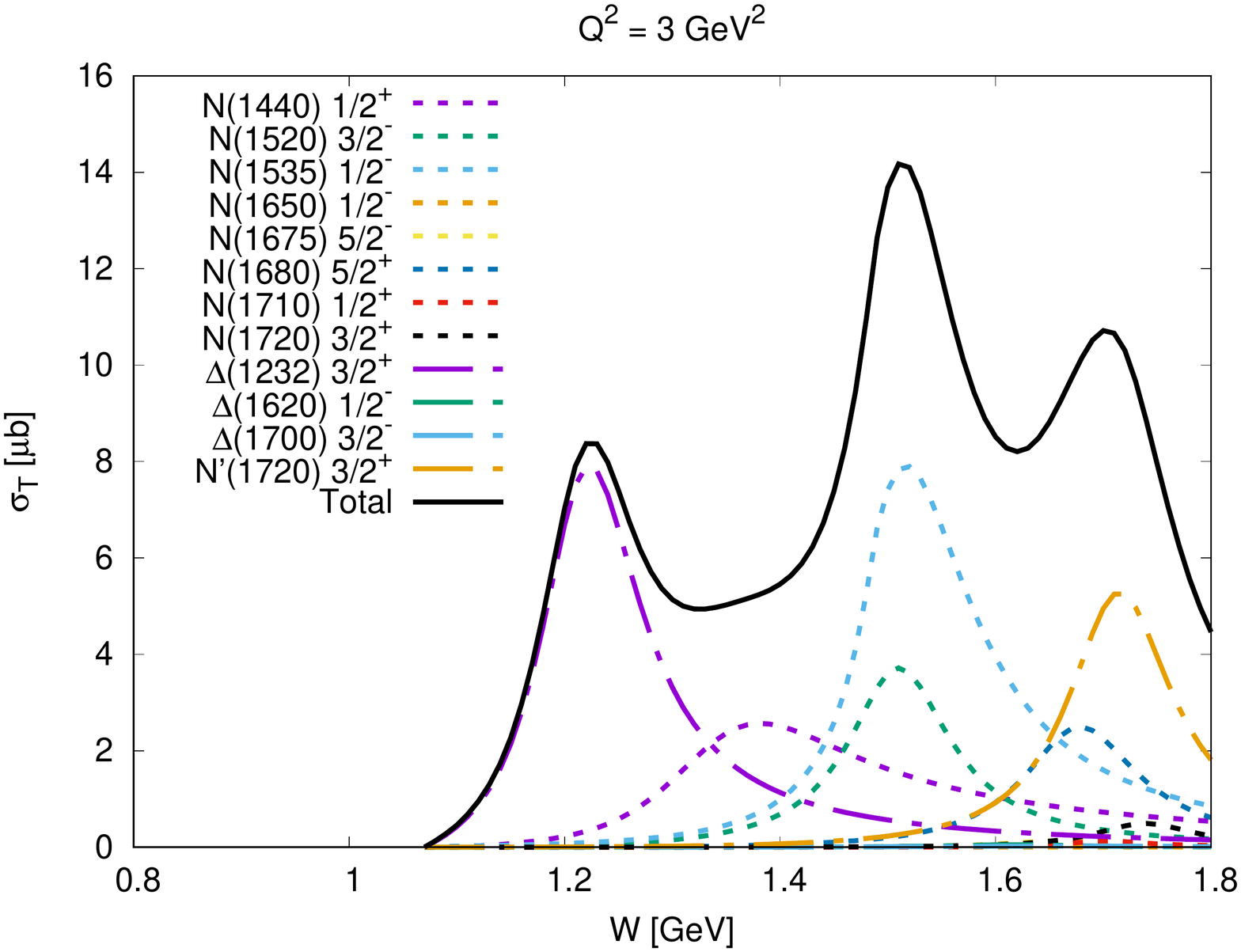}
\includegraphics[width=0.4\textwidth]{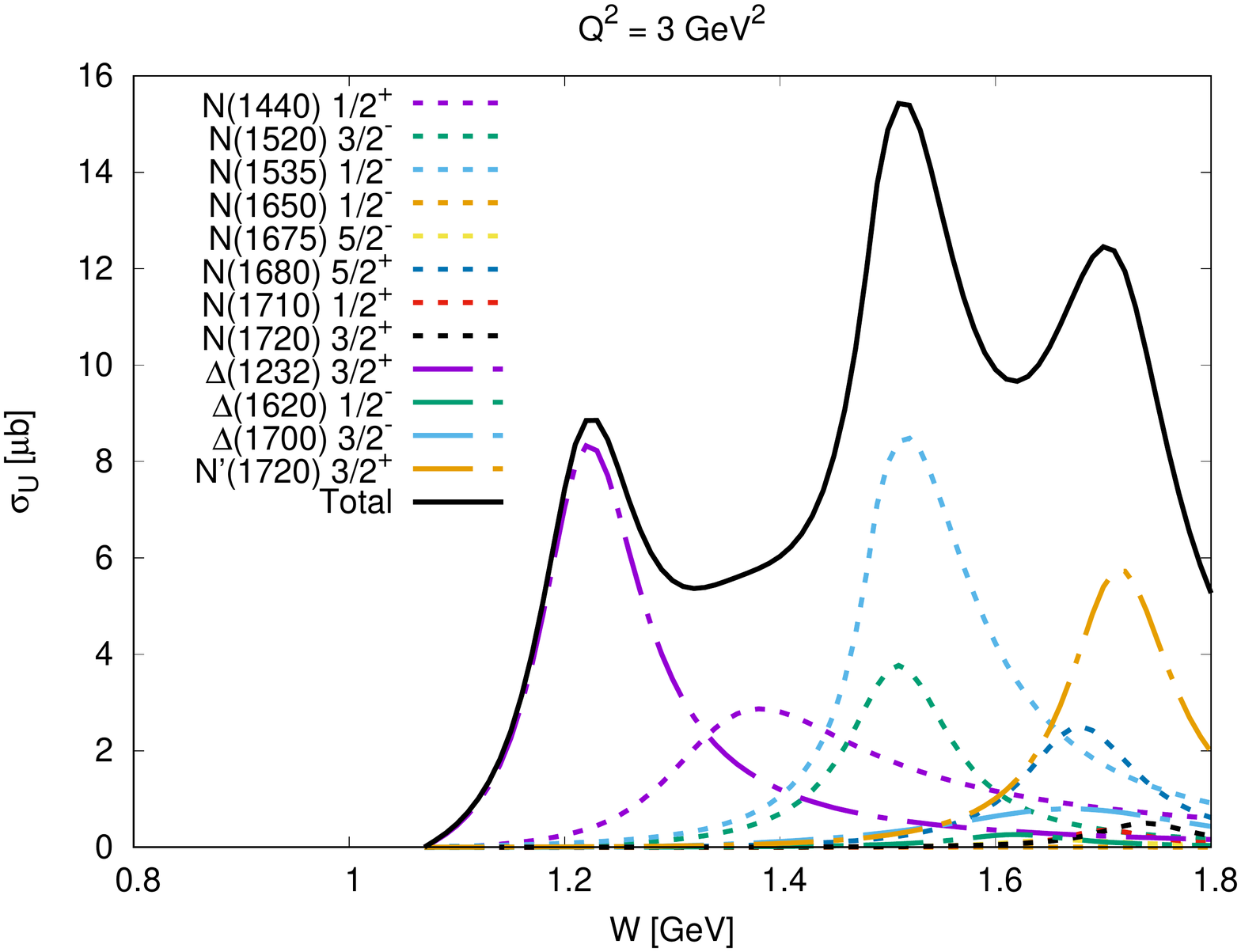}
\includegraphics[width=0.4\textwidth]{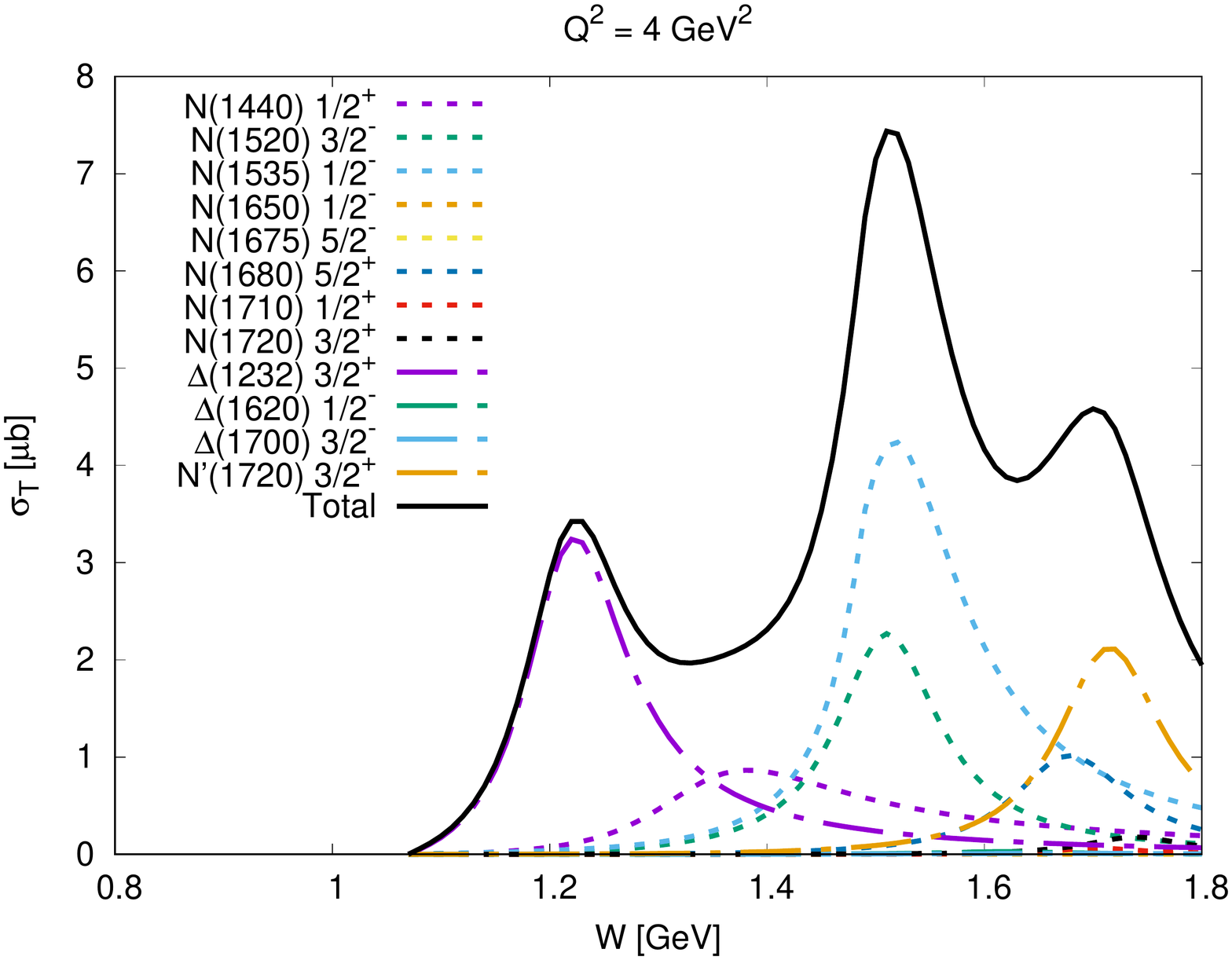}
\includegraphics[width=0.4\textwidth]{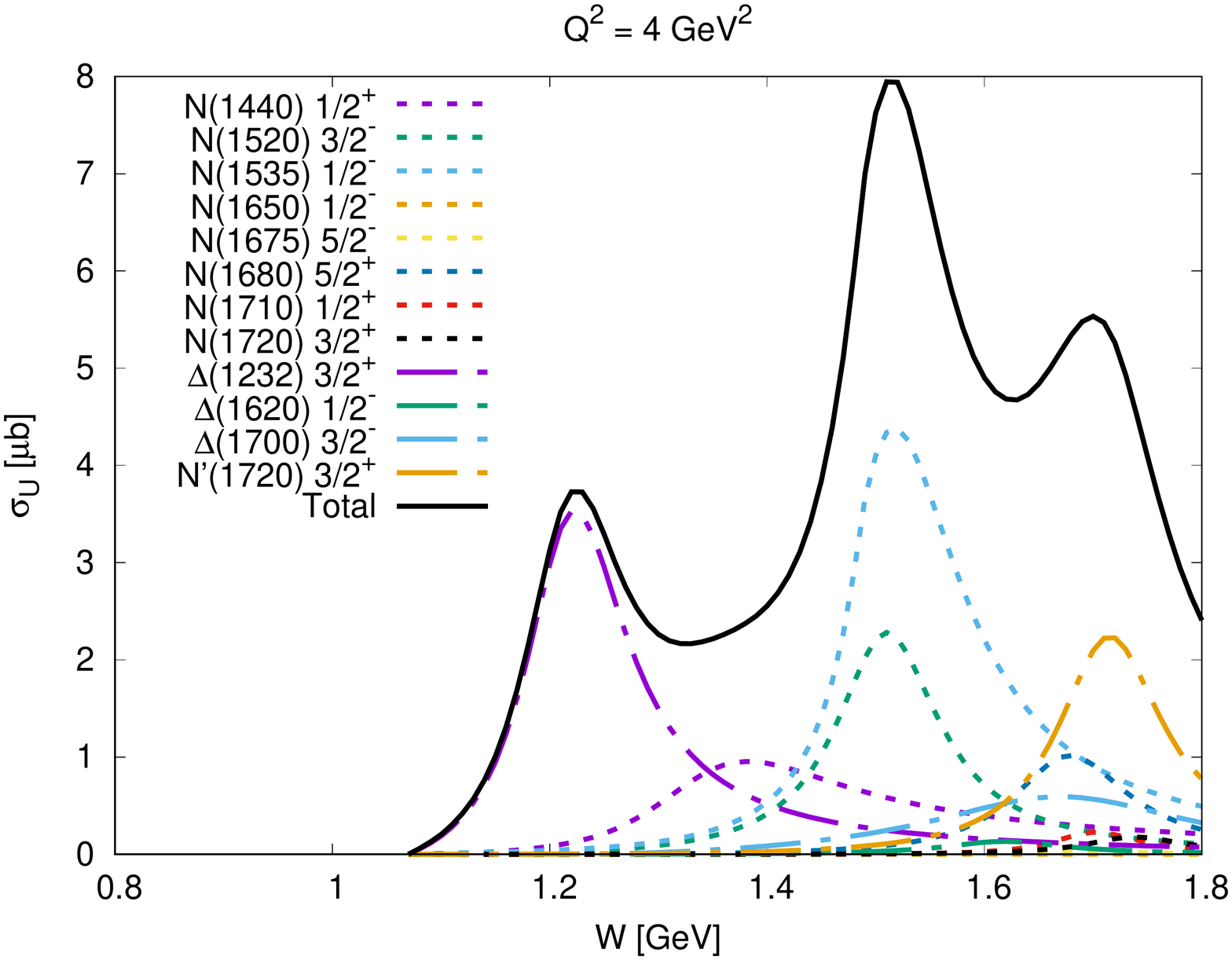}
\caption{Decomposition of the transverse (thick black curves in the left column) and unpolarized (thick black curves in the right column) resonant cross sections into the separate contributions of each resonance included in the model  at different representative $Q^2$, and at an electron beam energy of 10.6~GeV.}
\label{F:SingF2TLU}
\end{figure*}
Within the approach described in Section~\ref{S:form}, we evaluate the resonant contributions to inclusive virtual photon-proton unpolarized cross sections and to their transverse and longitudinal parts, as well as to the inclusive structure function $F_2$, in the range $1.07$~GeV$~\lesssim W\lesssim 1.8$~GeV that covers the three resonance regions. We also give predictions for inclusive electron scattering observables.

In Fig.~\ref{F:SingF2TLU}, we display the central values of the transverse and unpolarized cross sections for different $Q^2$, decomposed into the contributions from each individual resonance. Although the resonances clearly cluster into three regions, one can see that each of them displays tails that give important contributions also to the neighboring regions. Therefore, a reliable extraction of the resonance parameters requires the analysis of the observables measured in a wide $W$ interval overlapping with the neighboring resonance regions. 

 We first discuss the transverse cross section. The $\Delta(1232)~3/2^+$ represents a single contributor to the first resonance region, $1.07$~GeV$~\lesssim W \lesssim 1.4$~GeV, but a tail from the $N(1440)~1/2^+$ also affects the cross section there at $Q^2>2.5$~GeV$^2$. Furthermore, it becomes increasingly relevant at higher $Q^2$, since the $\Delta(1232)~3/2^+$ electrocouplings decrease with $Q^2$ much faster than those of the $N(1440)~1/2^+$~\cite{CLAS:SFDB}. 
 
 The transverse resonant cross sections in the second resonance region, $1.4$~GeV$~\lesssim W\lesssim 1.6$~GeV, are determined by the contributions from $N(1520)~3/2^-$, $N(1535)~1/2^-$ and the broad $N(1440)~1/2^+$. The contribution from the $N(1520)~3/2^-$ decreases with $Q^2$ faster than that from $N(1535)~1/2^-$, making the $N(1535)~1/2^-$ the largest contributor at $Q^2 > 2.0$~GeV$^2$. The slow decrease of the $N(1535)~1/2^-$ electrocouplings with $Q^2$ ~\cite{CLAS:coupsDB} results in an increase with $Q^2$ of its relative contribution to the transverse cross sections. 
 
 The third region, $1.6$~GeV$~\lesssim W \lesssim 1.8$~GeV, is composed by several overlapping resonances, the largest contributions coming from the $N(1680)~5/2^+$ and the $N^\prime(1720)~3/2^+$ candidate. The tail from the $N(1535)~1/2^-$ state becomes increasingly important in the generation of the resonant cross sections in the third resonance region at higher $Q^2$. Therefore, the knowledge on the electrocouplings of  the $N(1535)~1/2^-$ in the second region plays an important role in describing the third  region. The peak in the $W$-dependencies of the sum of contributions to the transverse cross sections at $W \approx 1.7$~GeV comes from the contribution of the candidate  $N^\prime(1720)~3/2^+$. When removing it, the peak becomes a shoulder for the whole range of $Q^2$ analysed.
 
 \begin{figure*}
\includegraphics[width=0.4\textwidth]{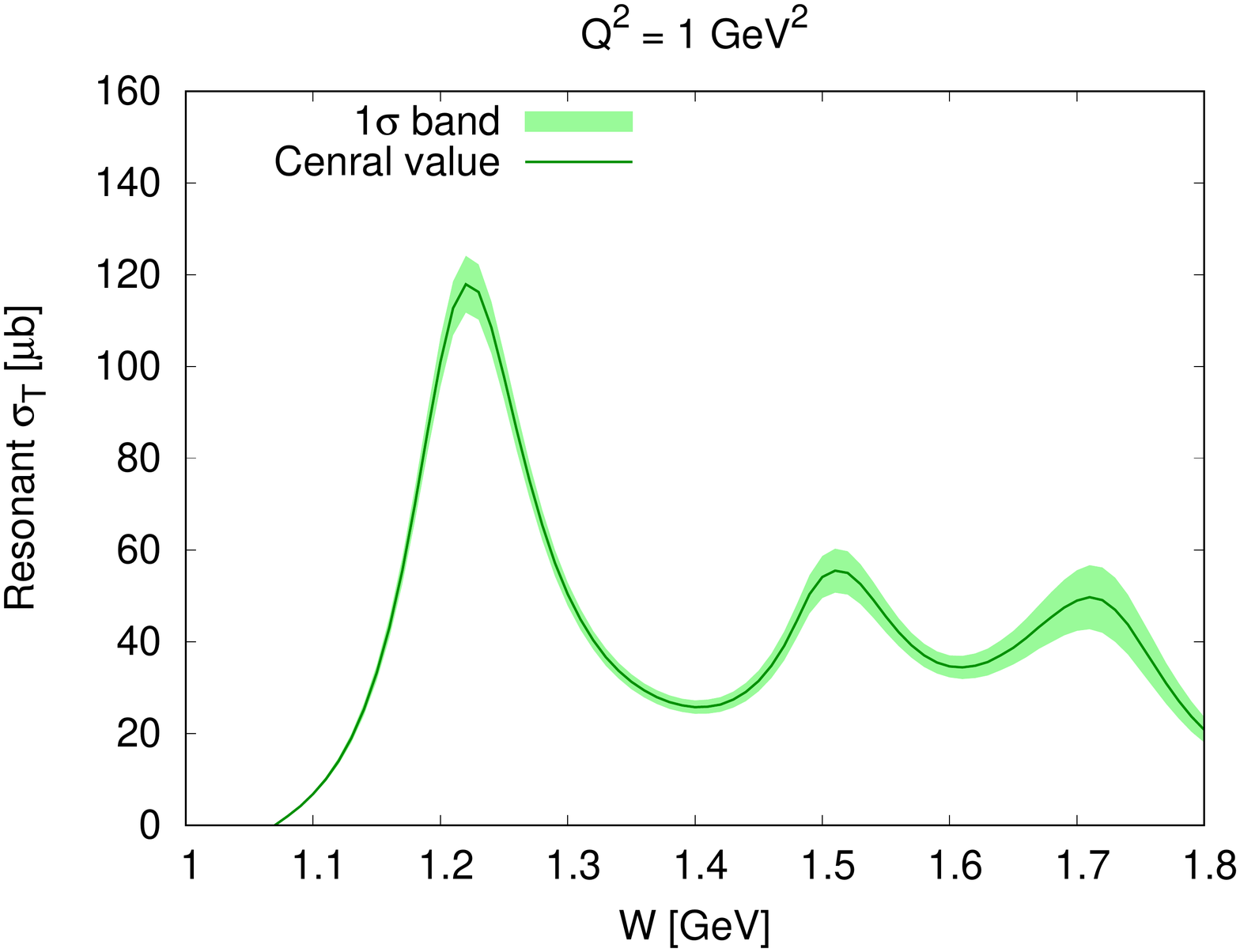}
\includegraphics[width=0.4\textwidth]{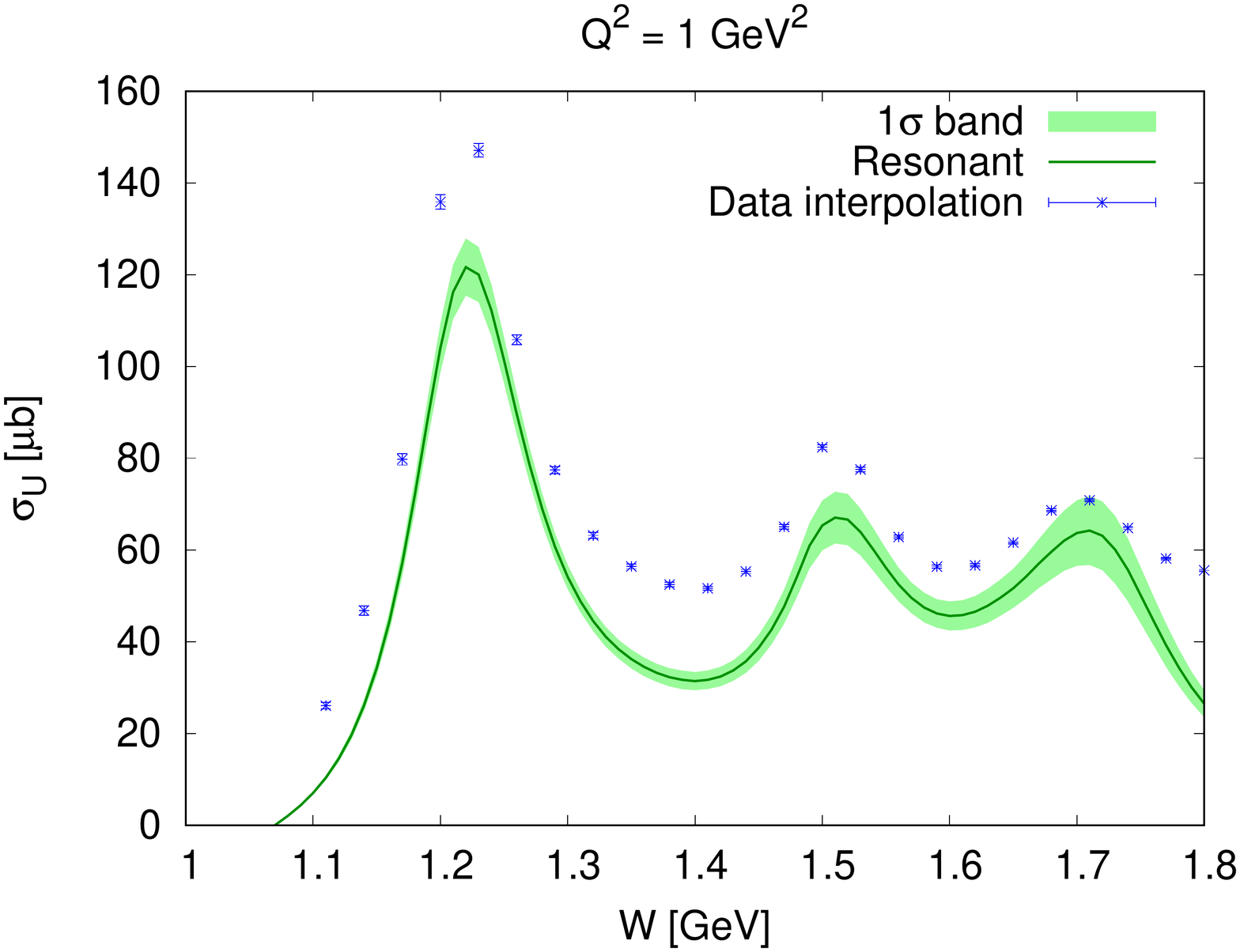}\\
\includegraphics[width=0.4\textwidth]{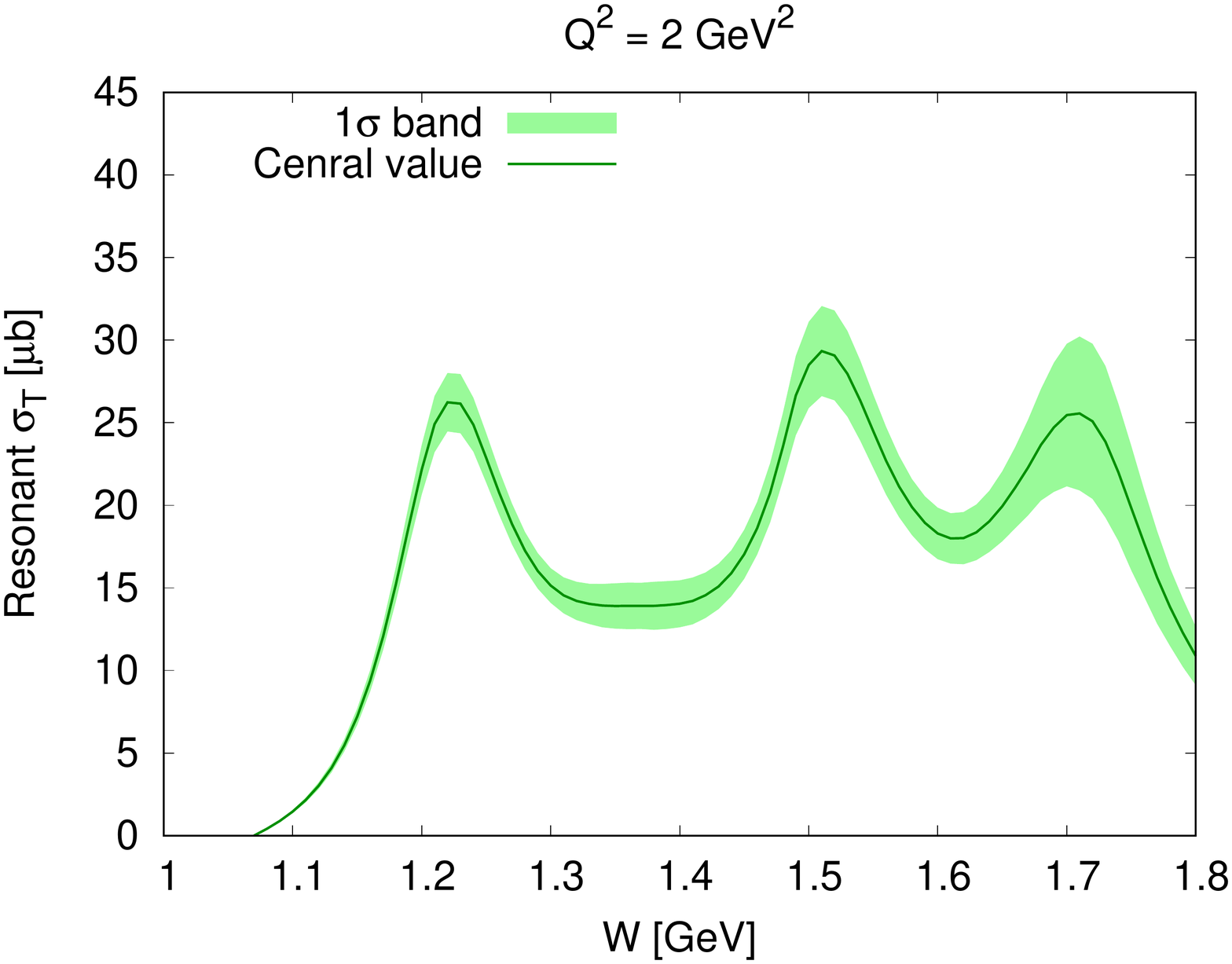}
\includegraphics[width=0.4\textwidth]{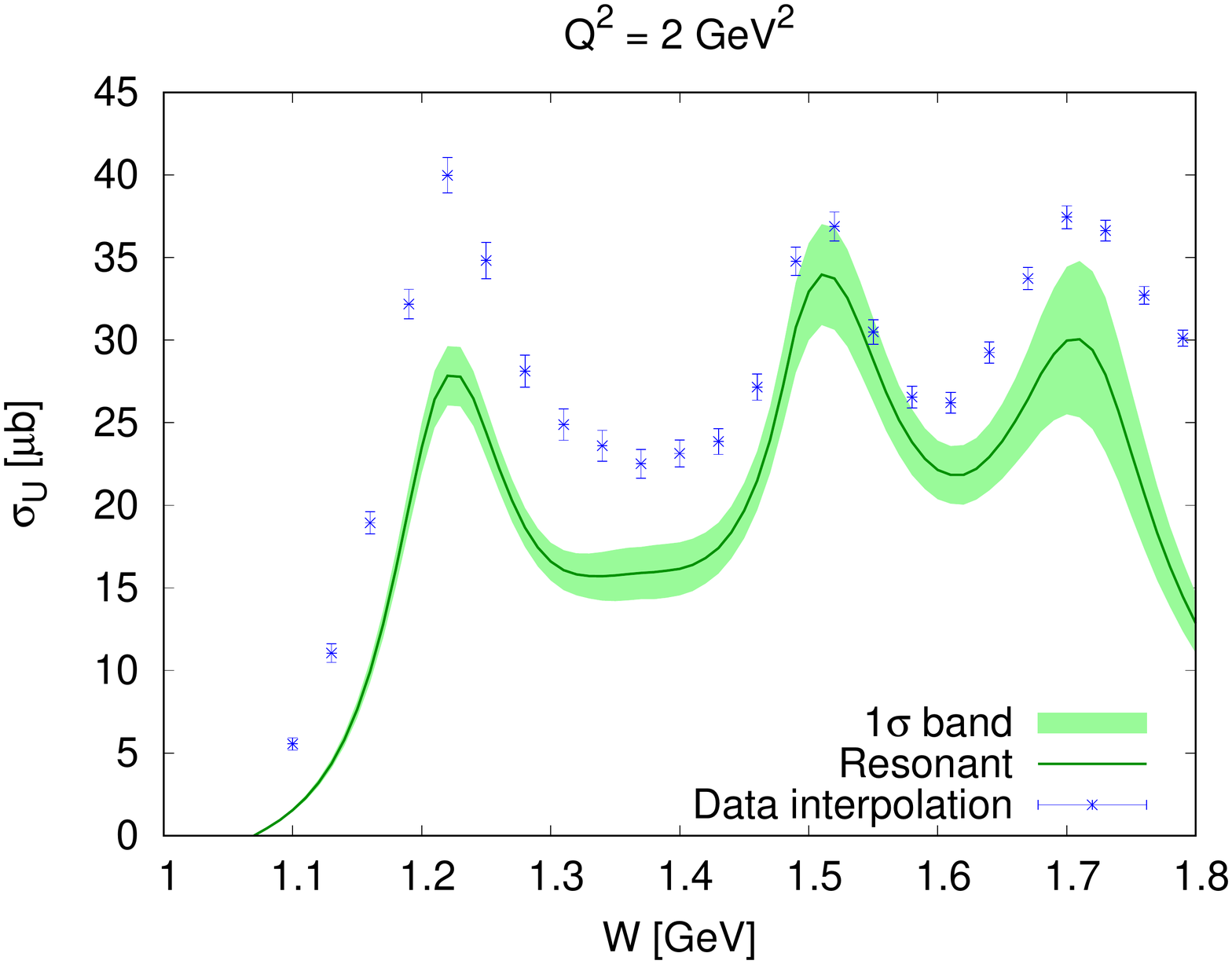}\\
\includegraphics[width=0.4\textwidth]{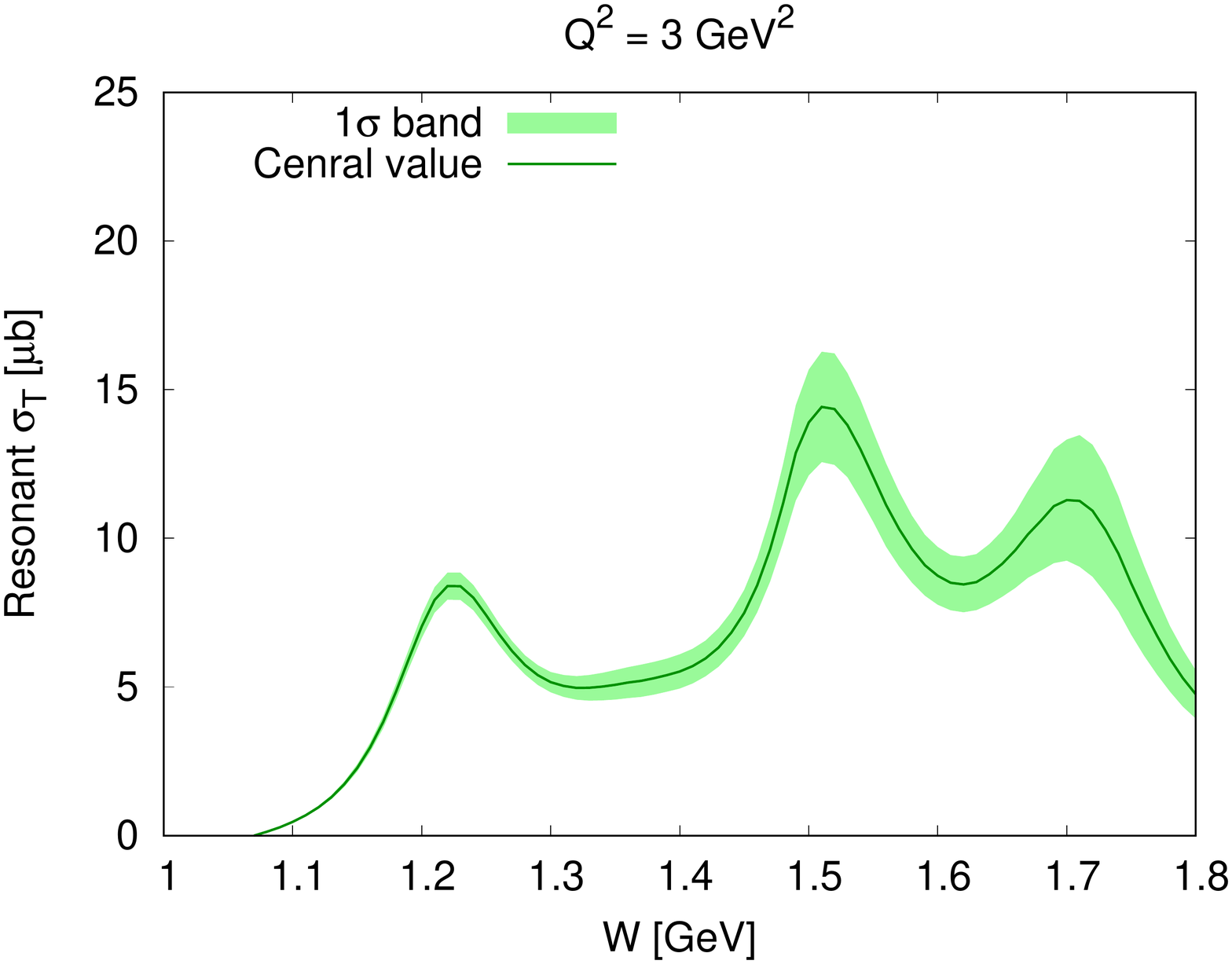}
\includegraphics[width=0.4\textwidth]{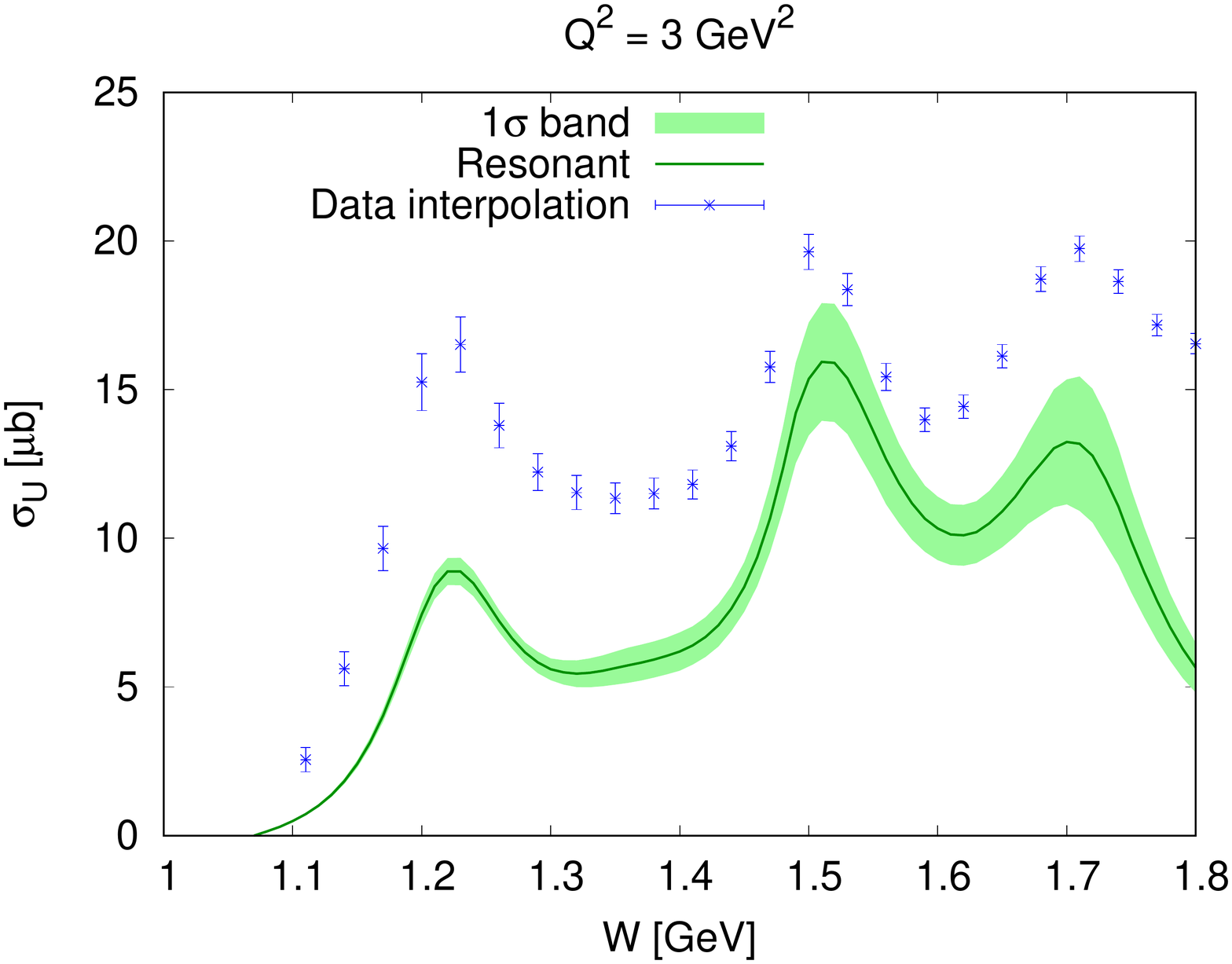}\\
\includegraphics[width=0.4\textwidth]{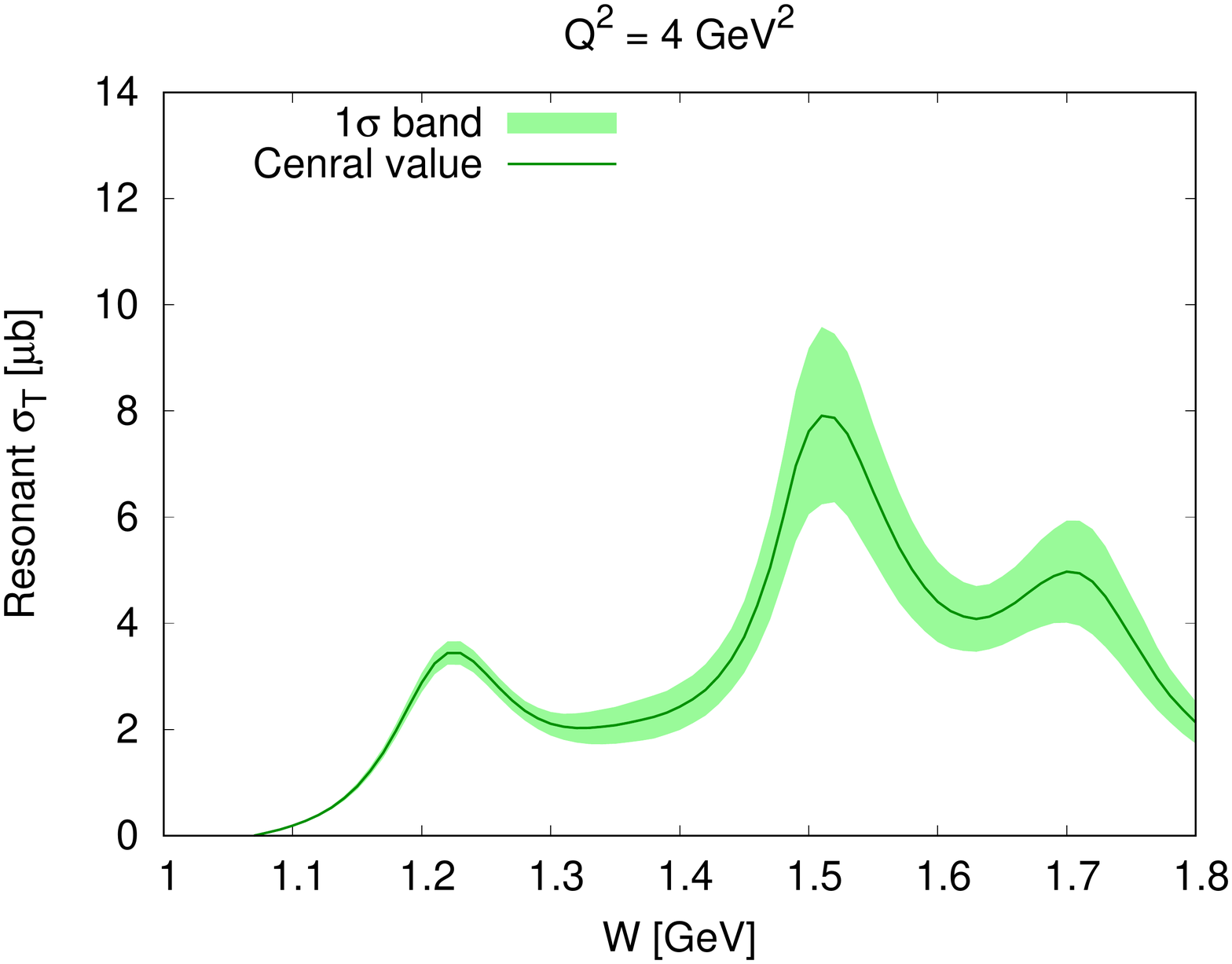}
\includegraphics[width=0.4\textwidth]{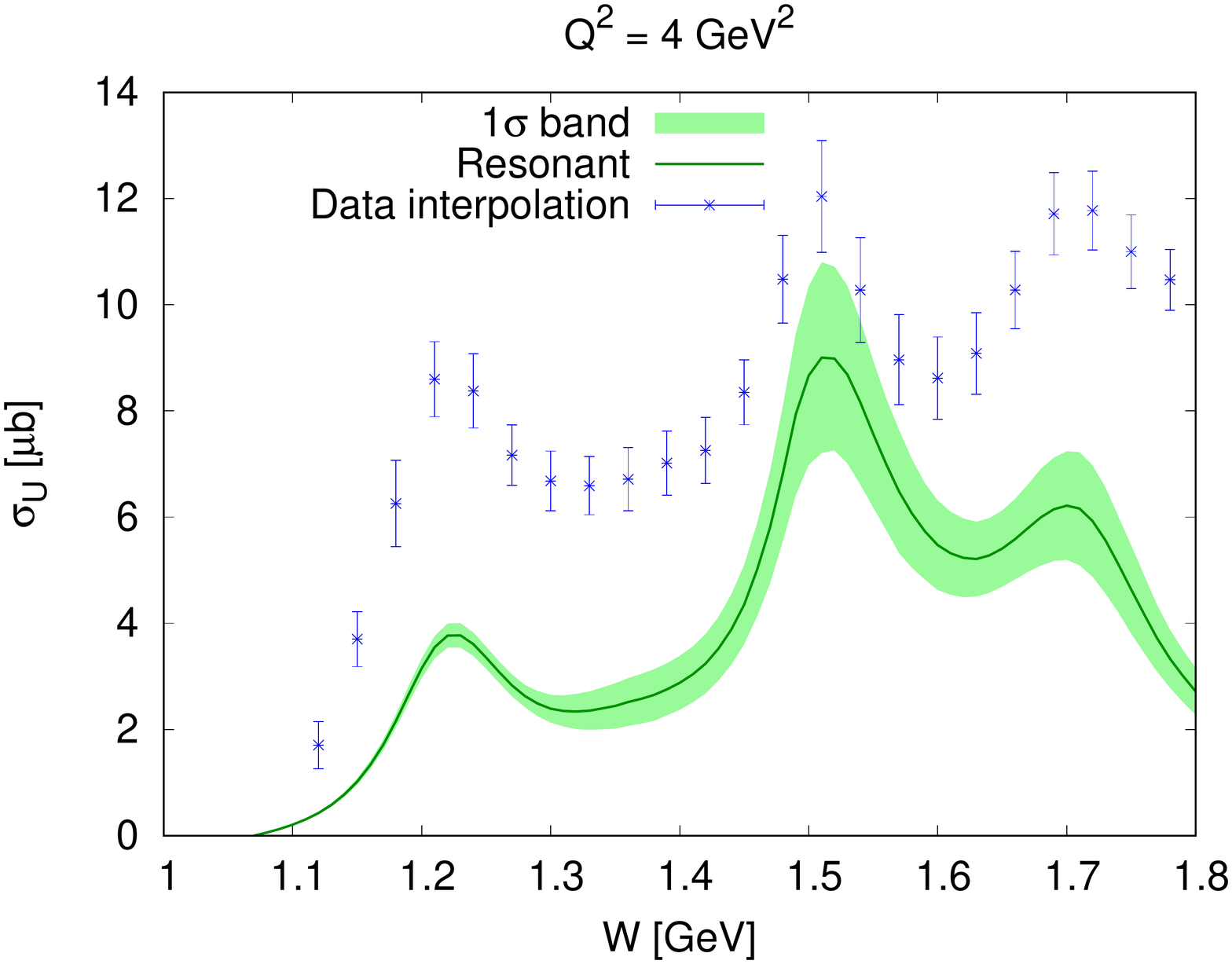}
\caption{Resonant contributions (green curves with uncertainty bands) to the transverse $\sigma_T$ (left column) and unpolarized $\sigma_U$ (right column) virtual photon-proton cross sections. The unpolarized cross sections are presented for the electron beam energy 10.6~GeV, and the data points and error bars are obtained in~\cite{CLAS:SFDB} by interpolating the CLAS and world measurements as described in Ref.~\cite{Golubenko:2019gxz}, with an updated version of $R_{LT}$, as described in the text.}
\label{F:sigsTLU}
\end{figure*}
 In Fig.~\ref{F:sigsTLU}, the total sum of all resonance contributions to   the  transverse,  $\sigma_{T}^R$ and unpolarized, $\sigma_{U}^R$ virtual photon-proton cross sections is shown for an electron beam energy of 10.6~GeV, and compared with representative examples of unpolarized cross section data~\cite{CLAS:SFDB,Golubenko:2019gxz}. 
 Note that the CLAS $F_2$ data were extracted from the measured differential cross sections via the parametrization in Ref.~\cite{Osipenko:2003bu} for the ratio $R_{LT}^\text{old}$. We opt to use an updated version for $R_{LT}^\text{new}$  as in Ref.~\cite{Tomalak:2015hva}, based on H1 and ZEUS data~\cite{Whitlow:1990gk,Dasu:1993vk,Arneodo:1996qe}. In order to do so, we use the interpolated $F_2^\text{old}$ data~\cite{CLAS:SFDB}, and transform them into the update version $F_2^\text{new}$ via the appropriate relation
\begin{align}
F_2^\text{new}=\frac{R_{LT}^\text{new}+1}{R_{LT}^\text{old}+1}\frac{\epsilon_T R_{LT}^\text{old}+1}{\epsilon_T R_{LT}^\text{new}+1}F_2^\text{old}.
\end{align}
The unpolarized cross section data are transformed accordingly. Note that the choice of the new $R_{LT}$ parametrization leads to a slight change in the unpolarized cross section data, especially in the second and third resonance regions and larger $Q^2$ bins. However, the difference is smaller than the 1$\sigma$ theory bands.

 We discussed above how the different resonances cluster into three regions, which correspond to the three peaks in $W$ observed in the unpolarized cross sections. The transverse resonant part gives the largest contribution
  to the resonant cross sections. The size of the longitudinal part increases with $W$, but overall does not exceed 30\% of the total cross section in the kinematical region shown in Fig.~\ref{F:sigsTLU}. Hence, the shapes of the resonant contributions to unpolarized and transverse cross sections are similar, and they both clearly show three separate peaks. 
  The unpolarized resonant cross sections show a pronounced evolution with $Q^2$: both the first and the third  regions show a stronger fall-off with $Q^2$ than the second  peak. In particular, in the first  region at $Q^2=1$~GeV$^2$, the resonant contribution is responsible for about 70\% of the cross section; at $Q^2=4$~GeV$^2$ it accounts for less than 40\%. In contrast, in the second  region the resonant contributions remain almost unchanged with $Q^2$, at the level of 80\%, mainly due to the slow evolution with $Q^2$ of the $N(1535)~1/2^-$ $A_{1/2}$ electrocoupling. In the third  region, the resonant contribution decreases from 90 \% to 50 \% within the aforementioned $Q^2$ range. This suggests that the different excited nucleon states display distinctively different structural features in the $Q^2$ evolution of their electrocouplings, further underlining the results in Fig.~\ref{F:SingF2TLU}. Therefore, in order to explore the strong QCD dynamics underlying the generation of the ground and excited nucleon states, the results on electrocouplings of all prominent nucleon resonances are needed. 

In the future our results on the resonant longitudinal contributions could be compared to the longitudinal virtual photon-proton cross sections. These can be inferred from the Hall~C data at Jefferson Lab on inclusive electron scattering cross sections~\cite{Liang:2004tj,Tvaskis:2016uxm}, which provide the information on the $R_{LT}$ ratio. The knowledge of the resonant contributions to the longitudinal cross section offers new opportunities in accessing gluon distributions  at $x$ within the resonance region~\cite{Altarelli:1978tq}.

\begin{figure*}
\includegraphics[width=0.4\textwidth]{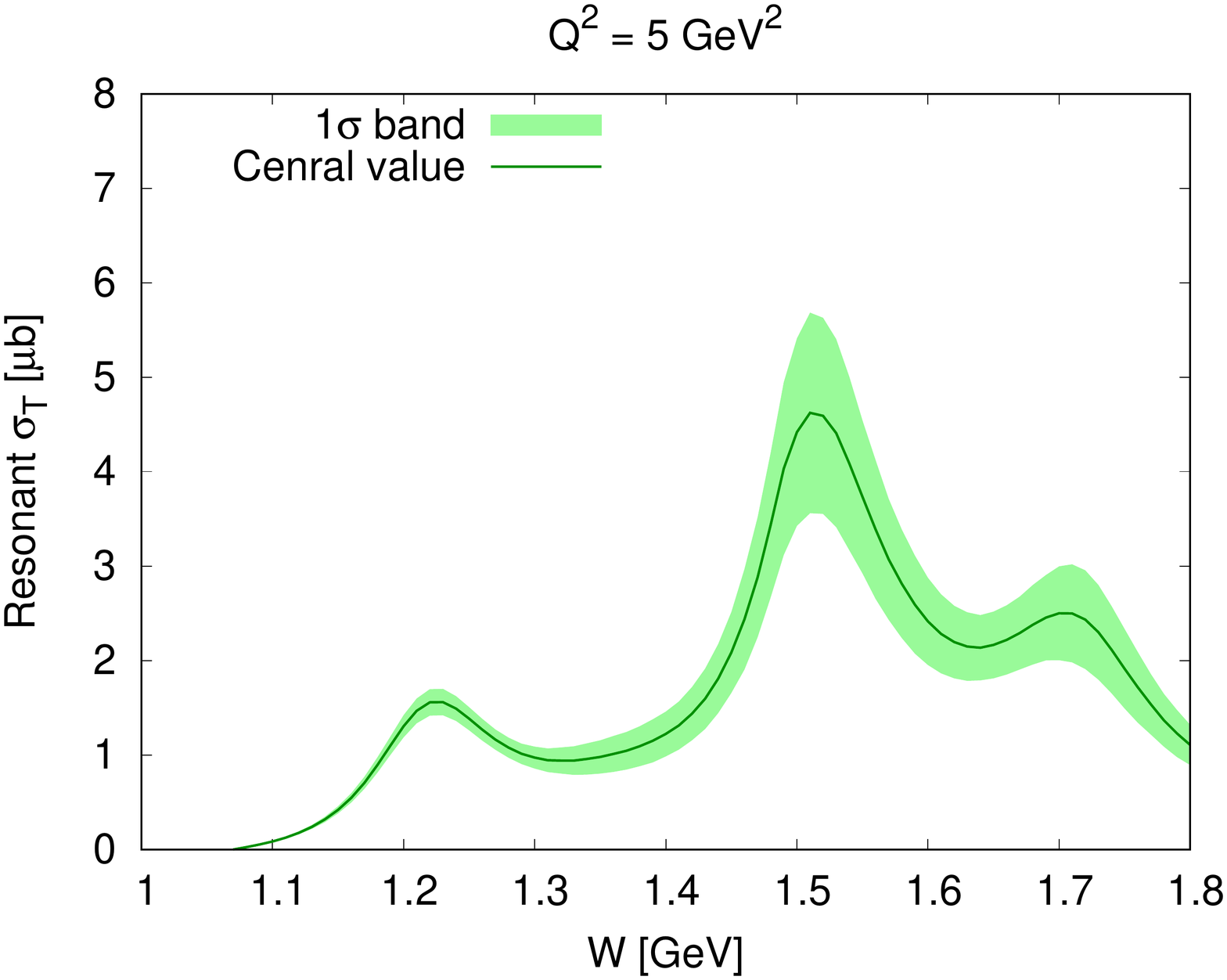}
\includegraphics[width=0.4\textwidth]{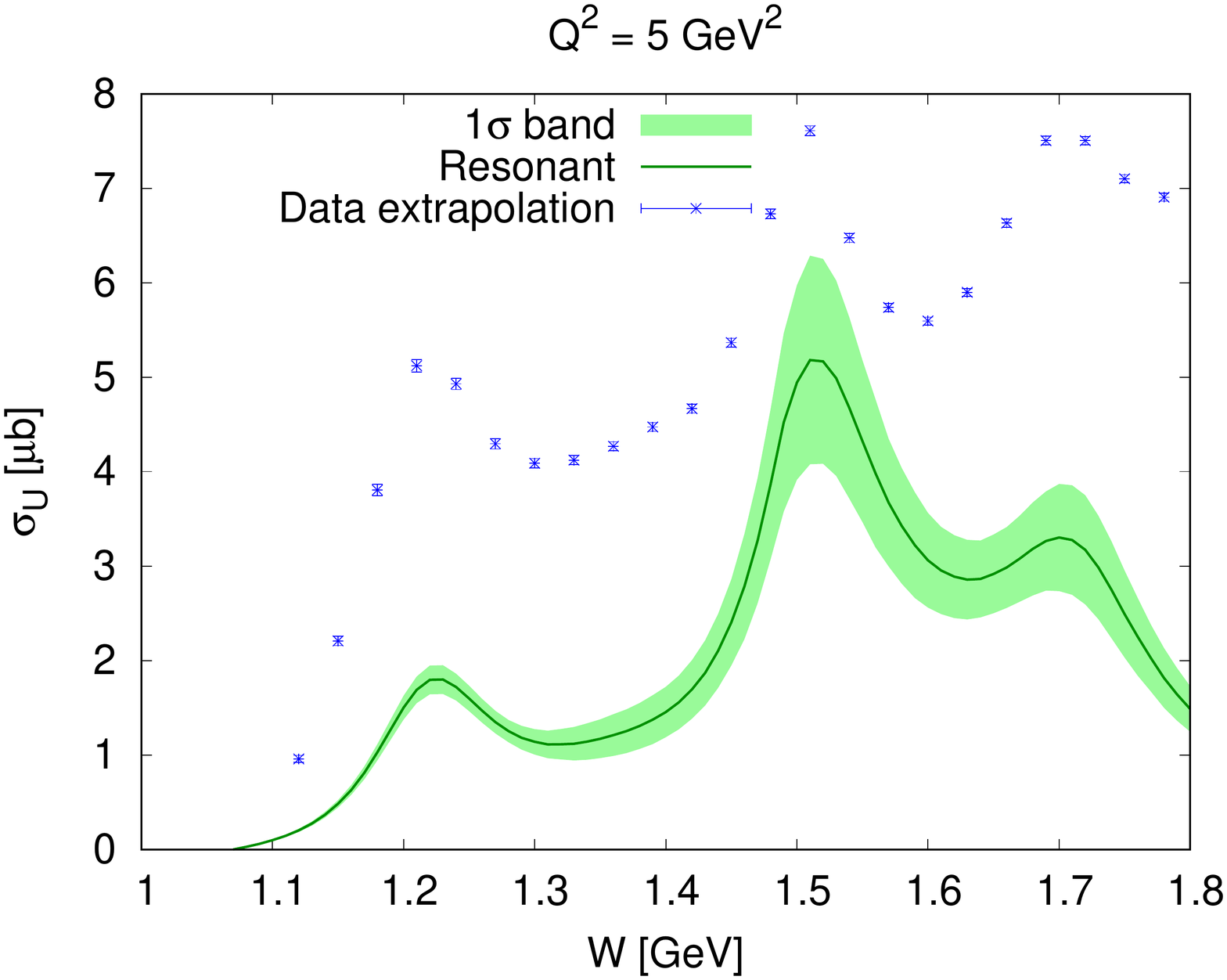}\\
\includegraphics[width=0.4\textwidth]{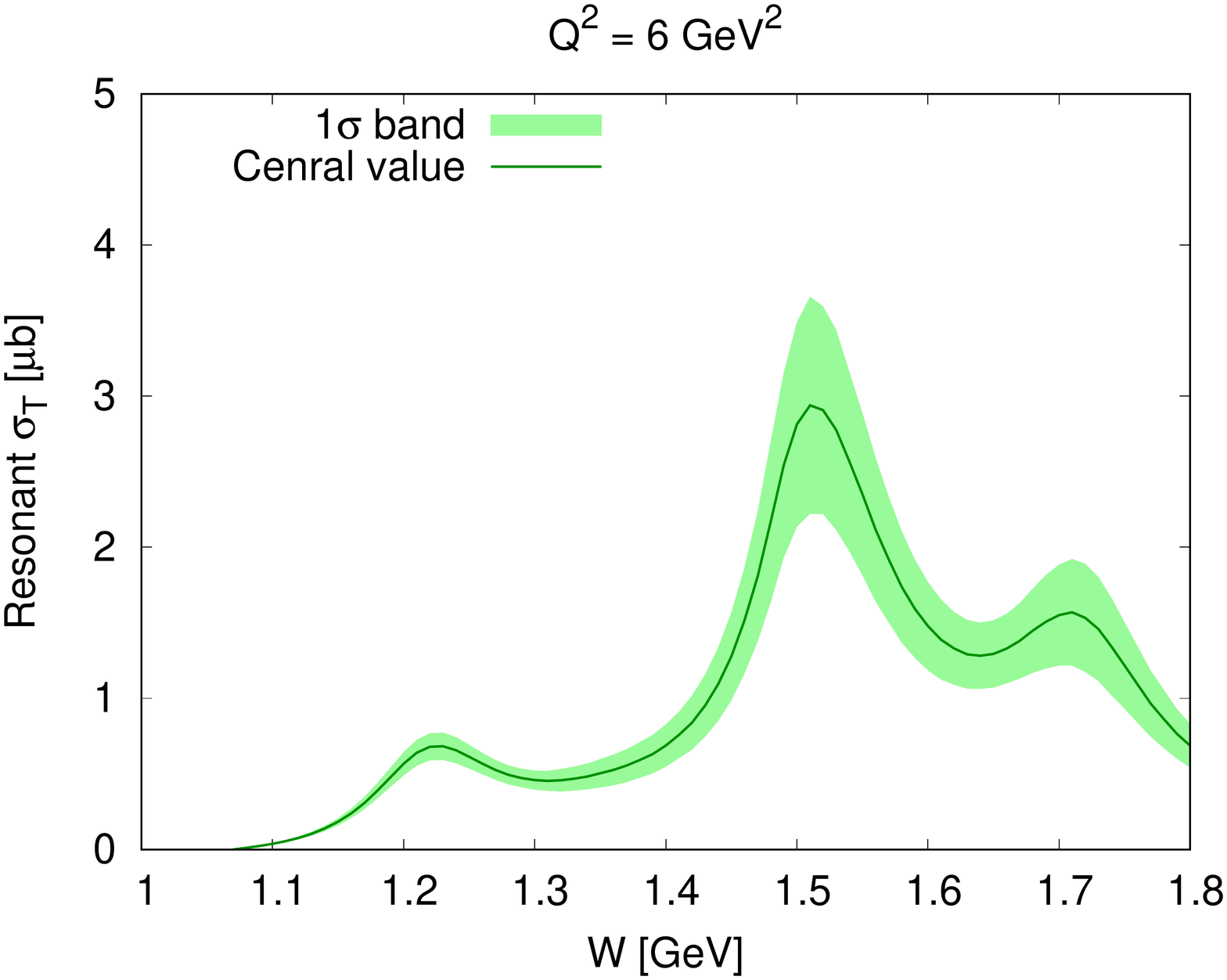}
\includegraphics[width=0.4\textwidth]{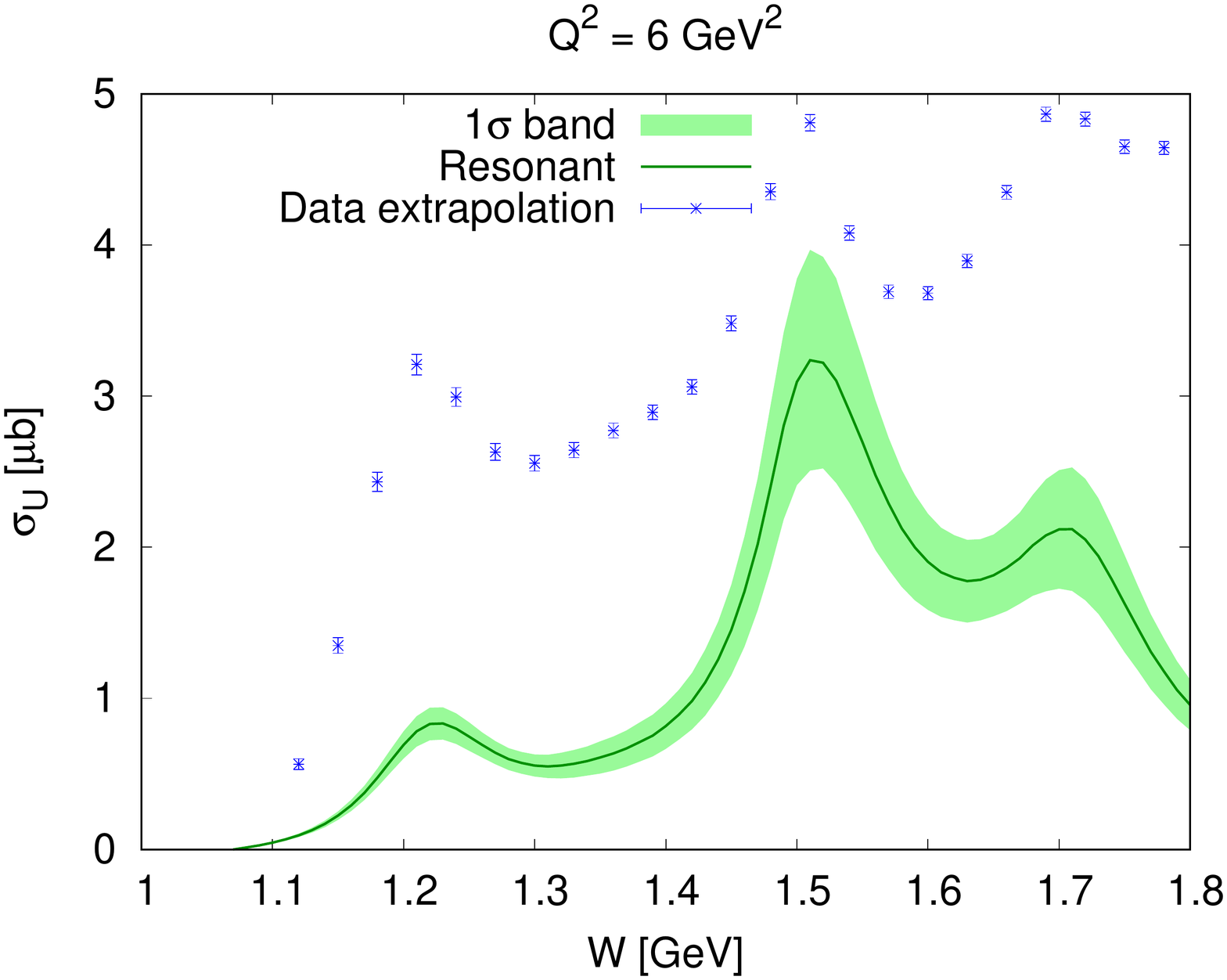}
\caption{Predicted inclusive virtual photon-proton cross sections (points with error bars) and resonant contributions at $Q^2=5.0$~GeV$^2$~(top) and $Q^2=6.0$~GeV$^2$~(bottom) in the kinematic area covered in the measurements with the CLAS12 detector~\cite{Burkert:2018nvj}. The error bars for the projected data shown in the plot are obtained for the statistics corresponding to the integrated luminosity collected with CLAS12 in the Spring 18 run and bin sizes $\Delta W=0.01$~GeV and $\Delta Q^2=0.1$~GeV$^2$. Further explanations as in Fig.~\ref{F:sigsTLU}.}
\label{F:sigsTLUprojected}
\end{figure*}
First experiments with the CLAS12 detector in Hall~B at Jefferson Lab started  in the Spring of 2018. The expected results will extend the available data on inclusive electron scattering in the resonance region to high photon virtualities of $Q^2>4.5\text{~GeV}^2$~\cite{Burkert:2018nvj}. The predicted virtual photon-proton cross sections based on the interpolation/extrapolation of the experimental results~\cite{CLAS:SFDB} and the resonant contributions estimated as described in Section~\ref{S:form} are shown in Fig.~\ref{F:sigsTLUprojected}. The uncertainties are computed for the expected inclusive electron scattering event statistics collected in the Spring 2018 run of integrated luminosity $12.8 ~
 \mu b^{-1}$~\cite{Golubenko:2019gxz}, and for the kinematic grid bin sizes $\Delta W=0.01$~GeV and $\Delta Q^2=0.1$~GeV$^2$. The expected statistical precision of the data is in the range from 0.2\% to 2.0\%, suggesting that the data uncertainties will be mostly determined by the measurement systematics. The computed resonant contributions allow us to elucidate the role of nucleon resonances in the measured inclusive electron scattering observables. The present paper thus offers a phenomenological tool for the analysis of the inclusive electron scattering data measured with CLAS12. This tool is available online~\cite{CLAS:SFDB}, allowing the interactive evaluation of inclusive electron scattering observables together with resonant contributions computed for integrated luminosities and kinematics grids as defined by the user.

\begin{figure*}
\includegraphics[width=0.45\textwidth]{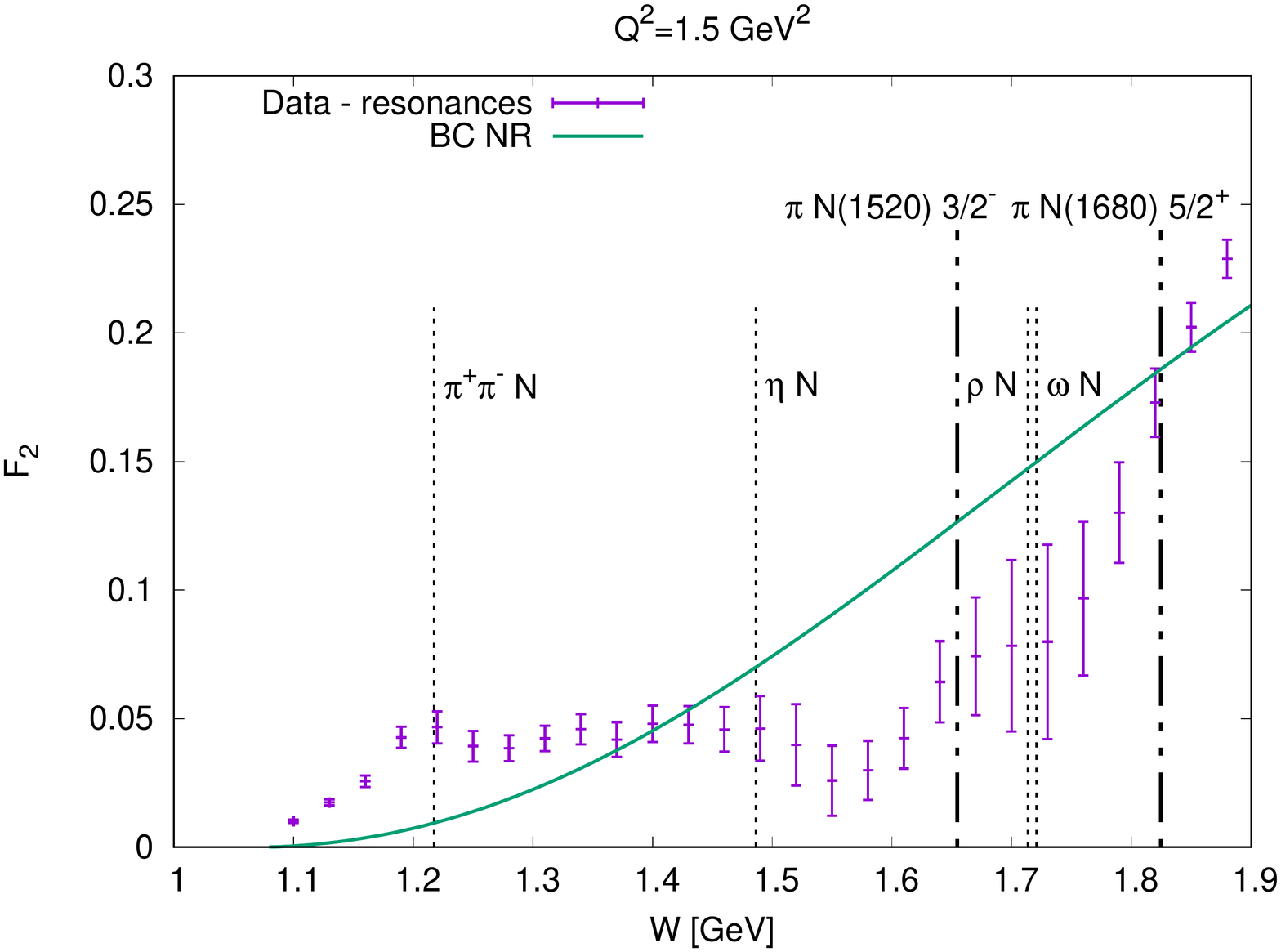}
\includegraphics[width=0.45\textwidth]{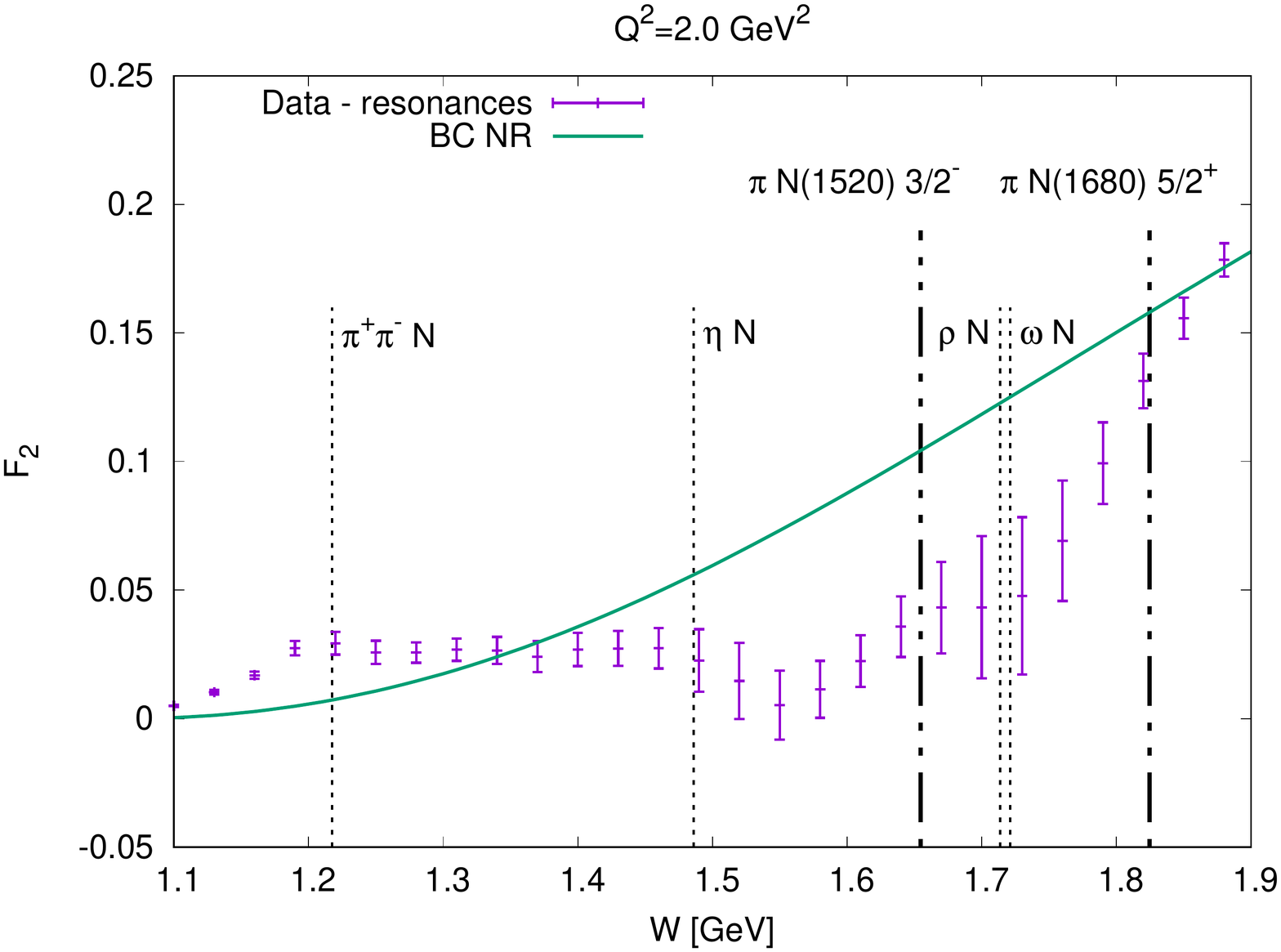}
\includegraphics[width=0.45\textwidth]{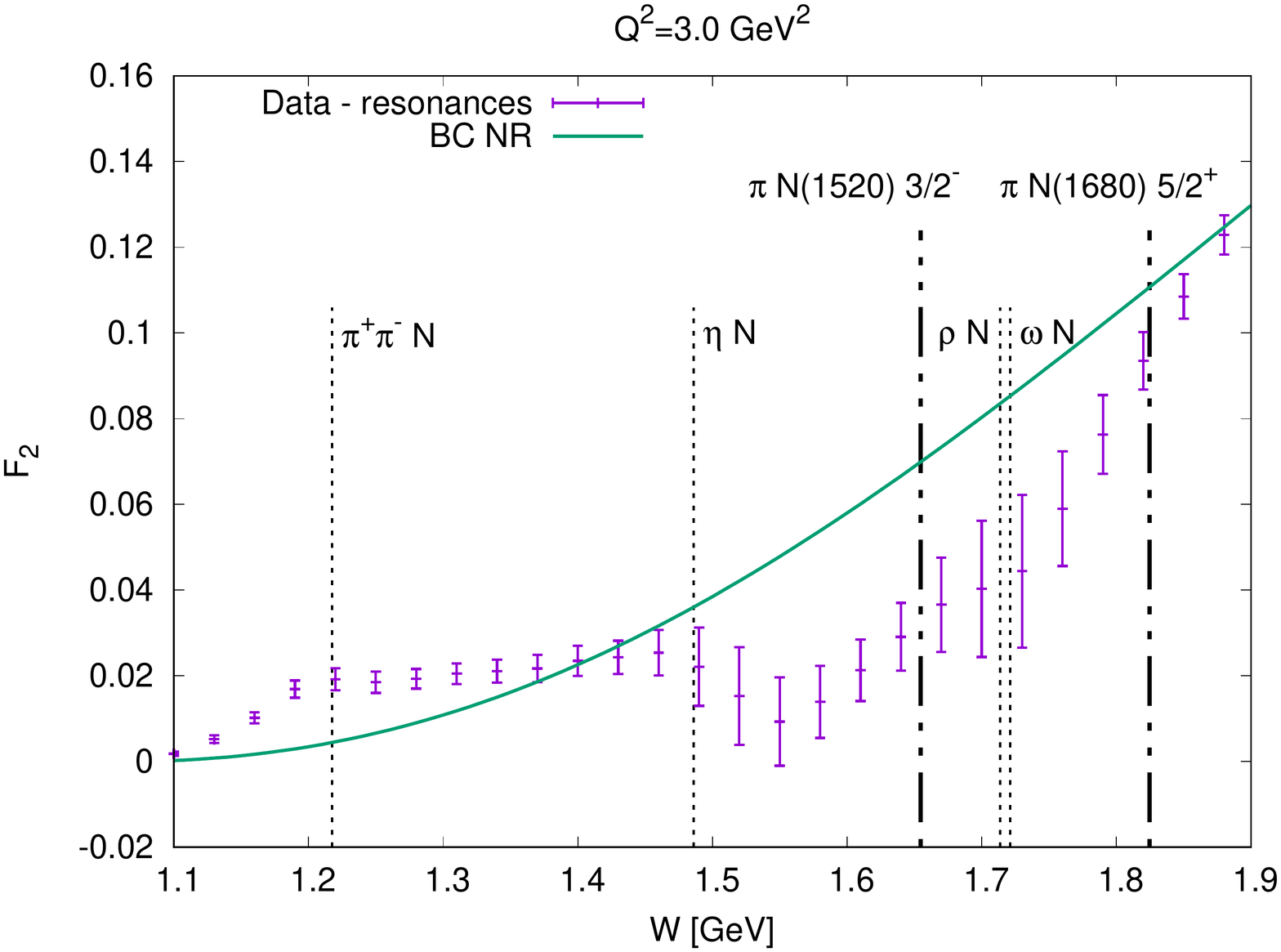}
\includegraphics[width=0.45\textwidth]{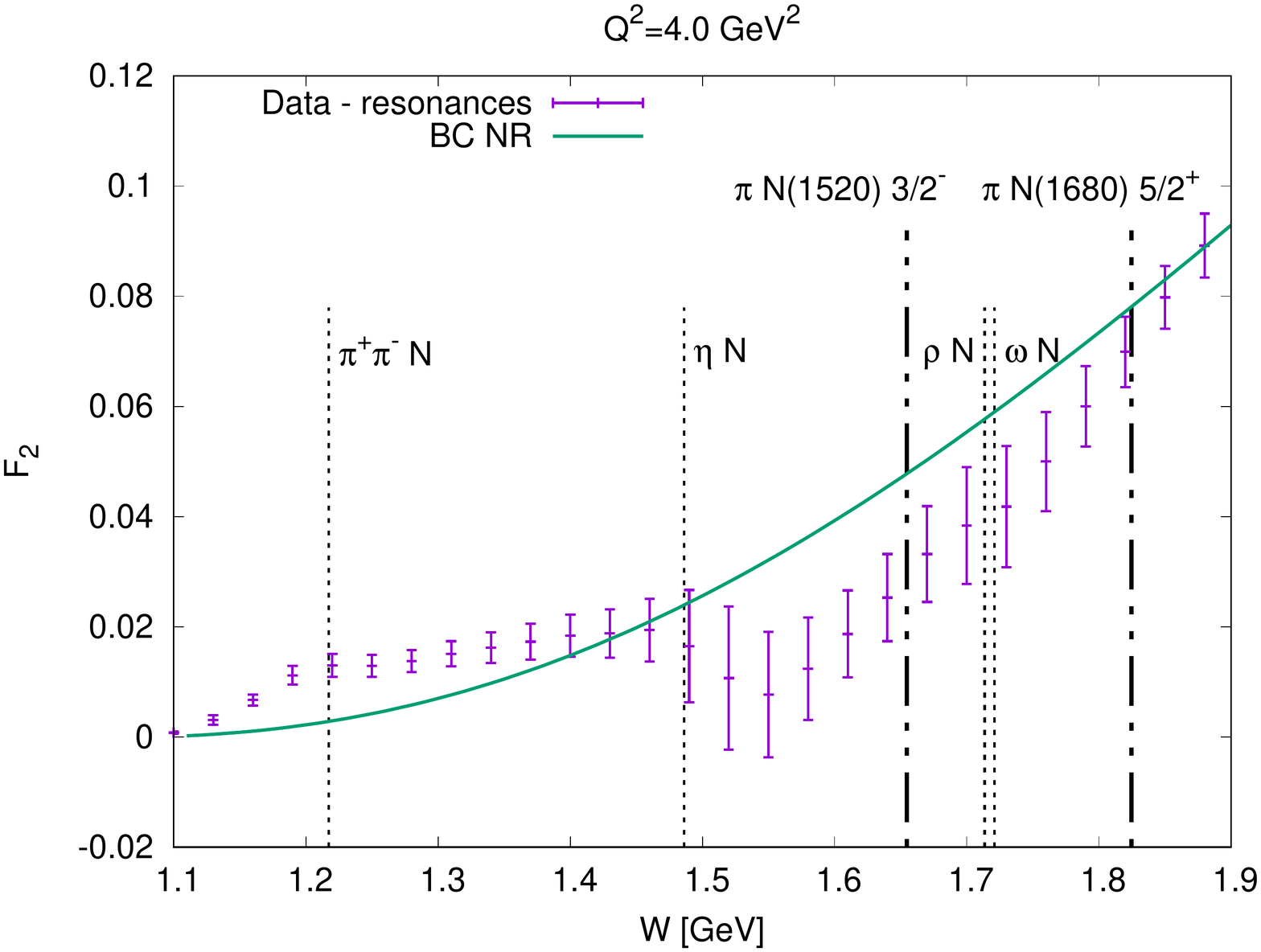}
\caption{Results (purple points with error bars) on the differences between the  $F_2$ structure functions estimated by interpolating the CLAS/world results~\cite{CLAS:SFDB,Golubenko:2019gxz} and the estimated resonant contributions, compared to the background as in Ref.~\cite{Christy:2007ve} (green curve). The dotted vertical lines show the opening of meson-nucleon electroproduction channels. The dash-dotted vertical lines show the opening of $\pi~N^*$ channels. The data and model uncertainties are propagated into the final error bars shown. The green curve is the background (NR) as in the model of Ref.~\cite{Christy:2007ve} (BC), which can be considered as the continuation of the background from the deep inelastic scattering region into the resonance region.}
\label{F:F2NR}
\end{figure*}
In our approach, the estimates for the resonant contributions are obtained from exclusive meson electroproduction data, which are independent from the inclusive electron scattering observables. This enables to evaluate the non-resonant contributions to inclusive electron scattering observables as the differences between the measured inclusive observables and the estimated resonant contributions. Of course, these estimates cannot duly isolate the interference terms between the resonant and non-resonant contributions for the exclusive channels. In Fig.~\ref{F:F2NR}, we show the thus obtained non-resonant contributions to the CLAS data on the inclusive structure function $F_2$~\cite{Osipenko:2003bu}. Unlike for the unpolarized cross sections, the choice of the new parametrization for $R_{LT}$ is barely noticeable in the $F_2$ data, since this observable is not very sensitive to this ratio. Overall, bearing in mind the large number of resonances and opening channels present, the resulting non-resonant contributions are rather smooth functions of $W$, especially at higher $Q^2$. We compare the contributions inferred from the data with the background model from Ref.~\cite{Christy:2007ve}, which can be considered as the continuation of the background from the deep inelastic scattering region into the resonance region. The non-resonant contributions determined by realistically accounting for the nucleon resonances, as described in Section~\ref{S:form}, demonstrate several structures and a sharp increase at $W$ from 1.6~GeV to 1.7~GeV seen in all $Q^2$-bins.  
One can observe several kinks in the $W$ dependence of the background for $F_2$. It appears, however, that each of them is associated with the opening of a  meson-baryon channel, namely $\pi\pi N$ at 1.21~GeV, $\eta N$ at 1.49~GeV,  and $\omega N$ at 1.72~GeV. We also show the $W$-values for the opening of the $\pi^+ N(1520)~3/2^-$, $\pi^+ N(1680)~5/2^+$ and $\rho N$ channels, calculated at the resonance {\bf central} masses. 
Because of the  appreciable decay widths ($\gtrsim 100$~MeV) of the unstable final states, instead of kinks these channels produce sharp growths in the $W$ dependence of the $F_2$ structure function at $1.6$~GeV~$\lesssim W \lesssim 1.7$~GeV, seen in all $Q^2$ bins.  All these features seen in the $W$ dependence 
 of the non-resonant contribution defined here are likely related to the manifestation of coupled channel effects.

\section{Summary}\label{S:sum}
In this work, we have developed a model for the evaluation of the resonant contributions to observables of inclusive electron scattering off protons. Due to the advances in exclusive reaction studies with CLAS~\cite{Aznauryan:2011qj,Mokeev:2018zxt}, we have experimental results on $\gamma_{v}pN^*$ electrocouplings of most of the excited nucleon states in the mass range below 1.8~GeV. For the first time, the resonant contributions to inclusive electron scattering observables have been evaluated from the experimental results on $\gamma_{v}pN^*$ resonance electrocouplings and the total decay widths. In particular, we computed the virtual photon and electron scattering cross sections and the $F_2$ structure function, at $W \leq 1.8$~GeV and 0.5~GeV$^2\leq Q^2\leq 6.0$~GeV$^2$. Our approach allows for the separation of the resonant part into longitudinal and transverse contributions.

Our studies elucidate the contributions from excited nucleon states to the three resonance regions. We observed substantial contributions from the resonance tails in the neighboring  regions. We found a nontrivial behaviour in the $Q^2$ evolution of the resonance structures: the first and third  peaks decrease strongly with $Q^2$, not only in their absolute value, but also relatively to the background; the second region decreases with $Q^2$, but it stays relatively constant with respect to the background. Such a behaviour underlines the essential differences in the structure of the excited nucleon states in the three regions. Studies of the $\gamma_{v}pN^*$ electrocouplings of all prominent nucleon resonances offer a unique way to explore the many facets of strong QCD in the generation of excited nucleons of different quantum numbers with distinctively different structural features.

By comparing the resonant contributions to the $F_2$ data from CLAS~\cite{Osipenko:2003bu}, we are therefore able to extract the separate contribution of the background as the difference between data on  $F_2$ inclusive structure functions and the resonant contributions. The thus evaluated background shows several kinks and a sharp increase at $W$ from 1.6~GeV to 1.7~GeV in all $Q^2$ bins covered by the CLAS data. This is related to the opening of different meson-baryon channels contributing to inclusive observables. 

The electrocouplings of nucleon resonances in the 1.8--2.0~GeV mass range and at 0.4~GeV$^2<Q^2<5.0$~GeV$^2$ will soon become available from data~\cite{Isupov:2017lnd,Markov:2018loh,Fedotov:2018oan,Trivedi:2018rgo}. We aim to use this work as a benchmark for describing low and high-energy data in wide $Q^2$ regions with one single combined resonance-background model. This is motivated by the CLAS12 endeavour on both exclusive and inclusive electron-induced reactions, which will extend the data base for comparison and constraints on the model towards high photon virtualities of $Q^2 > 4.5$~GeV$^2$. For exclusive reaction studies, the $Q^2$ coverage is to be extended to the largest interval ever achieved, from 0.05~GeV$^2$ to 12~GeV$^2$.

\begin{acknowledgments}
The authors thank I.~G.~Aznauryan, W.~Melnitchouk, J.~Qiu, and Ch.~Weiss for their valuable input. This material is based upon work supported by 
the U.S.~Department of Energy,
Office of Science, Office of Nuclear Physics 
under contracts  DE-AC05-06OR23177 and DE-FG02-87ER40365, 
U.S.~National Science Foundation under Grants 
No.~PHY-1415459 and No.~PHY-1513524,
PAPIIT-DGAPA (UNAM, Mexico) under Grant No.~IA101819,
CONACYT (Mexico) under Grant No.~251817, 
and the Skobeltsyn Nuclear Physics Institute 
and Physics Department at Moscow State University. This work was also supported by the Deutsche Forschungsgemeinschaft (DFG, German Research Foundation), in part through the Collaborative Research Center [The Low-Energy Frontier of the Standard Model, Projektnummer 204404729 - SFB 1044], and in part through the Cluster of Excellence [Precision Physics, Fundamental Interactions, and Structure of Matter] (PRISMA$^+$ EXC 2118/1) within the German Excellence Strategy (Project ID 39083149).
\end{acknowledgments}

\bibliographystyle{apsrev4-1-jpac}
\bibliography{output}{}

\begin{thebibliography}{73}%
\makeatletter
\providecommand \@ifxundefined [1]{%
 \@ifx{#1\undefined}
}%
\providecommand \@ifnum [1]{%
 \ifnum #1\expandafter \@firstoftwo
 \else \expandafter \@secondoftwo
 \fi
}%
\providecommand \@ifx [1]{%
 \ifx #1\expandafter \@firstoftwo
 \else \expandafter \@secondoftwo
 \fi
}%
\providecommand \natexlab [1]{#1}%
\providecommand \enquote  [1]{``#1''}%
\providecommand \bibnamefont  [1]{#1}%
\providecommand \bibfnamefont [1]{#1}%
\providecommand \citenamefont [1]{#1}%
\providecommand \href@noop [0]{\@secondoftwo}%
\providecommand \href [0]{\begingroup \@sanitize@url \@href}%
\providecommand \@href[1]{\@@startlink{#1}\@@href}%
\providecommand \@@href[1]{\endgroup#1\@@endlink}%
\providecommand \@sanitize@url [0]{\catcode `\\12\catcode `\$12\catcode
  `\&12\catcode `\#12\catcode `\^12\catcode `\_12\catcode `\%12\relax}%
\providecommand \@@startlink[1]{}%
\providecommand \@@endlink[0]{}%
\providecommand \url  [0]{\begingroup\@sanitize@url \@url }%
\providecommand \@url [1]{\endgroup\@href {#1}{\urlprefix }}%
\providecommand \urlprefix  [0]{URL }%
\providecommand \Eprint [0]{\href }%
\providecommand \doibase [0]{http://dx.doi.org/}%
\providecommand \selectlanguage [0]{\@gobble}%
\providecommand \bibinfo  [0]{\@secondoftwo}%
\providecommand \bibfield  [0]{\@secondoftwo}%
\providecommand \translation [1]{[#1]}%
\providecommand \BibitemOpen [0]{}%
\providecommand \bibitemStop [0]{}%
\providecommand \bibitemNoStop [0]{.\EOS\space}%
\providecommand \EOS [0]{\spacefactor3000\relax}%
\providecommand \BibitemShut  [1]{\csname bibitem#1\endcsname}%
\let\auto@bib@innerbib\@empty
\bibitem [{\citenamefont {Harland-Lang}\ \emph {et~al.}(2015)\citenamefont
  {Harland-Lang}, \citenamefont {Martin}, \citenamefont {Motylinski},\ and\
  \citenamefont {Thorne}}]{Harland-Lang:2014zoa}%
  \BibitemOpen
  \bibfield  {author} {\bibinfo {author} {\bibfnamefont {L.~A.}\ \bibnamefont
  {Harland-Lang}}, \bibinfo {author} {\bibfnamefont {A.~D.}\ \bibnamefont
  {Martin}}, \bibinfo {author} {\bibfnamefont {P.}~\bibnamefont {Motylinski}},
  \ and\ \bibinfo {author} {\bibfnamefont {R.~S.}\ \bibnamefont {Thorne}},\
  }\href {\doibase 10.1140/epjc/s10052-015-3397-6} {\bibfield  {journal}
  {\bibinfo  {journal} {Eur. Phys. J.}\ }\textbf {\bibinfo {volume} {C75}},\
  \bibinfo {pages} {204} (\bibinfo {year} {2015})},\ \Eprint
  {http://arxiv.org/abs/1412.3989}{\tt arXiv:1412.3989 [hep-ph]}\BibitemShut
  {NoStop}%
\bibitem [{\citenamefont {Dulat}\ \emph {et~al.}(2016)\citenamefont {Dulat},
  \citenamefont {Hou}, \citenamefont {Gao}, \citenamefont {Guzzi},
  \citenamefont {Huston}, \citenamefont {Nadolsky}, \citenamefont {Pumplin},
  \citenamefont {Schmidt}, \citenamefont {Stump},\ and\ \citenamefont
  {Yuan}}]{Dulat:2015mca}%
  \BibitemOpen
  \bibfield  {author} {\bibinfo {author} {\bibfnamefont {S.}~\bibnamefont
  {Dulat}}, \bibinfo {author} {\bibfnamefont {T.-J.}\ \bibnamefont {Hou}},
  \bibinfo {author} {\bibfnamefont {J.}~\bibnamefont {Gao}}, \bibinfo {author}
  {\bibfnamefont {M.}~\bibnamefont {Guzzi}}, \bibinfo {author} {\bibfnamefont
  {J.}~\bibnamefont {Huston}}, \bibinfo {author} {\bibfnamefont
  {P.}~\bibnamefont {Nadolsky}}, \bibinfo {author} {\bibfnamefont
  {J.}~\bibnamefont {Pumplin}}, \bibinfo {author} {\bibfnamefont
  {C.}~\bibnamefont {Schmidt}}, \bibinfo {author} {\bibfnamefont
  {D.}~\bibnamefont {Stump}}, \ and\ \bibinfo {author} {\bibfnamefont {C.~P.}\
  \bibnamefont {Yuan}},\ }\href {\doibase 10.1103/PhysRevD.93.033006}
  {\bibfield  {journal} {\bibinfo  {journal} {Phys. Rev.}\ }\textbf {\bibinfo
  {volume} {D93}},\ \bibinfo {pages} {033006} (\bibinfo {year} {2016})},\
  \Eprint {http://arxiv.org/abs/1506.07443}{\tt arXiv:1506.07443
  [hep-ph]}\BibitemShut {NoStop}%
\bibitem [{\citenamefont {Accardi}\ \emph {et~al.}(2016)\citenamefont
  {Accardi}, \citenamefont {Brady}, \citenamefont {Melnitchouk}, \citenamefont
  {Owens},\ and\ \citenamefont {Sato}}]{Accardi:2016qay}%
  \BibitemOpen
  \bibfield  {author} {\bibinfo {author} {\bibfnamefont {A.}~\bibnamefont
  {Accardi}}, \bibinfo {author} {\bibfnamefont {L.~T.}\ \bibnamefont {Brady}},
  \bibinfo {author} {\bibfnamefont {W.}~\bibnamefont {Melnitchouk}}, \bibinfo
  {author} {\bibfnamefont {J.~F.}\ \bibnamefont {Owens}}, \ and\ \bibinfo
  {author} {\bibfnamefont {N.}~\bibnamefont {Sato}},\ }\href {\doibase
  10.1103/PhysRevD.93.114017} {\bibfield  {journal} {\bibinfo  {journal} {Phys.
  Rev.}\ }\textbf {\bibinfo {volume} {D93}},\ \bibinfo {pages} {114017}
  (\bibinfo {year} {2016})},\ \Eprint {http://arxiv.org/abs/1602.03154}{\tt
  arXiv:1602.03154 [hep-ph]}\BibitemShut {NoStop}%
\bibitem [{\citenamefont {Alekhin}\ \emph {et~al.}(2017)\citenamefont
  {Alekhin}, \citenamefont {Bl\"umlein}, \citenamefont {Moch},\ and\
  \citenamefont {Placakyte}}]{Alekhin:2017kpj}%
  \BibitemOpen
  \bibfield  {author} {\bibinfo {author} {\bibfnamefont {S.}~\bibnamefont
  {Alekhin}}, \bibinfo {author} {\bibfnamefont {J.}~\bibnamefont {Bl\"umlein}},
  \bibinfo {author} {\bibfnamefont {S.}~\bibnamefont {Moch}}, \ and\ \bibinfo
  {author} {\bibfnamefont {R.}~\bibnamefont {Placakyte}},\ }\href {\doibase
  10.1103/PhysRevD.96.014011} {\bibfield  {journal} {\bibinfo  {journal} {Phys.
  Rev.}\ }\textbf {\bibinfo {volume} {D96}},\ \bibinfo {pages} {014011}
  (\bibinfo {year} {2017})},\ \Eprint {http://arxiv.org/abs/1701.05838}{\tt
  arXiv:1701.05838 [hep-ph]}\BibitemShut {NoStop}%
\bibitem [{\citenamefont {Ball}\ \emph {et~al.}(2017)\citenamefont {Ball} \emph
  {et~al.}}]{Ball:2017nwa}%
  \BibitemOpen
  \bibfield  {author} {\bibinfo {author} {\bibfnamefont {R.~D.}\ \bibnamefont
  {Ball}} \emph {et~al.} (\bibinfo {collaboration} {NNPDF} Collaboration),\
  }\href {\doibase 10.1140/epjc/s10052-017-5199-5} {\bibfield  {journal}
  {\bibinfo  {journal} {Eur. Phys. J.}\ }\textbf {\bibinfo {volume} {C77}},\
  \bibinfo {pages} {663} (\bibinfo {year} {2017})},\ \Eprint
  {http://arxiv.org/abs/1706.00428}{\tt arXiv:1706.00428 [hep-ph]}\BibitemShut
  {NoStop}%
\bibitem [{\citenamefont {Jimenez-Delgado}\ \emph {et~al.}(2013)\citenamefont
  {Jimenez-Delgado}, \citenamefont {Melnitchouk},\ and\ \citenamefont
  {Owens}}]{Jimenez-Delgado:2013sma}%
  \BibitemOpen
  \bibfield  {author} {\bibinfo {author} {\bibfnamefont {P.}~\bibnamefont
  {Jimenez-Delgado}}, \bibinfo {author} {\bibfnamefont {W.}~\bibnamefont
  {Melnitchouk}}, \ and\ \bibinfo {author} {\bibfnamefont {J.~F.}\ \bibnamefont
  {Owens}},\ }\href {\doibase 10.1088/0954-3899/40/9/093102} {\bibfield
  {journal} {\bibinfo  {journal} {J. Phys.}\ }\textbf {\bibinfo {volume}
  {G40}},\ \bibinfo {pages} {093102} (\bibinfo {year} {2013})},\ \Eprint
  {http://arxiv.org/abs/1306.6515}{\tt arXiv:1306.6515 [hep-ph]}\BibitemShut
  {NoStop}%
\bibitem [{\citenamefont {Gao}\ \emph {et~al.}(2018)\citenamefont {Gao},
  \citenamefont {Harland-Lang},\ and\ \citenamefont {Rojo}}]{Gao:2017yyd}%
  \BibitemOpen
  \bibfield  {author} {\bibinfo {author} {\bibfnamefont {J.}~\bibnamefont
  {Gao}}, \bibinfo {author} {\bibfnamefont {L.}~\bibnamefont {Harland-Lang}}, \
  and\ \bibinfo {author} {\bibfnamefont {J.}~\bibnamefont {Rojo}},\ }\href
  {\doibase 10.1016/j.physrep.2018.03.002} {\bibfield  {journal} {\bibinfo
  {journal} {Phys. Rept.}\ }\textbf {\bibinfo {volume} {742}},\ \bibinfo
  {pages} {1} (\bibinfo {year} {2018})},\ \Eprint
  {http://arxiv.org/abs/1709.04922}{\tt arXiv:1709.04922 [hep-ph]}\BibitemShut
  {NoStop}%
\bibitem [{\citenamefont {Deur}\ \emph {et~al.}(2018)\citenamefont {Deur},
  \citenamefont {Brodsky},\ and\ \citenamefont {De~T\'eramond}}]{Deur:2018roz}%
  \BibitemOpen
  \bibfield  {author} {\bibinfo {author} {\bibfnamefont {A.}~\bibnamefont
  {Deur}}, \bibinfo {author} {\bibfnamefont {S.~J.}\ \bibnamefont {Brodsky}}, \
  and\ \bibinfo {author} {\bibfnamefont {G.~F.}\ \bibnamefont
  {De~T\'eramond}},\ }\href@noop {} {\  (\bibinfo {year} {2018})},\ \Eprint
  {http://arxiv.org/abs/1807.05250}{\tt arXiv:1807.05250 [hep-ph]}\BibitemShut
  {NoStop}%
\bibitem [{\citenamefont {Osipenko}\ \emph {et~al.}(2003)\citenamefont
  {Osipenko} \emph {et~al.}}]{Osipenko:2003bu}%
  \BibitemOpen
  \bibfield  {author} {\bibinfo {author} {\bibfnamefont {M.}~\bibnamefont
  {Osipenko}} \emph {et~al.} (\bibinfo {collaboration} {CLAS} Collaboration),\
  }\href {\doibase 10.1103/PhysRevD.67.092001} {\bibfield  {journal} {\bibinfo
  {journal} {Phys. Rev.}\ }\textbf {\bibinfo {volume} {D67}},\ \bibinfo {pages}
  {092001} (\bibinfo {year} {2003})},\ \Eprint
  {http://arxiv.org/abs/hep-ph/0301204}{\tt arXiv:hep-ph/0301204
  [hep-ph]}\BibitemShut {NoStop}%
\bibitem [{\citenamefont {Malace}\ \emph {et~al.}(2009)\citenamefont {Malace}
  \emph {et~al.}}]{Malace:2009kw}%
  \BibitemOpen
  \bibfield  {author} {\bibinfo {author} {\bibfnamefont {S.~P.}\ \bibnamefont
  {Malace}} \emph {et~al.} (\bibinfo {collaboration} {Jefferson Lab E00-115}
  Collaboration),\ }\href {\doibase 10.1103/PhysRevC.80.035207} {\bibfield
  {journal} {\bibinfo  {journal} {Phys. Rev.}\ }\textbf {\bibinfo {volume}
  {C80}},\ \bibinfo {pages} {035207} (\bibinfo {year} {2009})},\ \Eprint
  {http://arxiv.org/abs/0905.2374}{\tt arXiv:0905.2374 [nucl-ex]}\BibitemShut
  {NoStop}%
\bibitem [{\citenamefont {Christy}\ and\ \citenamefont
  {Bosted}(2010)}]{Christy:2007ve}%
  \BibitemOpen
  \bibfield  {author} {\bibinfo {author} {\bibfnamefont {M.~E.}\ \bibnamefont
  {Christy}}\ and\ \bibinfo {author} {\bibfnamefont {P.~E.}\ \bibnamefont
  {Bosted}},\ }\href {\doibase 10.1103/PhysRevC.81.055213} {\bibfield
  {journal} {\bibinfo  {journal} {Phys. Rev.}\ }\textbf {\bibinfo {volume}
  {C81}},\ \bibinfo {pages} {055213} (\bibinfo {year} {2010})},\ \Eprint
  {http://arxiv.org/abs/0712.3731}{\tt arXiv:0712.3731 [hep-ph]}\BibitemShut
  {NoStop}%
\bibitem [{\citenamefont {Prok}\ \emph {et~al.}(2014)\citenamefont {Prok} \emph
  {et~al.}}]{Prok:2014ltt}%
  \BibitemOpen
  \bibfield  {author} {\bibinfo {author} {\bibfnamefont {Y.}~\bibnamefont
  {Prok}} \emph {et~al.} (\bibinfo {collaboration} {CLAS} Collaboration),\
  }\href {\doibase 10.1103/PhysRevC.90.025212} {\bibfield  {journal} {\bibinfo
  {journal} {Phys. Rev.}\ }\textbf {\bibinfo {volume} {C90}},\ \bibinfo {pages}
  {025212} (\bibinfo {year} {2014})},\ \Eprint
  {http://arxiv.org/abs/1404.6231}{\tt arXiv:1404.6231 [nucl-ex]}\BibitemShut
  {NoStop}%
\bibitem [{\citenamefont {Keppel}\ \emph {et~al.}(2000)\citenamefont {Keppel},
  \citenamefont {Niculescu} \emph {et~al.}}]{JLab:E00-002}%
  \BibitemOpen
  \bibfield  {author} {\bibinfo {author} {\bibfnamefont {C.~E.}\ \bibnamefont
  {Keppel}}, \bibinfo {author} {\bibfnamefont {M.~I.}\ \bibnamefont
  {Niculescu}},  \emph {et~al.},\ }\href@noop {} {} (\bibinfo {year} {2000}),\
  \bibinfo {note} {{JLab experiment E00-002}}\BibitemShut {NoStop}%
\bibitem [{\citenamefont {Liang}\ \emph {et~al.}(2004)\citenamefont {Liang}
  \emph {et~al.}}]{Liang:2004tj}%
  \BibitemOpen
  \bibfield  {author} {\bibinfo {author} {\bibfnamefont {Y.}~\bibnamefont
  {Liang}} \emph {et~al.} (\bibinfo {collaboration} {Jefferson Lab Hall C
  E94-110} Collaboration),\ }\href@noop {} {\  (\bibinfo {year} {2004})},\
  \Eprint {http://arxiv.org/abs/nucl-ex/0410027}{\tt arXiv:nucl-ex/0410027
  [nucl-ex]}\BibitemShut {NoStop}%
\bibitem [{\citenamefont {Golubenko}\ \emph {et~al.}(2019)\citenamefont
  {Golubenko}, \citenamefont {Chesnokov}, \citenamefont {Ishkhanov},\ and\
  \citenamefont {Mokeev}}]{Golubenko:2019gxz}%
  \BibitemOpen
  \bibfield  {author} {\bibinfo {author} {\bibfnamefont {A.~A.}\ \bibnamefont
  {Golubenko}}, \bibinfo {author} {\bibfnamefont {V.~V.}\ \bibnamefont
  {Chesnokov}}, \bibinfo {author} {\bibfnamefont {B.~S.}\ \bibnamefont
  {Ishkhanov}}, \ and\ \bibinfo {author} {\bibfnamefont {V.~I.}\ \bibnamefont
  {Mokeev}},\ }\href@noop {} {\  (\bibinfo {year} {2019})},\ \Eprint
  {http://arxiv.org/abs/1902.02900}{\tt arXiv:1902.02900 [hep-ex]}\BibitemShut
  {NoStop}%
\bibitem [{CLA({\natexlab{a}})}]{CLAS:DB}%
  \BibitemOpen
  \href {https://clasweb.jlab.org/physicsdb/} {\enquote {\bibinfo {title}
  {{CLAS Physics Database}},}\ }\bibinfo {note}
  {\href{https://clasweb.jlab.org/physicsdb/}{\newline
  https://clasweb.jlab.org/physicsdb/}}\BibitemShut {NoStop}%
\bibitem [{CLA({\natexlab{b}})}]{CLAS:SFDB}%
  \BibitemOpen
  \href {http://clas.sinp.msu.ru/strfun/} {\enquote {\bibinfo {title}
  {{Structure functions and cross-sections Database}},}\ }\bibinfo {note}
  {\href{http://clas.sinp.msu.ru/strfun/}{http://clas.sinp.msu.ru/strfun/}}\BibitemShut
  {NoStop}%
\bibitem [{\citenamefont {Ricco}\ \emph {et~al.}(1999)\citenamefont {Ricco},
  \citenamefont {Simula},\ and\ \citenamefont {Battaglieri}}]{Ricco:1998yr}%
  \BibitemOpen
  \bibfield  {author} {\bibinfo {author} {\bibfnamefont {G.}~\bibnamefont
  {Ricco}}, \bibinfo {author} {\bibfnamefont {S.}~\bibnamefont {Simula}}, \
  and\ \bibinfo {author} {\bibfnamefont {M.}~\bibnamefont {Battaglieri}},\
  }\href {\doibase 10.1016/S0550-3213(99)00302-8} {\bibfield  {journal}
  {\bibinfo  {journal} {Nucl. Phys.}\ }\textbf {\bibinfo {volume} {B555}},\
  \bibinfo {pages} {306} (\bibinfo {year} {1999})},\ \Eprint
  {http://arxiv.org/abs/hep-ph/9901360}{\tt arXiv:hep-ph/9901360
  [hep-ph]}\BibitemShut {NoStop}%
\bibitem [{\citenamefont {Tomalak}\ and\ \citenamefont
  {Vanderhaeghen}(2016)}]{Tomalak:2015hva}%
  \BibitemOpen
  \bibfield  {author} {\bibinfo {author} {\bibfnamefont {O.}~\bibnamefont
  {Tomalak}}\ and\ \bibinfo {author} {\bibfnamefont {M.}~\bibnamefont
  {Vanderhaeghen}},\ }\href {\doibase 10.1140/epjc/s10052-016-3966-3}
  {\bibfield  {journal} {\bibinfo  {journal} {Eur. Phys. J.}\ }\textbf
  {\bibinfo {volume} {C76}},\ \bibinfo {pages} {125} (\bibinfo {year}
  {2016})},\ \Eprint {http://arxiv.org/abs/1512.09113}{\tt arXiv:1512.09113
  [hep-ph]}\BibitemShut {NoStop}%
\bibitem [{\citenamefont {Whitlow}\ \emph {et~al.}(1990)\citenamefont
  {Whitlow}, \citenamefont {Rock}, \citenamefont {Bodek}, \citenamefont
  {Riordan},\ and\ \citenamefont {Dasu}}]{Whitlow:1990gk}%
  \BibitemOpen
  \bibfield  {author} {\bibinfo {author} {\bibfnamefont {L.~W.}\ \bibnamefont
  {Whitlow}}, \bibinfo {author} {\bibfnamefont {S.}~\bibnamefont {Rock}},
  \bibinfo {author} {\bibfnamefont {A.}~\bibnamefont {Bodek}}, \bibinfo
  {author} {\bibfnamefont {E.~M.}\ \bibnamefont {Riordan}}, \ and\ \bibinfo
  {author} {\bibfnamefont {S.}~\bibnamefont {Dasu}},\ }\href {\doibase
  10.1016/0370-2693(90)91176-C} {\bibfield  {journal} {\bibinfo  {journal}
  {Phys. Lett.}\ }\textbf {\bibinfo {volume} {B250}},\ \bibinfo {pages} {193}
  (\bibinfo {year} {1990})}\BibitemShut {NoStop}%
\bibitem [{\citenamefont {Dasu}\ \emph {et~al.}(1994)\citenamefont {Dasu} \emph
  {et~al.}}]{Dasu:1993vk}%
  \BibitemOpen
  \bibfield  {author} {\bibinfo {author} {\bibfnamefont {S.}~\bibnamefont
  {Dasu}} \emph {et~al.},\ }\href {\doibase 10.1103/PhysRevD.49.5641}
  {\bibfield  {journal} {\bibinfo  {journal} {Phys. Rev.}\ }\textbf {\bibinfo
  {volume} {D49}},\ \bibinfo {pages} {5641} (\bibinfo {year}
  {1994})}\BibitemShut {NoStop}%
\bibitem [{\citenamefont {Arneodo}\ \emph {et~al.}(1997)\citenamefont {Arneodo}
  \emph {et~al.}}]{Arneodo:1996qe}%
  \BibitemOpen
  \bibfield  {author} {\bibinfo {author} {\bibfnamefont {M.}~\bibnamefont
  {Arneodo}} \emph {et~al.} (\bibinfo {collaboration} {New Muon}
  Collaboration),\ }\href {\doibase 10.1016/S0550-3213(96)00538-X} {\bibfield
  {journal} {\bibinfo  {journal} {Nucl. Phys.}\ }\textbf {\bibinfo {volume}
  {B483}},\ \bibinfo {pages} {3} (\bibinfo {year} {1997})},\ \Eprint
  {http://arxiv.org/abs/hep-ph/9610231}{\tt arXiv:hep-ph/9610231
  [hep-ph]}\BibitemShut {NoStop}%
\bibitem [{\citenamefont {Burkert}(2018)}]{Burkert:2018nvj}%
  \BibitemOpen
  \bibfield  {author} {\bibinfo {author} {\bibfnamefont {V.~D.}\ \bibnamefont
  {Burkert}},\ }\href {\doibase 10.1146/annurev-nucl-101917-021129} {\bibfield
  {journal} {\bibinfo  {journal} {Ann. Rev. Nucl. Part. Sci.}\ }\textbf
  {\bibinfo {volume} {68}},\ \bibinfo {pages} {405} (\bibinfo {year}
  {2018})}\BibitemShut {NoStop}%
\bibitem [{\citenamefont {Feynman}(1972)}]{Feynman:1972xm}%
  \BibitemOpen
  \bibfield  {author} {\bibinfo {author} {\bibfnamefont {R.~P.}\ \bibnamefont
  {Feynman}},\ }\href@noop {} {\bibfield  {journal} {\bibinfo  {journal} {Conf.
  Proc.}\ }\textbf {\bibinfo {volume} {C720611}},\ \bibinfo {pages} {75}
  (\bibinfo {year} {1972})}\BibitemShut {NoStop}%
\bibitem [{\citenamefont {Farrar}\ and\ \citenamefont
  {Jackson}(1975)}]{Farrar:1975yb}%
  \BibitemOpen
  \bibfield  {author} {\bibinfo {author} {\bibfnamefont {G.~R.}\ \bibnamefont
  {Farrar}}\ and\ \bibinfo {author} {\bibfnamefont {D.~R.}\ \bibnamefont
  {Jackson}},\ }\href {\doibase 10.1103/PhysRevLett.35.1416} {\bibfield
  {journal} {\bibinfo  {journal} {Phys. Rev. Lett.}\ }\textbf {\bibinfo
  {volume} {35}},\ \bibinfo {pages} {1416} (\bibinfo {year}
  {1975})}\BibitemShut {NoStop}%
\bibitem [{\citenamefont {Close}\ and\ \citenamefont
  {Thomas}(1988)}]{Close:1988br}%
  \BibitemOpen
  \bibfield  {author} {\bibinfo {author} {\bibfnamefont {F.~E.}\ \bibnamefont
  {Close}}\ and\ \bibinfo {author} {\bibfnamefont {A.~W.}\ \bibnamefont
  {Thomas}},\ }\href {\doibase 10.1016/0370-2693(88)90530-8} {\bibfield
  {journal} {\bibinfo  {journal} {Phys. Lett.}\ }\textbf {\bibinfo {volume}
  {B212}},\ \bibinfo {pages} {227} (\bibinfo {year} {1988})}\BibitemShut
  {NoStop}%
\bibitem [{\citenamefont {Leader}\ \emph {et~al.}(2002)\citenamefont {Leader},
  \citenamefont {Sidorov},\ and\ \citenamefont {Stamenov}}]{Leader:2001kh}%
  \BibitemOpen
  \bibfield  {author} {\bibinfo {author} {\bibfnamefont {E.}~\bibnamefont
  {Leader}}, \bibinfo {author} {\bibfnamefont {A.~V.}\ \bibnamefont {Sidorov}},
  \ and\ \bibinfo {author} {\bibfnamefont {D.~B.}\ \bibnamefont {Stamenov}},\
  }\href {\doibase 10.1007/s100520200901} {\bibfield  {journal} {\bibinfo
  {journal} {Eur. Phys. J.}\ }\textbf {\bibinfo {volume} {C23}},\ \bibinfo
  {pages} {479} (\bibinfo {year} {2002})},\ \Eprint
  {http://arxiv.org/abs/hep-ph/0111267}{\tt arXiv:hep-ph/0111267
  [hep-ph]}\BibitemShut {NoStop}%
\bibitem [{\citenamefont {Nocera}(2015)}]{Nocera:2014uea}%
  \BibitemOpen
  \bibfield  {author} {\bibinfo {author} {\bibfnamefont {E.~R.}\ \bibnamefont
  {Nocera}},\ }\href {\doibase 10.1016/j.physletb.2015.01.021} {\bibfield
  {journal} {\bibinfo  {journal} {Phys. Lett.}\ }\textbf {\bibinfo {volume}
  {B742}},\ \bibinfo {pages} {117} (\bibinfo {year} {2015})},\ \Eprint
  {http://arxiv.org/abs/1410.7290}{\tt arXiv:1410.7290 [hep-ph]}\BibitemShut
  {NoStop}%
\bibitem [{\citenamefont {Ji}(2013)}]{Ji:2013dva}%
  \BibitemOpen
  \bibfield  {author} {\bibinfo {author} {\bibfnamefont {X.}~\bibnamefont
  {Ji}},\ }\href {\doibase 10.1103/PhysRevLett.110.262002} {\bibfield
  {journal} {\bibinfo  {journal} {Phys. Rev. Lett.}\ }\textbf {\bibinfo
  {volume} {110}},\ \bibinfo {pages} {262002} (\bibinfo {year} {2013})},\
  \Eprint {http://arxiv.org/abs/1305.1539}{\tt arXiv:1305.1539
  [hep-ph]}\BibitemShut {NoStop}%
\bibitem [{\citenamefont {Radyushkin}(2017)}]{Radyushkin:2017cyf}%
  \BibitemOpen
  \bibfield  {author} {\bibinfo {author} {\bibfnamefont {A.~V.}\ \bibnamefont
  {Radyushkin}},\ }\href {\doibase 10.1103/PhysRevD.96.034025} {\bibfield
  {journal} {\bibinfo  {journal} {Phys. Rev.}\ }\textbf {\bibinfo {volume}
  {D96}},\ \bibinfo {pages} {034025} (\bibinfo {year} {2017})},\ \Eprint
  {http://arxiv.org/abs/1705.01488}{\tt arXiv:1705.01488 [hep-ph]}\BibitemShut
  {NoStop}%
\bibitem [{\citenamefont {Ma}\ and\ \citenamefont {Qiu}(2018)}]{Ma:2017pxb}%
  \BibitemOpen
  \bibfield  {author} {\bibinfo {author} {\bibfnamefont {Y.-Q.}\ \bibnamefont
  {Ma}}\ and\ \bibinfo {author} {\bibfnamefont {J.-W.}\ \bibnamefont {Qiu}},\
  }\href {\doibase 10.1103/PhysRevLett.120.022003} {\bibfield  {journal}
  {\bibinfo  {journal} {Phys. Rev. Lett.}\ }\textbf {\bibinfo {volume} {120}},\
  \bibinfo {pages} {022003} (\bibinfo {year} {2018})},\ \Eprint
  {http://arxiv.org/abs/1709.03018}{\tt arXiv:1709.03018 [hep-ph]}\BibitemShut
  {NoStop}%
\bibitem [{\citenamefont {Lin}\ \emph {et~al.}(2018)\citenamefont {Lin} \emph
  {et~al.}}]{Lin:2017snn}%
  \BibitemOpen
  \bibfield  {author} {\bibinfo {author} {\bibfnamefont {H.-W.}\ \bibnamefont
  {Lin}} \emph {et~al.},\ }\href {\doibase 10.1016/j.ppnp.2018.01.007}
  {\bibfield  {journal} {\bibinfo  {journal} {Prog. Part. Nucl. Phys.}\
  }\textbf {\bibinfo {volume} {100}},\ \bibinfo {pages} {107} (\bibinfo {year}
  {2018})},\ \Eprint {http://arxiv.org/abs/1711.07916}{\tt arXiv:1711.07916
  [hep-ph]}\BibitemShut {NoStop}%
\bibitem [{\citenamefont {Bloom}\ and\ \citenamefont
  {Gilman}(1970)}]{Bloom:1970xb}%
  \BibitemOpen
  \bibfield  {author} {\bibinfo {author} {\bibfnamefont {E.~D.}\ \bibnamefont
  {Bloom}}\ and\ \bibinfo {author} {\bibfnamefont {F.~J.}\ \bibnamefont
  {Gilman}},\ }\href {\doibase 10.1103/PhysRevLett.25.1140} {\bibfield
  {journal} {\bibinfo  {journal} {Phys. Rev. Lett.}\ }\textbf {\bibinfo
  {volume} {25}},\ \bibinfo {pages} {1140} (\bibinfo {year}
  {1970})}\BibitemShut {NoStop}%
\bibitem [{\citenamefont {Melnitchouk}(2005)}]{Melnitchouk:2005ye}%
  \BibitemOpen
  \bibfield  {author} {\bibinfo {author} {\bibfnamefont {W.}~\bibnamefont
  {Melnitchouk}},\ }\href
  {http://www1.jlab.org/Ul/publications/view_pub.cfm?pub_id=6477} {\  (\bibinfo
  {year} {2005})},\ \bibinfo {note}
  {\href{http://www1.jlab.org/Ul/publications/view_pub.cfm?pub_id=6477}{JLAB-THY-06-459}}\BibitemShut
  {NoStop}%
\bibitem [{\citenamefont {Melnitchouk}\ \emph {et~al.}(2005)\citenamefont
  {Melnitchouk}, \citenamefont {Ent},\ and\ \citenamefont
  {Keppel}}]{Melnitchouk:2005zr}%
  \BibitemOpen
  \bibfield  {author} {\bibinfo {author} {\bibfnamefont {W.}~\bibnamefont
  {Melnitchouk}}, \bibinfo {author} {\bibfnamefont {R.}~\bibnamefont {Ent}}, \
  and\ \bibinfo {author} {\bibfnamefont {C.}~\bibnamefont {Keppel}},\ }\href
  {\doibase 10.1016/j.physrep.2004.10.004} {\bibfield  {journal} {\bibinfo
  {journal} {Phys. Rept.}\ }\textbf {\bibinfo {volume} {406}},\ \bibinfo
  {pages} {127} (\bibinfo {year} {2005})},\ \Eprint
  {http://arxiv.org/abs/hep-ph/0501217}{\tt arXiv:hep-ph/0501217
  [hep-ph]}\BibitemShut {NoStop}%
\bibitem [{\citenamefont {Christy}\ and\ \citenamefont
  {Melnitchouk}(2011)}]{Christy:2011cv}%
  \BibitemOpen
  \bibfield  {author} {\bibinfo {author} {\bibfnamefont {M.~E.}\ \bibnamefont
  {Christy}}\ and\ \bibinfo {author} {\bibfnamefont {W.}~\bibnamefont
  {Melnitchouk}},\ }\href {\doibase 10.1088/1742-6596/299/1/012004} {\bibfield
  {journal} {\bibinfo  {journal} {J. Phys. Conf. Ser.}\ }\textbf {\bibinfo
  {volume} {299}},\ \bibinfo {pages} {012004} (\bibinfo {year} {2011})},\
  \Eprint {http://arxiv.org/abs/1104.0239}{\tt arXiv:1104.0239
  [nucl-ex]}\BibitemShut {NoStop}%
\bibitem [{\citenamefont {Aznauryan}\ and\ \citenamefont
  {Burkert}(2012{\natexlab{a}})}]{Aznauryan:2011qj}%
  \BibitemOpen
  \bibfield  {author} {\bibinfo {author} {\bibfnamefont {I.~G.}\ \bibnamefont
  {Aznauryan}}\ and\ \bibinfo {author} {\bibfnamefont {V.~D.}\ \bibnamefont
  {Burkert}},\ }\href {\doibase 10.1016/j.ppnp.2011.08.001} {\bibfield
  {journal} {\bibinfo  {journal} {Prog. Part. Nucl. Phys.}\ }\textbf {\bibinfo
  {volume} {67}},\ \bibinfo {pages} {1} (\bibinfo {year}
  {2012}{\natexlab{a}})},\ \Eprint {http://arxiv.org/abs/1109.1720}{\tt
  arXiv:1109.1720 [hep-ph]}\BibitemShut {NoStop}%
\bibitem [{\citenamefont {Park}\ \emph {et~al.}(2015)\citenamefont {Park} \emph
  {et~al.}}]{Park:2014yea}%
  \BibitemOpen
  \bibfield  {author} {\bibinfo {author} {\bibfnamefont {K.}~\bibnamefont
  {Park}} \emph {et~al.} (\bibinfo {collaboration} {CLAS} Collaboration),\
  }\href {\doibase 10.1103/PhysRevC.91.045203} {\bibfield  {journal} {\bibinfo
  {journal} {Phys. Rev.}\ }\textbf {\bibinfo {volume} {C91}},\ \bibinfo {pages}
  {045203} (\bibinfo {year} {2015})},\ \Eprint
  {http://arxiv.org/abs/1412.0274}{\tt arXiv:1412.0274 [nucl-ex]}\BibitemShut
  {NoStop}%
\bibitem [{\citenamefont {Mokeev}\ \emph
  {et~al.}(2016{\natexlab{a}})\citenamefont {Mokeev}, \citenamefont
  {Aznauryan}, \citenamefont {Burkert},\ and\ \citenamefont
  {Gothe}}]{Mokeev:2015moa}%
  \BibitemOpen
  \bibfield  {author} {\bibinfo {author} {\bibfnamefont {V.~I.}\ \bibnamefont
  {Mokeev}}, \bibinfo {author} {\bibfnamefont {I.}~\bibnamefont {Aznauryan}},
  \bibinfo {author} {\bibfnamefont {V.}~\bibnamefont {Burkert}}, \ and\
  \bibinfo {author} {\bibfnamefont {R.}~\bibnamefont {Gothe}},\ }\href
  {\doibase 10.1051/epjconf/201611301013} {\bibfield  {journal} {\bibinfo
  {journal} {EPJ Web Conf.}\ }\textbf {\bibinfo {volume} {113}},\ \bibinfo
  {pages} {01013} (\bibinfo {year} {2016}{\natexlab{a}})},\ \Eprint
  {http://arxiv.org/abs/1508.04088}{\tt arXiv:1508.04088 [nucl-ex]}\BibitemShut
  {NoStop}%
\bibitem [{\citenamefont {Mokeev}(2016)}]{Mokeev:2016hqv}%
  \BibitemOpen
  \bibfield  {author} {\bibinfo {author} {\bibfnamefont {V.~I.}\ \bibnamefont
  {Mokeev}},\ }\href {\doibase 10.1007/s00601-016-1127-8} {\bibfield  {journal}
  {\bibinfo  {journal} {Few Body Syst.}\ }\textbf {\bibinfo {volume} {57}},\
  \bibinfo {pages} {909} (\bibinfo {year} {2016})},\ \Eprint
  {http://arxiv.org/abs/1602.04160}{\tt arXiv:1602.04160 [nucl-ex]}\BibitemShut
  {NoStop}%
\bibitem [{\citenamefont {Burkert}\ and\ \citenamefont
  {Roberts}(2019)}]{Burkert:2019bhp}%
  \BibitemOpen
  \bibfield  {author} {\bibinfo {author} {\bibfnamefont {V.~D.}\ \bibnamefont
  {Burkert}}\ and\ \bibinfo {author} {\bibfnamefont {C.~D.}\ \bibnamefont
  {Roberts}},\ }\href {\doibase 10.1103/RevModPhys.91.011003} {\bibfield
  {journal} {\bibinfo  {journal} {Rev. Mod. Phys.}\ }\textbf {\bibinfo {volume}
  {91}},\ \bibinfo {pages} {011003} (\bibinfo {year} {2019})},\ \Eprint
  {http://arxiv.org/abs/1710.02549}{\tt arXiv:1710.02549 [nucl-ex]}\BibitemShut
  {NoStop}%
\bibitem [{\citenamefont {Mokeev}(2018)}]{Mokeev:2018zxt}%
  \BibitemOpen
  \bibfield  {author} {\bibinfo {author} {\bibfnamefont {V.~I.}\ \bibnamefont
  {Mokeev}} (\bibinfo {collaboration} {CLAS} Collaboration),\ }\href {\doibase
  10.1007/s00601-018-1369-8} {\bibfield  {journal} {\bibinfo  {journal} {Few
  Body Syst.}\ }\textbf {\bibinfo {volume} {59}},\ \bibinfo {pages} {46}
  (\bibinfo {year} {2018})},\ \Eprint {http://arxiv.org/abs/1801.09750}{\tt
  arXiv:1801.09750 [nucl-ex]}\BibitemShut {NoStop}%
\bibitem [{\citenamefont {Burkert}\ \emph {et~al.}(2019)\citenamefont
  {Burkert}, \citenamefont {Mokeev},\ and\ \citenamefont
  {Ishkhanov}}]{Burkert:2019opk}%
  \BibitemOpen
  \bibfield  {author} {\bibinfo {author} {\bibfnamefont {V.~D.}\ \bibnamefont
  {Burkert}}, \bibinfo {author} {\bibfnamefont {V.~I.}\ \bibnamefont {Mokeev}},
  \ and\ \bibinfo {author} {\bibfnamefont {B.~S.}\ \bibnamefont {Ishkhanov}},\
  }\href@noop {} {\  (\bibinfo {year} {2019})},\ \Eprint
  {http://arxiv.org/abs/1901.09709}{\tt arXiv:1901.09709 [nucl-ex]}\BibitemShut
  {NoStop}%
\bibitem [{\citenamefont {Aznauryan}(2003)}]{Aznauryan:2002gd}%
  \BibitemOpen
  \bibfield  {author} {\bibinfo {author} {\bibfnamefont {I.~G.}\ \bibnamefont
  {Aznauryan}},\ }\href {\doibase 10.1103/PhysRevC.67.015209} {\bibfield
  {journal} {\bibinfo  {journal} {Phys. Rev.}\ }\textbf {\bibinfo {volume}
  {C67}},\ \bibinfo {pages} {015209} (\bibinfo {year} {2003})},\ \Eprint
  {http://arxiv.org/abs/nucl-th/0206033}{\tt arXiv:nucl-th/0206033
  [nucl-th]}\BibitemShut {NoStop}%
\bibitem [{\citenamefont {Mokeev}\ \emph {et~al.}(2009)\citenamefont {Mokeev},
  \citenamefont {Burkert}, \citenamefont {Lee}, \citenamefont {Elouadrhiri},
  \citenamefont {Fedotov},\ and\ \citenamefont {Ishkhanov}}]{Mokeev:2008iw}%
  \BibitemOpen
  \bibfield  {author} {\bibinfo {author} {\bibfnamefont {V.~I.}\ \bibnamefont
  {Mokeev}}, \bibinfo {author} {\bibfnamefont {V.~D.}\ \bibnamefont {Burkert}},
  \bibinfo {author} {\bibfnamefont {T.-S.~H.}\ \bibnamefont {Lee}}, \bibinfo
  {author} {\bibfnamefont {L.}~\bibnamefont {Elouadrhiri}}, \bibinfo {author}
  {\bibfnamefont {G.~V.}\ \bibnamefont {Fedotov}}, \ and\ \bibinfo {author}
  {\bibfnamefont {B.~S.}\ \bibnamefont {Ishkhanov}},\ }\href {\doibase
  10.1103/PhysRevC.80.045212} {\bibfield  {journal} {\bibinfo  {journal} {Phys.
  Rev.}\ }\textbf {\bibinfo {volume} {C80}},\ \bibinfo {pages} {045212}
  (\bibinfo {year} {2009})},\ \Eprint {http://arxiv.org/abs/0809.4158}{\tt
  arXiv:0809.4158 [hep-ph]}\BibitemShut {NoStop}%
\bibitem [{\citenamefont {Aznauryan}\ \emph {et~al.}(2009)\citenamefont
  {Aznauryan} \emph {et~al.}}]{Aznauryan:2009mx}%
  \BibitemOpen
  \bibfield  {author} {\bibinfo {author} {\bibfnamefont {I.~G.}\ \bibnamefont
  {Aznauryan}} \emph {et~al.} (\bibinfo {collaboration} {CLAS} Collaboration),\
  }\href {\doibase 10.1103/PhysRevC.80.055203} {\bibfield  {journal} {\bibinfo
  {journal} {Phys. Rev.}\ }\textbf {\bibinfo {volume} {C80}},\ \bibinfo {pages}
  {055203} (\bibinfo {year} {2009})},\ \Eprint
  {http://arxiv.org/abs/0909.2349}{\tt arXiv:0909.2349 [nucl-ex]}\BibitemShut
  {NoStop}%
\bibitem [{\citenamefont {Mokeev}\ \emph {et~al.}(2012)\citenamefont {Mokeev}
  \emph {et~al.}}]{Mokeev:2012vsa}%
  \BibitemOpen
  \bibfield  {author} {\bibinfo {author} {\bibfnamefont {V.~I.}\ \bibnamefont
  {Mokeev}} \emph {et~al.} (\bibinfo {collaboration} {CLAS} Collaboration),\
  }\href {\doibase 10.1103/PhysRevC.86.035203} {\bibfield  {journal} {\bibinfo
  {journal} {Phys. Rev.}\ }\textbf {\bibinfo {volume} {C86}},\ \bibinfo {pages}
  {035203} (\bibinfo {year} {2012})},\ \Eprint
  {http://arxiv.org/abs/1205.3948}{\tt arXiv:1205.3948 [nucl-ex]}\BibitemShut
  {NoStop}%
\bibitem [{\citenamefont {Mokeev}\ \emph
  {et~al.}(2016{\natexlab{b}})\citenamefont {Mokeev} \emph
  {et~al.}}]{Mokeev:2015lda}%
  \BibitemOpen
  \bibfield  {author} {\bibinfo {author} {\bibfnamefont {V.~I.}\ \bibnamefont
  {Mokeev}} \emph {et~al.},\ }\href {\doibase 10.1103/PhysRevC.93.025206}
  {\bibfield  {journal} {\bibinfo  {journal} {Phys. Rev.}\ }\textbf {\bibinfo
  {volume} {C93}},\ \bibinfo {pages} {025206} (\bibinfo {year}
  {2016}{\natexlab{b}})},\ \Eprint {http://arxiv.org/abs/1509.05460}{\tt
  arXiv:1509.05460 [nucl-ex]}\BibitemShut {NoStop}%
\bibitem [{\citenamefont {Dugger}\ \emph {et~al.}(2009)\citenamefont {Dugger}
  \emph {et~al.}}]{Dugger:2009pn}%
  \BibitemOpen
  \bibfield  {author} {\bibinfo {author} {\bibfnamefont {M.}~\bibnamefont
  {Dugger}} \emph {et~al.} (\bibinfo {collaboration} {CLAS} Collaboration),\
  }\href {\doibase 10.1103/PhysRevC.79.065206} {\bibfield  {journal} {\bibinfo
  {journal} {Phys. Rev.}\ }\textbf {\bibinfo {volume} {C79}},\ \bibinfo {pages}
  {065206} (\bibinfo {year} {2009})},\ \Eprint
  {http://arxiv.org/abs/0903.1110}{\tt arXiv:0903.1110 [hep-ex]}\BibitemShut
  {NoStop}%
\bibitem [{\citenamefont {Golovatch}\ \emph {et~al.}(2019)\citenamefont
  {Golovatch} \emph {et~al.}}]{Golovatch:2018hjk}%
  \BibitemOpen
  \bibfield  {author} {\bibinfo {author} {\bibfnamefont {E.}~\bibnamefont
  {Golovatch}} \emph {et~al.} (\bibinfo {collaboration} {CLAS} Collaboration),\
  }\href {\doibase 10.1016/j.physletb.2018.10.013} {\bibfield  {journal}
  {\bibinfo  {journal} {Phys. Lett.}\ }\textbf {\bibinfo {volume} {B788}},\
  \bibinfo {pages} {371} (\bibinfo {year} {2019})},\ \Eprint
  {http://arxiv.org/abs/1806.01767}{\tt arXiv:1806.01767 [nucl-ex]}\BibitemShut
  {NoStop}%
\bibitem [{\citenamefont {Thompson}\ \emph {et~al.}(2001)\citenamefont
  {Thompson} \emph {et~al.}}]{Thompson:2000by}%
  \BibitemOpen
  \bibfield  {author} {\bibinfo {author} {\bibfnamefont {R.}~\bibnamefont
  {Thompson}} \emph {et~al.} (\bibinfo {collaboration} {CLAS} Collaboration),\
  }\href {\doibase 10.1103/PhysRevLett.86.1702} {\bibfield  {journal} {\bibinfo
   {journal} {Phys. Rev. Lett.}\ }\textbf {\bibinfo {volume} {86}},\ \bibinfo
  {pages} {1702} (\bibinfo {year} {2001})},\ \Eprint
  {http://arxiv.org/abs/hep-ex/0011029}{\tt arXiv:hep-ex/0011029
  [hep-ex]}\BibitemShut {NoStop}%
\bibitem [{\citenamefont {Denizli}\ \emph {et~al.}(2007)\citenamefont {Denizli}
  \emph {et~al.}}]{Denizli:2007tq}%
  \BibitemOpen
  \bibfield  {author} {\bibinfo {author} {\bibfnamefont {H.}~\bibnamefont
  {Denizli}} \emph {et~al.} (\bibinfo {collaboration} {CLAS} Collaboration),\
  }\href {\doibase 10.1103/PhysRevC.76.015204} {\bibfield  {journal} {\bibinfo
  {journal} {Phys. Rev.}\ }\textbf {\bibinfo {volume} {C76}},\ \bibinfo {pages}
  {015204} (\bibinfo {year} {2007})},\ \Eprint
  {http://arxiv.org/abs/0704.2546}{\tt arXiv:0704.2546 [nucl-ex]}\BibitemShut
  {NoStop}%
\bibitem [{\citenamefont {Dalton}\ \emph {et~al.}(2009)\citenamefont {Dalton}
  \emph {et~al.}}]{Dalton:2008aa}%
  \BibitemOpen
  \bibfield  {author} {\bibinfo {author} {\bibfnamefont {M.~M.}\ \bibnamefont
  {Dalton}} \emph {et~al.},\ }\href {\doibase 10.1103/PhysRevC.80.015205}
  {\bibfield  {journal} {\bibinfo  {journal} {Phys. Rev.}\ }\textbf {\bibinfo
  {volume} {C80}},\ \bibinfo {pages} {015205} (\bibinfo {year} {2009})},\
  \Eprint {http://arxiv.org/abs/0804.3509}{\tt arXiv:0804.3509
  [hep-ex]}\BibitemShut {NoStop}%
\bibitem [{CLA({\natexlab{c}})}]{CLAS:coupsDB}%
  \BibitemOpen
  \href {https://userweb.jlab.org/~mokeev/resonance_electrocouplings/}
  {\enquote {\bibinfo {title} {Nucleon resonances electro- and
  photocouplings},}\ }\bibinfo {note}
  {\href{https://userweb.jlab.org/~mokeev/resonance_electrocouplings/}{\newline
  https://userweb.jlab.org/$\sim$mokeev/
  \mbox{resonance\_electrocouplings/}}}\BibitemShut {NoStop}%
\bibitem [{CLA({\natexlab{d}})}]{CLAS:coups}%
  \BibitemOpen
  \href {https://userweb.jlab.org/~isupov/couplings/} {\enquote {\bibinfo
  {title} {Fits of the resonances electrocouplings},}\ }\bibinfo {note}
  {\href{https://userweb.jlab.org/~isupov/couplings/}{\newline
  https://userweb.jlab.org/$\sim$isupov/couplings/}}\BibitemShut {NoStop}%
\bibitem [{\citenamefont {Tanabashi}\ \emph {et~al.}(2018)\citenamefont
  {Tanabashi} \emph {et~al.}}]{Tanabashi:2018oca}%
  \BibitemOpen
  \bibfield  {author} {\bibinfo {author} {\bibfnamefont {M.}~\bibnamefont
  {Tanabashi}} \emph {et~al.} (\bibinfo {collaboration} {Particle Data Group}
  Collaboration),\ }\href {\doibase 10.1103/PhysRevD.98.030001} {\bibfield
  {journal} {\bibinfo  {journal} {Phys. Rev.}\ }\textbf {\bibinfo {volume}
  {D98}},\ \bibinfo {pages} {030001} (\bibinfo {year} {2018})}\BibitemShut
  {NoStop}%
\bibitem [{\citenamefont {Drechsel}\ \emph {et~al.}(2003)\citenamefont
  {Drechsel}, \citenamefont {Pasquini},\ and\ \citenamefont
  {Vanderhaeghen}}]{Drechsel:2002ar}%
  \BibitemOpen
  \bibfield  {author} {\bibinfo {author} {\bibfnamefont {D.}~\bibnamefont
  {Drechsel}}, \bibinfo {author} {\bibfnamefont {B.}~\bibnamefont {Pasquini}},
  \ and\ \bibinfo {author} {\bibfnamefont {M.}~\bibnamefont {Vanderhaeghen}},\
  }\href {\doibase 10.1016/S0370-1573(02)00636-1} {\bibfield  {journal}
  {\bibinfo  {journal} {Phys. Rept.}\ }\textbf {\bibinfo {volume} {378}},\
  \bibinfo {pages} {99} (\bibinfo {year} {2003})},\ \Eprint
  {http://arxiv.org/abs/hep-ph/0212124}{\tt arXiv:hep-ph/0212124
  [hep-ph]}\BibitemShut {NoStop}%
\bibitem [{\citenamefont {Isupov}\ \emph {et~al.}(2017)\citenamefont {Isupov}
  \emph {et~al.}}]{Isupov:2017lnd}%
  \BibitemOpen
  \bibfield  {author} {\bibinfo {author} {\bibfnamefont {E.~L.}\ \bibnamefont
  {Isupov}} \emph {et~al.} (\bibinfo {collaboration} {CLAS} Collaboration),\
  }\href {\doibase 10.1103/PhysRevC.96.025209} {\bibfield  {journal} {\bibinfo
  {journal} {Phys. Rev.}\ }\textbf {\bibinfo {volume} {C96}},\ \bibinfo {pages}
  {025209} (\bibinfo {year} {2017})},\ \Eprint
  {http://arxiv.org/abs/1705.01901}{\tt arXiv:1705.01901 [nucl-ex]}\BibitemShut
  {NoStop}%
\bibitem [{\citenamefont {Fedotov}\ \emph {et~al.}(2018)\citenamefont {Fedotov}
  \emph {et~al.}}]{Fedotov:2018oan}%
  \BibitemOpen
  \bibfield  {author} {\bibinfo {author} {\bibfnamefont {G.~V.}\ \bibnamefont
  {Fedotov}} \emph {et~al.} (\bibinfo {collaboration} {CLAS} Collaboration),\
  }\href {\doibase 10.1103/PhysRevC.98.025203} {\bibfield  {journal} {\bibinfo
  {journal} {Phys. Rev.}\ }\textbf {\bibinfo {volume} {C98}},\ \bibinfo {pages}
  {025203} (\bibinfo {year} {2018})},\ \Eprint
  {http://arxiv.org/abs/1804.05136}{\tt arXiv:1804.05136 [nucl-ex]}\BibitemShut
  {NoStop}%
\bibitem [{\citenamefont {Trivedi}(2019)}]{Trivedi:2018rgo}%
  \BibitemOpen
  \bibfield  {author} {\bibinfo {author} {\bibfnamefont {A.}~\bibnamefont
  {Trivedi}},\ }\href {\doibase 10.1007/s00601-018-1471-y} {\bibfield
  {journal} {\bibinfo  {journal} {Few Body Syst.}\ }\textbf {\bibinfo {volume}
  {60}} (\bibinfo {year} {2019}),\ 10.1007/s00601-018-1471-y}\BibitemShut
  {NoStop}%
\bibitem [{\citenamefont {Frolov}\ \emph {et~al.}(1999)\citenamefont {Frolov}
  \emph {et~al.}}]{Frolov:1998pw}%
  \BibitemOpen
  \bibfield  {author} {\bibinfo {author} {\bibfnamefont {V.~V.}\ \bibnamefont
  {Frolov}} \emph {et~al.},\ }\href {\doibase 10.1103/PhysRevLett.82.45}
  {\bibfield  {journal} {\bibinfo  {journal} {Phys. Rev. Lett.}\ }\textbf
  {\bibinfo {volume} {82}},\ \bibinfo {pages} {45} (\bibinfo {year} {1999})},\
  \Eprint {http://arxiv.org/abs/hep-ex/9808024}{\tt arXiv:hep-ex/9808024
  [hep-ex]}\BibitemShut {NoStop}%
\bibitem [{\citenamefont {Armstrong}\ \emph {et~al.}(1999)\citenamefont
  {Armstrong} \emph {et~al.}}]{Armstrong:1998wg}%
  \BibitemOpen
  \bibfield  {author} {\bibinfo {author} {\bibfnamefont {C.~S.}\ \bibnamefont
  {Armstrong}} \emph {et~al.} (\bibinfo {collaboration} {Jefferson Lab E94014}
  Collaboration),\ }\href {\doibase 10.1103/PhysRevD.60.052004} {\bibfield
  {journal} {\bibinfo  {journal} {Phys. Rev.}\ }\textbf {\bibinfo {volume}
  {D60}},\ \bibinfo {pages} {052004} (\bibinfo {year} {1999})},\ \Eprint
  {http://arxiv.org/abs/nucl-ex/9811001}{\tt arXiv:nucl-ex/9811001
  [nucl-ex]}\BibitemShut {NoStop}%
\bibitem [{\citenamefont {Laveissiere}\ \emph {et~al.}(2004)\citenamefont
  {Laveissiere} \emph {et~al.}}]{Laveissiere:2003jf}%
  \BibitemOpen
  \bibfield  {author} {\bibinfo {author} {\bibfnamefont {G.}~\bibnamefont
  {Laveissiere}} \emph {et~al.} (\bibinfo {collaboration} {JLab Hall A}
  Collaboration),\ }\href {\doibase 10.1103/PhysRevC.69.045203} {\bibfield
  {journal} {\bibinfo  {journal} {Phys. Rev.}\ }\textbf {\bibinfo {volume}
  {C69}},\ \bibinfo {pages} {045203} (\bibinfo {year} {2004})},\ \Eprint
  {http://arxiv.org/abs/nucl-ex/0308009}{\tt arXiv:nucl-ex/0308009
  [nucl-ex]}\BibitemShut {NoStop}%
\bibitem [{\citenamefont {Sparveris}\ \emph {et~al.}(2005)\citenamefont
  {Sparveris} \emph {et~al.}}]{Sparveris:2004jn}%
  \BibitemOpen
  \bibfield  {author} {\bibinfo {author} {\bibfnamefont {N.~F.}\ \bibnamefont
  {Sparveris}} \emph {et~al.} (\bibinfo {collaboration} {OOPS} Collaboration),\
  }\href {\doibase 10.1103/PhysRevLett.94.022003} {\bibfield  {journal}
  {\bibinfo  {journal} {Phys. Rev. Lett.}\ }\textbf {\bibinfo {volume} {94}},\
  \bibinfo {pages} {022003} (\bibinfo {year} {2005})},\ \Eprint
  {http://arxiv.org/abs/nucl-ex/0408003}{\tt arXiv:nucl-ex/0408003
  [nucl-ex]}\BibitemShut {NoStop}%
\bibitem [{\citenamefont {Kelly}\ \emph {et~al.}(2007)\citenamefont {Kelly}
  \emph {et~al.}}]{Kelly:2005jy}%
  \BibitemOpen
  \bibfield  {author} {\bibinfo {author} {\bibfnamefont {J.~J.}\ \bibnamefont
  {Kelly}} \emph {et~al.},\ }\href {\doibase 10.1103/PhysRevC.75.025201}
  {\bibfield  {journal} {\bibinfo  {journal} {Phys. Rev.}\ }\textbf {\bibinfo
  {volume} {C75}},\ \bibinfo {pages} {025201} (\bibinfo {year} {2007})},\
  \Eprint {http://arxiv.org/abs/nucl-ex/0509004}{\tt arXiv:nucl-ex/0509004
  [nucl-ex]}\BibitemShut {NoStop}%
\bibitem [{\citenamefont {Stave}\ \emph {et~al.}(2008)\citenamefont {Stave}
  \emph {et~al.}}]{Stave:2008aa}%
  \BibitemOpen
  \bibfield  {author} {\bibinfo {author} {\bibfnamefont {S.}~\bibnamefont
  {Stave}} \emph {et~al.} (\bibinfo {collaboration} {A1} Collaboration),\
  }\href {\doibase 10.1103/PhysRevC.78.025209} {\bibfield  {journal} {\bibinfo
  {journal} {Phys. Rev.}\ }\textbf {\bibinfo {volume} {C78}},\ \bibinfo {pages}
  {025209} (\bibinfo {year} {2008})},\ \Eprint
  {http://arxiv.org/abs/0803.2476}{\tt arXiv:0803.2476 [hep-ex]}\BibitemShut
  {NoStop}%
\bibitem [{\citenamefont {Villano}\ \emph {et~al.}(2009)\citenamefont {Villano}
  \emph {et~al.}}]{Villano:2009sn}%
  \BibitemOpen
  \bibfield  {author} {\bibinfo {author} {\bibfnamefont {A.~N.}\ \bibnamefont
  {Villano}} \emph {et~al.},\ }\href {\doibase 10.1103/PhysRevC.80.035203}
  {\bibfield  {journal} {\bibinfo  {journal} {Phys. Rev.}\ }\textbf {\bibinfo
  {volume} {C80}},\ \bibinfo {pages} {035203} (\bibinfo {year} {2009})},\
  \Eprint {http://arxiv.org/abs/0906.2839}{\tt arXiv:0906.2839
  [nucl-ex]}\BibitemShut {NoStop}%
\bibitem [{\citenamefont {Tiator}\ \emph {et~al.}(2011)\citenamefont {Tiator},
  \citenamefont {Drechsel}, \citenamefont {Kamalov},\ and\ \citenamefont
  {Vanderhaeghen}}]{Tiator:2011pw}%
  \BibitemOpen
  \bibfield  {author} {\bibinfo {author} {\bibfnamefont {L.}~\bibnamefont
  {Tiator}}, \bibinfo {author} {\bibfnamefont {D.}~\bibnamefont {Drechsel}},
  \bibinfo {author} {\bibfnamefont {S.~S.}\ \bibnamefont {Kamalov}}, \ and\
  \bibinfo {author} {\bibfnamefont {M.}~\bibnamefont {Vanderhaeghen}},\ }\href
  {\doibase 10.1140/epjst/e2011-01488-9} {\bibfield  {journal} {\bibinfo
  {journal} {Eur. Phys. J. ST}\ }\textbf {\bibinfo {volume} {198}},\ \bibinfo
  {pages} {141} (\bibinfo {year} {2011})},\ \Eprint
  {http://arxiv.org/abs/1109.6745}{\tt arXiv:1109.6745 [nucl-th]}\BibitemShut
  {NoStop}%
\bibitem [{\citenamefont {Aznauryan}\ and\ \citenamefont
  {Burkert}(2012{\natexlab{b}})}]{Aznauryan:2012ec}%
  \BibitemOpen
  \bibfield  {author} {\bibinfo {author} {\bibfnamefont {I.~G.}\ \bibnamefont
  {Aznauryan}}\ and\ \bibinfo {author} {\bibfnamefont {V.~D.}\ \bibnamefont
  {Burkert}},\ }\href {\doibase 10.1103/PhysRevC.85.055202} {\bibfield
  {journal} {\bibinfo  {journal} {Phys. Rev.}\ }\textbf {\bibinfo {volume}
  {C85}},\ \bibinfo {pages} {055202} (\bibinfo {year} {2012}{\natexlab{b}})},\
  \Eprint {http://arxiv.org/abs/1201.5759}{\tt arXiv:1201.5759
  [hep-ph]}\BibitemShut {NoStop}%
\bibitem [{\citenamefont {Mokeev}\ and\ \citenamefont
  {Aznauryan}(2014)}]{Mokeev:2013kka}%
  \BibitemOpen
  \bibfield  {author} {\bibinfo {author} {\bibfnamefont {V.~I.}\ \bibnamefont
  {Mokeev}}\ and\ \bibinfo {author} {\bibfnamefont {I.~G.}\ \bibnamefont
  {Aznauryan}},\ }\href {\doibase 10.1142/S2010194514600805} {\bibfield
  {journal} {\bibinfo  {journal} {Int. J. Mod. Phys. Conf. Ser.}\ }\textbf
  {\bibinfo {volume} {26}},\ \bibinfo {pages} {1460080} (\bibinfo {year}
  {2014})},\ \Eprint {http://arxiv.org/abs/1310.1101}{\tt arXiv:1310.1101
  [nucl-ex]}\BibitemShut {NoStop}%
\bibitem [{\citenamefont {Tvaskis}\ \emph {et~al.}(2018)\citenamefont {Tvaskis}
  \emph {et~al.}}]{Tvaskis:2016uxm}%
  \BibitemOpen
  \bibfield  {author} {\bibinfo {author} {\bibfnamefont {V.}~\bibnamefont
  {Tvaskis}} \emph {et~al.},\ }\href {\doibase 10.1103/PhysRevC.97.045204}
  {\bibfield  {journal} {\bibinfo  {journal} {Phys. Rev.}\ }\textbf {\bibinfo
  {volume} {C97}},\ \bibinfo {pages} {045204} (\bibinfo {year} {2018})},\
  \Eprint {http://arxiv.org/abs/1606.02614}{\tt arXiv:1606.02614
  [nucl-ex]}\BibitemShut {NoStop}%
\bibitem [{\citenamefont {Altarelli}\ and\ \citenamefont
  {Martinelli}(1978)}]{Altarelli:1978tq}%
  \BibitemOpen
  \bibfield  {author} {\bibinfo {author} {\bibfnamefont {G.}~\bibnamefont
  {Altarelli}}\ and\ \bibinfo {author} {\bibfnamefont {G.}~\bibnamefont
  {Martinelli}},\ }\href {\doibase 10.1016/0370-2693(78)90109-0} {\bibfield
  {journal} {\bibinfo  {journal} {Phys. Lett.}\ }\textbf {\bibinfo {volume}
  {76B}},\ \bibinfo {pages} {89} (\bibinfo {year} {1978})}\BibitemShut
  {NoStop}%
\bibitem [{\citenamefont {Markov}\ \emph {et~al.}(2018)\citenamefont {Markov},
  \citenamefont {Joo}, \citenamefont {Ungaro}, \citenamefont {Smith},\ and\
  \citenamefont {Mokeev}}]{Markov:2018loh}%
  \BibitemOpen
  \bibfield  {author} {\bibinfo {author} {\bibfnamefont {N.}~\bibnamefont
  {Markov}}, \bibinfo {author} {\bibfnamefont {K.}~\bibnamefont {Joo}},
  \bibinfo {author} {\bibfnamefont {M.}~\bibnamefont {Ungaro}}, \bibinfo
  {author} {\bibfnamefont {L.~C.}\ \bibnamefont {Smith}}, \ and\ \bibinfo
  {author} {\bibfnamefont {V.}~\bibnamefont {Mokeev}},\ }\href {\doibase
  10.1007/s00601-018-1448-x} {\bibfield  {journal} {\bibinfo  {journal} {Few
  Body Syst.}\ }\textbf {\bibinfo {volume} {59}},\ \bibinfo {pages} {134}
  (\bibinfo {year} {2018})}\BibitemShut {NoStop}%
\end{thebibliography}%
\end{document}